%% file: main.tex
\documentclass{aastex631}

\usepackage{natbib}
\usepackage{color}
\usepackage{apjfonts}
\usepackage{xfrac}
\usepackage{ulem}
\usepackage{soul}
\usepackage{amsmath,xspace}
\usepackage{hyperref}
\hypersetup{linkcolor=red,citecolor=blue,filecolor=cyan,urlcolor=magenta}

\usepackage{graphicx}
\usepackage[space]{grffile}
\usepackage{latexsym}
\usepackage{textcomp}
\usepackage{longtable}
\usepackage{CJKutf8}

\usepackage{multirow,booktabs}
\usepackage{amsfonts,amsmath,amssymb}
\usepackage{natbib}
\usepackage{url}
\usepackage[T1]{fontenc}
\usepackage{ae,aecompl}

\newif\iflatexml\latexmlfalse
\DeclareGraphicsExtensions{.pdf,.PDF,.png,.PNG,.jpg,.JPG,.jpeg,.JPEG}

\usepackage[utf8]{inputenc}
\usepackage[english]{babel}
\usepackage{graphicx}

\newcommand{\kms}{${\rm km\, s^{-1}}$}

\newcommand{\Ntargets}{11,554}
\newcommand{\Ngoodspec}{11,416} 
\newcommand{\Nstars}{10,414} 
\newcommand{\NAnd}{7,527} 
\newcommand{\NAndStars}{7,438} 
\newcommand{\NAndHIIPN}{43} 
\newcommand{\Nclusters}{136} 
\newcommand{\NQSO}{184} 
\newcommand{\Ngalaxies}{683} 
\newcommand{\NVI}{3,150} 

\def\update#1{\noindent{\color{black}\bf #1 }}

\shorttitle{DESI Observations of M31}
\shortauthors{DESI MWS Team}

\begin{document}
\begin{CJK*}{UTF8}{gbsn}

\title{DESI Observations of the Andromeda Galaxy: Revealing the Immigration History of our Nearest Neighbor}

\input{author_list}


\begin{abstract}
We present Dark Energy Spectroscopic Instrument (DESI) observations of the inner halo of M31, which reveal the kinematics of a recent
merger—a galactic immigration event—in exquisite detail. Of the 11,416 sources studied in 3.75~hr
of on-sky exposure time, 7,438 are M31 sources with well-measured radial velocities. The observations
reveal intricate coherent kinematic structure in the positions and velocities of individual stars: streams,
wedges, and chevrons. While hints of coherent structures have been previously detected in M31, this
is the first time they have been seen with such detail and clarity in a galaxy beyond the Milky Way. 
We find clear kinematic evidence for shell structures in the Giant Stellar Stream, the Northeast Shelf and Western Shelf regions. 
The kinematics are remarkably similar to the predictions of dynamical models constructed to explain the spatial morphology of the inner halo. 
The results are consistent with the interpretation that much of the substructure in the inner halo of M31 is produced
by a single galactic immigration event 1–2 Gyr ago. Significant numbers of metal-rich stars ([Fe/H]~$>-0.5$) are present
in all of the detected substructures, suggesting that the immigrating galaxy had an extended star
formation history. We also investigate the ability of the shells and Giant Stellar Stream to constrain
the gravitational potential of M31, and estimate the mass within a projected radius of 125 kpc to be
$\log_{10}\, M_{\rm NFW}(<125\,{\rm kpc})/M_\odot = 11.80_{-0.10}^{+0.12}$.
The results herald a new era in our ability to study stars on a
galactic scale and the immigration histories of galaxies.

\end{abstract}

\section{Introduction}
\label{sec:introduction}


The histories of galaxies 
have much in common with 
that of the United States: in both cases, waves of immigration (of stars, people) have added to the existing inhabitants. In the process of galaxy assembly, smaller galaxies are expected to fall into larger galaxies and disperse their stars in a hierarchical merging process \citep{bullock2001,bullock_johnston_2005,Cooper2010}. 
How do we know this? In the case of immigration to the US, numerous documents, such as government records, can be used to reconstruct the historical movements of individuals and therefore large-scale migration patterns. 
Although no such records are available for galaxies, we can nevertheless reconstruct their immigration histories from the motions of their individual stars.
Migrating stars merge into galaxies on cosmic timescales and we can expect to observe stars on their migration paths today; the record of their immigration ancestry preserved in phase space even for migration events that began billions of years ago. 
Discerning migration events (i.e., to identify coherent structure in the positions and motions of stars on galactic scales) requires measurements of large stellar samples over large areas. 
Previously prohibitive, such studies are now straightforward with the advent of 
highly multiplexed multi-object spectroscopy on telescopes with wide fields of view. 

M31, our closest large galactic neighbor, has a mass comparable to that of the Milky Way. Our location in the Milky Way  
offers a fortuitous vantage point from which to observe galactic migration in action in M31. 
While the Milky Way gives us 
an up-close (``on-stage'') view of 
the dynamics 
of a large spiral galaxy, our position within the disk of the Milky Way obscures large portions of the Galaxy from our view. In contrast, with our external (``upper balcony'') perspective on M31, it is straightforward to survey the entire galaxy for clues to its 
immigration history. 


The expected observational signatures of galactic migration include 
debris streams, shells, rings, and plumes, the expected outcomes of 
merger interactions 
between large galaxies and their companions
%
\citep[e.g.,][]{bullock2001,bullock_johnston_2005,McConnachie2009,Cooper2010,Martinez-Delgado2010,Pop2018}.
The detailed kinematic study of these features can help us reconstruct the assembly history of a galaxy as well as enable dynamical measurements of its mass distribution \citep[e.g.,][]{Merrifield1998,Ibata2004}.

Both the Milky Way and M31 show signs of mergers. 
Photometric and kinematic studies of the Milky Way reveal complex substructure suggesting that the vast majority of the stars in the halo may have been accreted in past mergers \citep{bell2008,schlaufman_etal_2012,Naidu_etal_2020} with the inner halo dominated by one or more massive satellite galaxies 
that happened 
more than 8~Gyr ago \citep{Belokurov2018,Helmi2018,Gallart2019,Bonaca2020,Kraken2020,XiangRix2022}. In addition, the Milky Way is currently in the process of assimilating the Sagittarius Dwarf Galaxy, a merger that had its first passage through the Milky Way disk about 5.7~Gy ago \citep[e.g.,][]{Ibata1994,Ruiz-Lara2020}.


Similarly, photometric observations of the M31 stellar halo suggest that our large neighbor has had a complex merger history: its halo shows a high degree of asymmetry, with spatially and chemically coherent structures spread out over its entire extent \citep[e.g.,][]{Ibata2004,Ferguson2016,McConnachie2018}. In particular, the inner halo of M31 contains prominent tidal features, including the Giant Stellar Stream \citep[GSS;][]{Ibata2001,Ibata2004}, which extends 100~kpc to the southeast, and the Northeast and Western Shelves---diffuse but sharp-edged, fan-shaped extensions to the Northeast and West of the center of M31 respectively \citep[e.g.,][]{Ferguson2016}, 
structures that have been interpreted as tidal debris from a companion galaxy that merged with M31 relatively recently  \citep[e.g.,][]{Ibata2004, font2006, Fardal2006, Fardal2007, Fardal2008, Fardal2012, Fardal2013, Mori2008, Sadoun2014, Hammer2018, dsouza2018, Kirihara2017a,Milosevic2022}.


Spectroscopy of individual stars can greatly enhance our ability to identify migration patterns through the measurement of line-of-sight radial velocities and metallicities.
The disk and halo of 
M31 have been the focus of numerous spectroscopic studies, especially over the last two decades. Most studies of the M31 halo have used the DEIMOS instrument \citep{DEIMOS} on the Keck II Telescope for pencil-beam surveys in various 
regions, catching tantalizing glimpses of complex kinematic structure. These studies have determined that the GSS is a relatively metal rich \citep[{[Fe/H]$\approx-0.8$}; e.g.,][]{Gilbert2019,Gilbert2020,Escala2021}, kinematically cold feature \citep[velocity dispersion $11\pm3$~km/s; e.g.,][]{Ibata2004} within a larger metal poor halo, 
and have 
revealed an additional cold velocity structure in the region of the GSS \citep{Kalirai2006a,Gilbert2009b}. 
The kinematics of the GSS have been used to estimate an enclosed total galaxy mass of $7.5\times 10^{11} M_\odot$ within 125~kpc \citep{Ibata2004}.
%
The average metallicity of the M31 halo (often derived photometrically)
appears to decrease with radius \citep{Kalirai2006b,Ibata2014,Gilbert2014,Gilbert2020,Escala2021}, suggesting that much of the inner stellar halo is a mixture of relatively more metal-rich accreted satellite galaxies into the underlying, more metal-poor halo.
The kinematically cold substructures like the GSS are found to be more metal rich than the surrounding dynamically hot stellar population 
\citep{Gilbert2019}, which can be understood if they are produced by fairly massive (and therefore metal-rich) progenitors.





The Dark Energy Spectroscopic Instrument \citep[DESI;][]{DESI_Instr_Overview2022} on the Mayall 4m telescope at KPNO provides a unique opportunity to advance our understanding of the M31 system. DESI's $3.2^\circ$ diameter field of view and high multiplex capability ($\approx$5000 fibers) are 
well matched to the density on the sky of 
the brightest constituents of M31's inner halo: its asymptotic giant branch (AGB) stars, those at the tip of the red giant branch, and luminous blue stars, stellar clusters, HII regions, and planetary nebulae. Here we present new DESI observations of $\sim$11,000 stars towards M31 that clearly demonstrate that high-quality stellar kinematics can be acquired efficiently over the wide field of view needed to provide unique insights into the migration history of this galaxy.

This paper is organized as follows. In \S~\ref{sec:data} we describe the M31 observations and the pipeline reductions. In \S~\ref{sec:results} we present the position-velocity data for the observed sources, revealing  complex kinematic structures. We also provide a brief description of the stellar spectroscopic metallicity measurements and their spatial distribution. In \S~\ref{sec: simulations}, we compare our observations to results from cosmological simulations and explore a more tailored N-body model that demonstrates that much of the observed kinematic structure can result from a single encounter. In \S~\ref{sec:MassConstraints}, we use the observations to constrain the mass of the M31 system.
In \S~\ref{sec:discussion} we compare our results to those of previous studies and model predictions, discuss the nature of the progenitor galaxy responsible for the observed kinematic substructure and the constraints we can place on the mass of M31 from these data. We present our conclusions in \S~\ref{sec:conclusion}. The Appendices present tables of the redshifts of 
non-M31 sources, i.e., higher-redshift galaxies and Milky Way stars, measured by our DESI observations. 
Throughout this paper we adopt the M31 line-of-sight velocity of $-$300~\kms\ (based on the value of $-$300$\pm$4~\kms\ reported by \citealt{McConnachie2012}; see also the Third Reference Catalog of Bright Galaxies, \citealt{rc3}), and a distance to M31 of 785$\pm$25~kpc \citep{McConnachie2005}, which results in a scale of $\approx13.7$~kpc/deg. We assume that the galaxy disk is centered at 
(RA,Dec) =  (10.6847$^\circ$, 41.26875$^\circ$) and viewed at an inclination of 77$^\circ$ to the line of sight and at a sky position angle of PA=$38^\circ$
\citep[see, e.g.,][]{WalterbosKennicutt1987,mackey2019}. 
We define the ellipse containing the disk of M31 to have semi-major and semi-minor axes of 1.5$^\circ$ and 0.337$^\circ$ respectively. 
While heliocentric velocities are presented in the tables, in all figures and discussion we convert all velocities to the Galactic Standard of Rest (GSR; assumes a Solar velocity of [12.9, 245.6, 7.78]~\kms\ in [x,y,z]) and also reference velocities to a M31-centric frame by adding 113.656~km~s$^{-1}$ (i.e., the equivalent of adding 300~\kms\ to their heliocentric velocities). 


\section{Observations and Data}
\label{sec:data}

\subsection{Target Selection}
\label{sec:targetselection}

The goal of this initial short M31 campaign with DESI, a fiber-fed spectrograph on a 4m diameter telescope, was to determine whether the instrument was capable of measuring stellar radial velocities and spectroscopic metallicities for M31 halo stars. 
Since the DESI Legacy Imaging Surveys \citep[hereinafter LS;][]{Dey2019} in the South Galactic Cap only extends south of Dec $\lesssim33^\circ$ and does not include the region around M31, our primary target selection was based on the source catalogs from the PAndAS survey \citep{McConnachie2018}, 
$g$- and $i$-band survey covering a $>400$~deg$^2$ region around M31 and M33, which we cross matched with the {\it Gaia} DR2 \citep{gaia,GaiaDR2summary}, and CatWISE2020 \citep{catwise2021} catalogs. While the PAndAS data contain $>10\sigma$ photometry for stars to $g\approx25, i\approx24$, our target selection was restricted to stars brighter than $z=21.5$~mag to ensure measurements of sufficient signal-to-noise ratio in about 90~min of effective exposure time with DESI. Since the PAndAS catalog does not include $z$-band measurements (which were needed for estimating exposure times and target selection), we constructed an estimate of the DESI Legacy Imaging Surveys $z$-band magnitude using the following relation:
\begin{align}
    (g-z) \,= \,0.15\,{\rm max}((g-i) - 1.8, 0)\, +\, 2.21\,  + \, 1.27 \, ((g-i)-1.8)
\end{align} 
This relation was derived by cross-matching point sources in PAndAS and LS DR9 in the region where they overlap and fitting a broken linear function to $(g-z)$ vs $(g-i)$. The [16,84] percentiles of the residuals in ($g-z$) are [$-0.05,0.11$] mag for sources with $i\le 21$~mag.

As M31 is centered at a Galactic latitude of $b=-21.6^\circ$ and this work targets relatively bright stars with $z<21$, the main contamination to stellar target samples is from Milky Way disk and halo stars. Prior spectroscopic surveys have primarily selected targets using colors in the region spanned by the red giant branch (RGB) isochrones at the distance of M31, and using photometry in the DDO51 intermediate band filter to separate M31 red giants from Milky Way dwarf stars \citep[e.g.,][]{Guhathakurta2006}. 
The resulting samples tend to have $(V-I) \lesssim 2$, which is generally appropriate given the location of the metal poor ([Fe/H]~$\lesssim-0.5$) isochrones that define the bulk of the M31 halo. 


Our 
target selection for M31 stars took a different approach, where we primarily focused on maximizing the number of targeted M31 stars with the help of machine-learning-driven classification.
%
We 
constructed  separate selections based on Random Forest classifications optimized for:
\begin{itemize}
    \item a bright ($z<19$~mag) M31 disk selection (M31 Disk Bright); 
    \item a faint ($19\le z\lesssim 21.5$~mag) M31 disk selection (M31 Disk Faint);  and
    \item a faint ($z\lesssim 21.5$~mag) halo selection, tuned to select targets in the Giant Stellar Stream (M31 Stream Faint).
    \end{itemize}

The Random Forest classification \citep{Breiman2001} approach uses an ensemble of decision trees  constructed from training data. 
Our classification relies on the following inputs: $g$ and $i$ photometry from the PAndAS catalog; the proper motion (PMRA, PMDEC, PMRA\_ERROR, PMDEC\_ERROR), parallax (PARALLAX, PARALLAX$\_$ERROR), and photometric (PHOT\_G\_MEAN\_MAG, PHOT\_BP\_MEAN\_MAG, PHOT\_RP\_MEAN\_MAG) data from {\it Gaia} DR2; 
along with the {\it WISE} W1 and W2 photometry (W1MPRO, W2MPRO) from the CatWISE2020 catalogs. When these quantities were unavailable (i.e., for sources too faint for {\it Gaia} or {\it WISE}), 
placeholder values were used (i.e., 99.99). We did not use the PAndAS morphology flags in the Random Forest selection. 

Each classifier is trained on a set of stars labeled as either an M31 member or a background/foreground star.
Since we do not have an unambiguous classification for every star (as an M31 member or non-member) we use a statistical decontamination approach. 
Specifically we consider two areas around M31, one centered on an object of interest (i.e., the disk or the GSS), and another far enough away that it would not have many M31 stars. We then remove the (likely) MW contaminants from 
the first field  by picking a nearest neighbor in data-space for each star in the background field (with appropriate scaling to the areas of the field).
We are left with a list of objects that are quite likely M31 members in the first field, and background stars in the second field.
This provides us with the training set for the random forest.
We use a standard cross-validation technique to choose the best tuning parameters of the random forest classifier (such as the tree depth and minimum leaf sizes) and obtain the probabilities for each star that it belongs to
M31, $P_{\rm M31}$.
We then select targets with $P_{\rm M31}>P_{\rm cut}$, where the minimum probability $P_{\rm cut}$ is chosen to ensure a high enough target density to match the DESI fiber density.

While the Random Forest results in a fairly complex selection, most of the faint ($z>19$ mag) targets are approximately bounded by the polygon defined by the points
$((g-i),i)$=([2.0, 2.4, 3.05, 4.0, 4.0, 2.0], [22.0, 21.67, 21.67, 20.8, 22.9, 22.0]). 

\begin{figure}[ht]
    \centering
    \includegraphics[width=\textwidth]{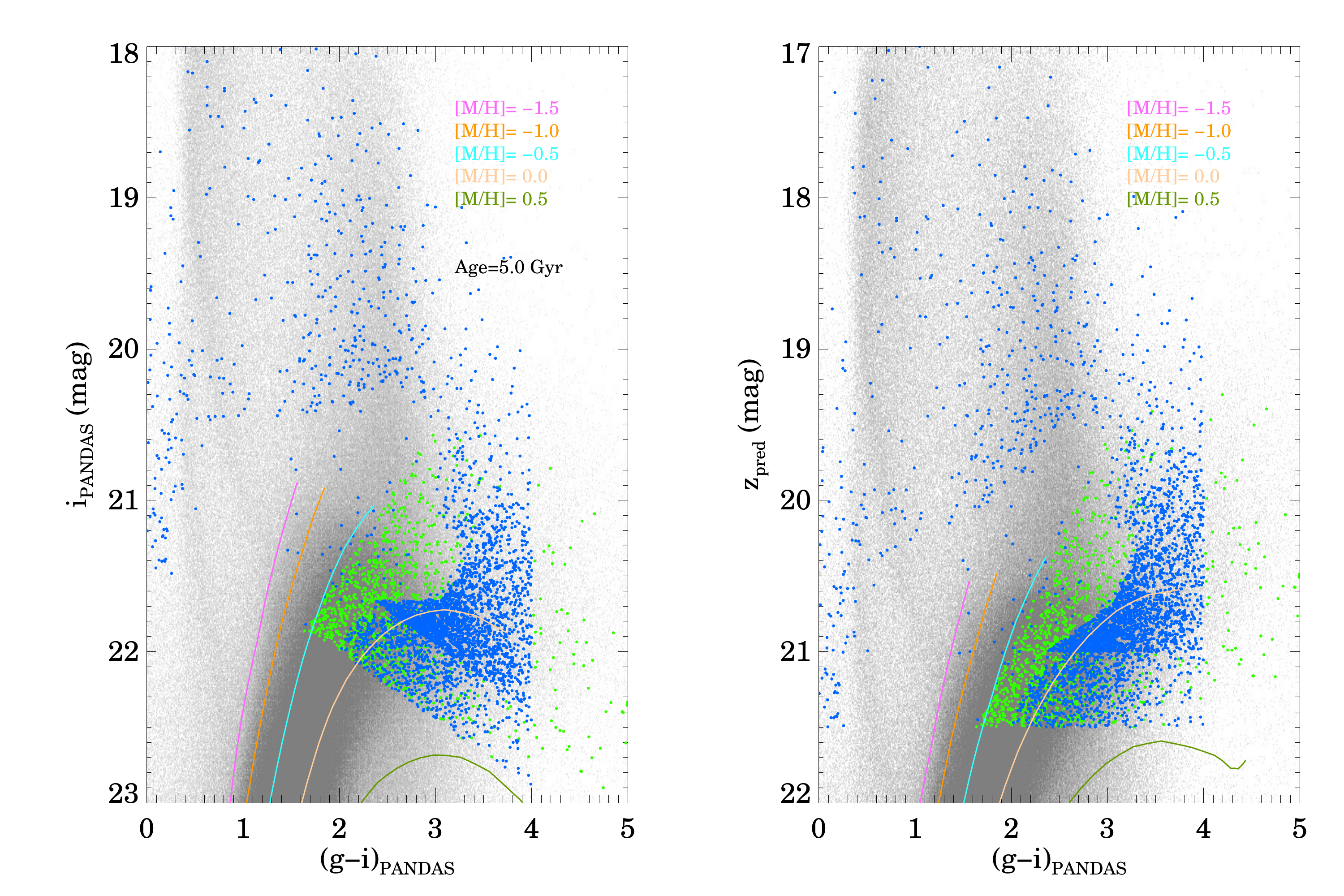}
    \caption{Color-magnitude diagrams for the targets chosen for spectroscopy. The background greyscale in both panels is all stars from the PAndAS catalog within 5$^\circ$ of M31. Representative targets derived from the random forest selection are shown by blue dots and ones chosen by the backup selection are the green dots. The colored lines denote the positions of the PARSEC isochrones \citep{PARSEC2012,PARSEC2013} for an age of 5~Gyr and for metallicities [M/H]=[$-1.5,-1,-0.5,0,+0.5$] at the distance of M31.}
    \label{fig:colormagsel}
\end{figure}

\begin{figure}
    \centering
    \includegraphics{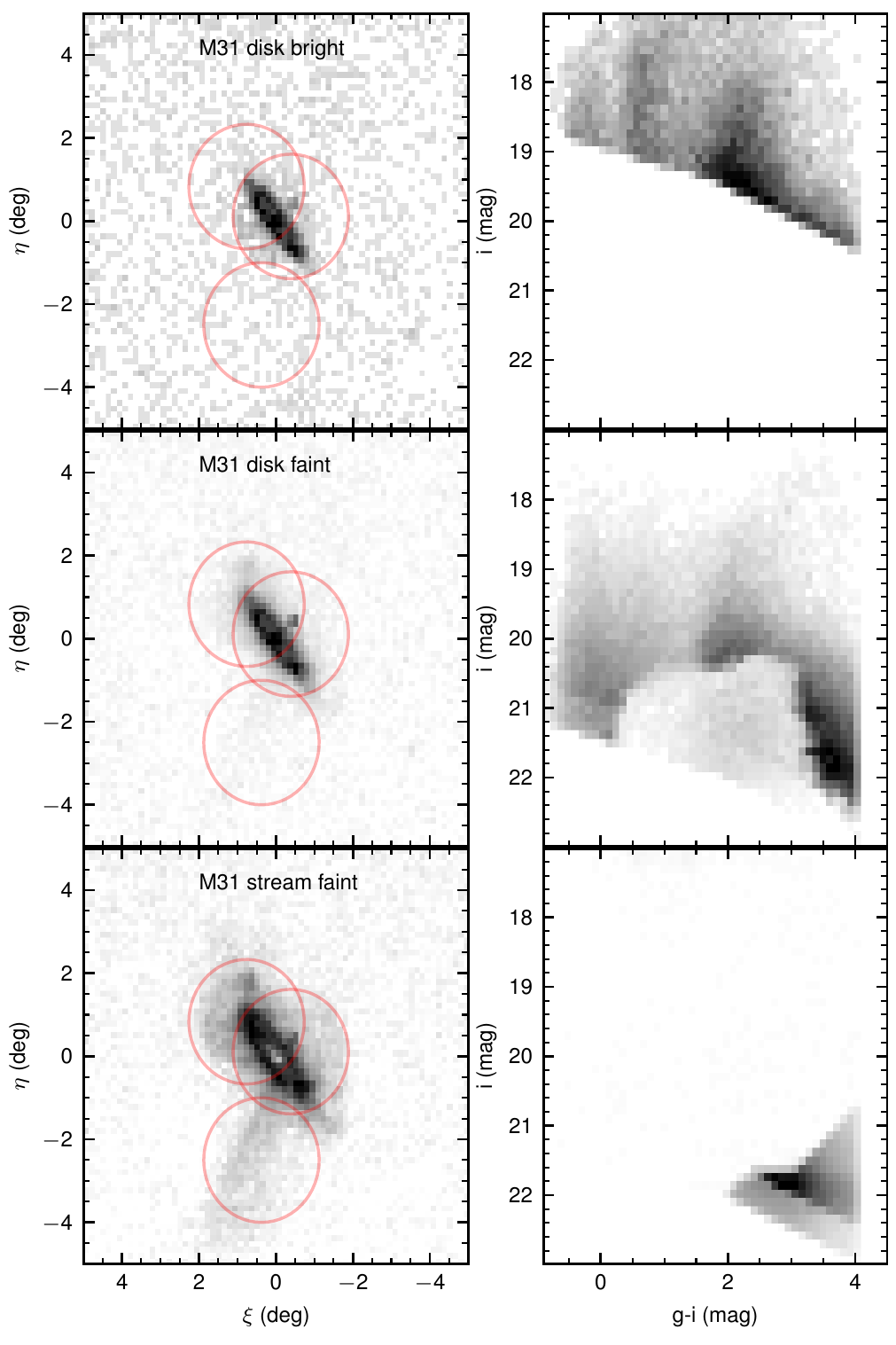}
    \caption{The selection of the M31 targets in our 3 main groups (M31 Bright, M31 Faint and M31 Stream Faint). The left panels show the density of selected targets on the sky. The red circles show the field locations. The right panels show the colour-magnitude distribution of the targets in the form of a Hess diagram.}
    \label{fig:colormagsel1}
\end{figure}

The resulting samples for the halo and disk are shown by the filled circles in Figure~\ref{fig:colormagsel}. Our selection is biased to redder regions in $(g-i)$ relative to the selections used by the previous Keck/DEIMOS campaigns. We therefore sample primarily the metal-rich and older RGB and redder AGB stars and do not sample the metal-poor regions well. Despite this bias, the Random Forest approach is ``optimal'' in the sense of minimizing the contamination by Milky Way stars and background galaxies. 
Figure~\ref{fig:colormagsel1} shows the density of the selected sources on the sky as well as color-magnitude distributions for the three selections (M31 Disk Faint, M31 Disk Bright, and M31 Stream Faint). 

In the outer regions of the M31 halo, the Random Forest selection results in a target density that underfills the DESI fibers. Hence, we supplemented the Random Forest selection with a simple selection to define backup, or filler targets:
$$z\le 21.5\quad {\rm\ and}\quad 20.5 \le i\le 24.5 $$
$$[23.5-(g-i)] \le i \le [14.5+5(g-i)]$$
$$(g-i) \le 5.0$$

In the disk field (which was originally selected for DESI first light observations), the filler targets included known bright targets---HII regions, planetary nebulae (PNe), globular clusters, luminous blue variables (LBVs)---many of which have spectroscopic information from past studies and can be used as a check on the DESI radial velocities. 
HII region and PNe sources were selected from the compilation of \citet{Sanders2012}. 
Globular cluster candidate sources were selected from the compilation of \citet{mackey2019} and from Version 5 of the Revised Bologna Catalog \citep[RBCv5;][]{galleti2007,galleti2014cat}. 
A small number of bright variable sources identified in the Zwicky Transient Survey Catalogs using the ANTARES time-domain event broker \citep{Matheson2021} were also included, as were bright sources from the SPLASH survey \citep{Guhathakurta2006,Dorman2012,Dorman2015}.
In the 
M31 halo, the existing spectroscopy at magnitudes DESI can reach ($z\lesssim 21.5$~mag) is more limited, but we included all known cluster and variable sources as potential targets.

Finally, we complemented the list of M31 targets with background QSO candidates selected using data from the {\it WISE} and {\it Gaia} satellites. Background QSOs are invaluable probes of the interstellar and circumgalactic medium around galaxies, and all prior studies have only yielded confirmed redshifts for $\sim100$ QSOs. We used a simple {\it WISE} selection (described in Appendix~\ref{appendix:qso}) to select bright ($G\le20.5$~mag) QSOs (with a sky surface density of $\approx1.8~{\rm deg}^{-2}$) around M31. We vetted this selection using spectroscopically confirmed QSOs from the study of \cite{massey2019} and the LAMOST surveys \citep{Huo2010,Huo2013,Huo2015}.

All these targets were prepared were assigned unique TARGETIDs and prepared for inclusion in the DESI Secondary Target Program. The technical details of DESI target selection, such as the unique TARGETID associated with a target, the different phases of DESI targeting, and how targeting bits can be used to isolate targets from different DESI programs are described in \citet{myers22a}.

\subsection{Observations}
\label{sec:observations}

DESI is a wide-field, fiber-fed multi-object spectroscopic instrument mounted on the Mayall 4m Telescope of the Kitt Peak National Observatory. With a 3.2$^\circ$ diameter field of view populated by 5020 robotically positioned fibers, DESI offers an unprecedented (and currently unmatched) capability for wide-field astrophysical surveys. Details of the DESI instrument, operational plan, and science mission are presented in \citet{DESI_Tech_FDR,DESI_Science_FDR} and 
\citet{DESI_Instr_Overview2022}. Briefly, the  $\approx$1.5~arcsec diameter DESI fibers feed ten three-arm spectrographs which provide continuous coverage over the wide wavelength range 3600\AA\ to 9800\AA\ with a resolving power $R\equiv\lambda/\Delta\lambda$ varying from $\approx$2000 in the blue to 5500 in the red. The three spectrograph arms span the wavelength ranges 3600--5930\AA\ (blue or B), 5600--7720\AA\ (red or R), and 7470--9800 (near-IR, hereafter NIR or Z). DESI is very efficient: its total system throughput varies from 20\% at 3800\AA\ to nearly 50\% at 8500\AA\ (not including fiber aperture losses or atmospheric extinction) and has an overhead of less than 2~minutes between exposures \citep[for details see][]{DESI_Instr_Overview2022}. Technical details of DESI operations, such as the unique TILEID associated with a tile (i.e., a specific fiber assignment configuration centered at a given sky location), and how DESI observations are planned and proceed, are detailed in \citet{schlafly22a}.


DESI ``tiles'' were constructed, each incorporating targets from all three selections described above. DESI observations of M31 were obtained in 2021 January (TILEIDs 80713 and 80715, covering the optical disk of the galaxy) and 2022 January (TILEID 82634, positioned on the Giant Stellar Stream; and 82635, targeting the Northeast Shelf; see Table~\ref{tab:desiobs}). 
The 2021 January data (on M31's disk) were taken during the early Survey Validation phase of DESI observations, when the instrument was not fully operational and observing procedures were being tested. Tile 80713 was observed on the night of 2021 January 10 in mediocre observing conditions for an effective exposure time\footnote{The DESI effective exposure time corresponds to the time required to reach the observed signal-to-noise ratio under the ``standard’’ observing conditions of a dark sky with ideal transparency and median seeing of 1.1\arcsec\ at an airmass of 1.0 (i.e., at zenith). See \citet{Guy2022} for details.}
of $t_{\rm eff} = 758$~sec, but the bulk of the fibers were not positioned correctly due to a bug, and the observation resulted in usable spectra for only 730 targets. The tile was redesigned (with all the same targets) as TILEID 80715, and successfully observed on the night of 2021 January 15 for $t_{\rm eff}=1906$~sec.  During these observations, Petal \# 3 (i.e., the 36$^\circ$ pie-shaped focal-plane wedge containing 500 fibers spanning the position angle range $270^\circ<PA<306^\circ$) was non-functional. As a result, no data were obtained on a portion 
of the Western Shelf region of the M31 inner halo during these observations.


DESI observed the tile centered on the Giant Stellar Stream (TILEID=82634) on the night of 2022 January 3. These observations were obtained under excellent conditions: dark, clear skies with seeing of 1\arcsec, and an effective exposure time of 1.5 hr was reached in 63~minutes. The tile centered on the Northeast Shelf (TILEID=82635) was observed on the nights of 2022 January 21 and 2022 January 27, under somewhat poorer conditions. 

In summary, DESI observed a total of three tiles with a total effective exposure time of $\approx 3.75$~hr. The 
tiles 82634 and 82635 were each observed for an effective time of $\approx$1.5~hr. 

\begin{deluxetable}{clcccccl}
\tablecaption{DESI Observations of M31\tablenotemark{a}}
\label{tab:desiobs}
\tablehead{
\colhead{Obs Date}  &  
\colhead{Tile ID} & 
\colhead{RA$_{\rm cen}$} & 
\colhead{DEC$_{\rm cen}$} & 
\colhead{Exposure Time} & 
\colhead{$t_{\rm eff}$} & 
\colhead{$N_{\rm targ}$} & 
\colhead{Comments}
}
\startdata
    2021-01-10  & 80713 & 10.170 & +41.380 & 2700 & 758 & 730\tablenotemark{b} &  Petal 3 non-functional; limited fiber reach \\
    2021-01-15  & 80715\tablenotemark{c} & 10.170 & +41.380 & 2700 & 1906 & 3130 & Petal 3 non-functional; limited fiber reach\\
    2022-01-03  & 82634 & 11.185 & +38.768 &  3809 & 5400 & 4215  & \\
    2022-01-21  & 82635 & 11.700 & +42.100 & 2960 & 1800 & 4263  & \\
    2022-01-27  & 82635 & 11.700 & +42.100 & 5003 & 3600 & 4263  & \\
\enddata
\tablenotetext{a}{Columns are: (1) Local date (at Kitt Peak) for the observation (in yyyy-mm-dd format); (2) DESI Tile Identification Number; (3,4) RA and Dec in J2000 for the center of the tile; (5) the on-sky actual exposure time; (6) the effective exposure time; (7) the number of sources successfully targeted during the observation; (8) comments on the observation.}
\tablenotetext{b}{The bulk of fibers in these observations did not reach their targets because of an error in the fiber assignment file.}
\tablenotetext{c}{Tile ID 80715 is a duplicate of 80713 (i.e., identical targets), with the errors in the 80713 assignment file corrected.}
\end{deluxetable}

\subsection{Data Reduction}
\label{sec:datareduction}

 
 The data were processed using the standard initial data reduction pipeline corresponding to the internal data release ``Fuji''
 \citep{Guy2022}.
 There were however several modifications required to process the M31 data. Initially the targeting for tile 80713 did not have correctly identified flux calibration standards as it was located outside the LS footprint. As a result, it could not be processed with the default DESI pipeline parameters, and we therefore manually identified a set of stars as flux standards through color-magnitude selection in {\it Gaia} G/BP/RP bands and provided the TARGETIDs of these new flux standards to the spectroscopic pipeline. Subsequent to the first observations of the 80713 tile, DESI targeting is now able to correctly deal with fields outside the LS footprint and the standards are selected purely through {\it Gaia} photometry, with no custom flux calibration standards needed.
 
We visually inspected the spectra using the ``Prospect'' tool\footnote{\url{https://github.com/desihub/prospect}} created by E. Armengaud \citep[for further details please see][]{alexander22a,lan22a}. An initial visual inspection (VI) revealed that spectra with low quality flags (i.e., $0\le {\rm VI\_QUALITY} \le 2$) are located near the disk of the galaxy where the sky subtraction is poor due to the sky fibers being contaminated by emission lines and continuum light from the M31 disk. While DESI observations typically reserve 50-100 fibers for sky observations (``sky fibers''), the pipeline can successfully subtract the sky with minimal additional noise or systematic issues using as few as 10 sky fibers. 
We therefore examined each of the sky fibers, identified ones with the lowest median flux\footnote{We  selected sky fibers which satisfied $(\bar{s}_i - \bar{s}_{<70})/\bar{s}_{<70} \le 0.2$, where $\bar{s}_i$ is the median value of the sky in sky fiber $i$ measured in the wavelength region $\lambda\lambda6000-7000$\AA, and  $\bar{s}_{<70}$ is the similarly measured median sky value measured across all the sky fibers after rejecting the 30\% of the fibers with the highest skies. This procedure resulted in $\ge10$ sky fibers per petal which could be used for sky subtraction.}, and then reran the pipeline reductions using this subset. This re-reduction corrected the bulk of the problems with the sky subtraction. 
 
After the initial pipeline data reduction, the data were then processed through the redshift and stellar radial velocity/parameters pipelines. 
The initial catalog of redshifts was obtained with the Redrock package\footnote{\url{https://github.com/desihub/redrock}} \citep{Bailey2012,Bailey_Redrock:inprep}
which estimates redshifts by fitting a set of eigenspectra to the DESI spectra. The eigenspectra are constructed from star, galaxy, and QSO templates and are optimized for determining the velocities of galaxies 
over a wide range in redshift (from $-1100$~\kms\ to $z=6$).
To determine the radial velocities and stellar parameters, we also used the  Radial Velocity  pipeline (RVS) that is built on the RVSpecFit code\footnote{\url{https://github.com/segasai/rvspecfit}} \citep{Koposov2011,rvspecfit} and is used by the DESI Milky Way Survey (MWS). Details about the RVS pipeline and its outputs are provided in the MWS overview paper \citep{Cooper2022}, while here we provide a  brief summary. The stellar models for the fitting are built using the interpolated  PHOENIX stellar atmosphere models (spanning effective temperatures $2300\le T_{\rm eff} \le 15000$K) from \citet{Phoenix_2013} convolved to DESI resolution. These models  are fit  simultaneously to all three arms of the DESI spectra by optimizing the combined $\chi^2$. The spectra are not continuum normalized; instead we fit the spectra directly with functions of the form $T(\lambda)P(\lambda)$ where $T(\lambda)$ is the interpolated stellar template from the PHOENIX models and $P(\lambda)$ is a polynomial that takes care of potential flux calibration and/or normalization differences between the data and the model. The model fit provides estimates of the stellar atmospheric parameters $\log g$, $T_{\rm eff}$, [Fe/H] and [$\alpha$/Fe] together with heliocentric radial velocities in the range $\vert V_{\rm los}\vert\le 1500$~\kms.

\begin{figure}[th]
    \centering
    \includegraphics[width=\textwidth]{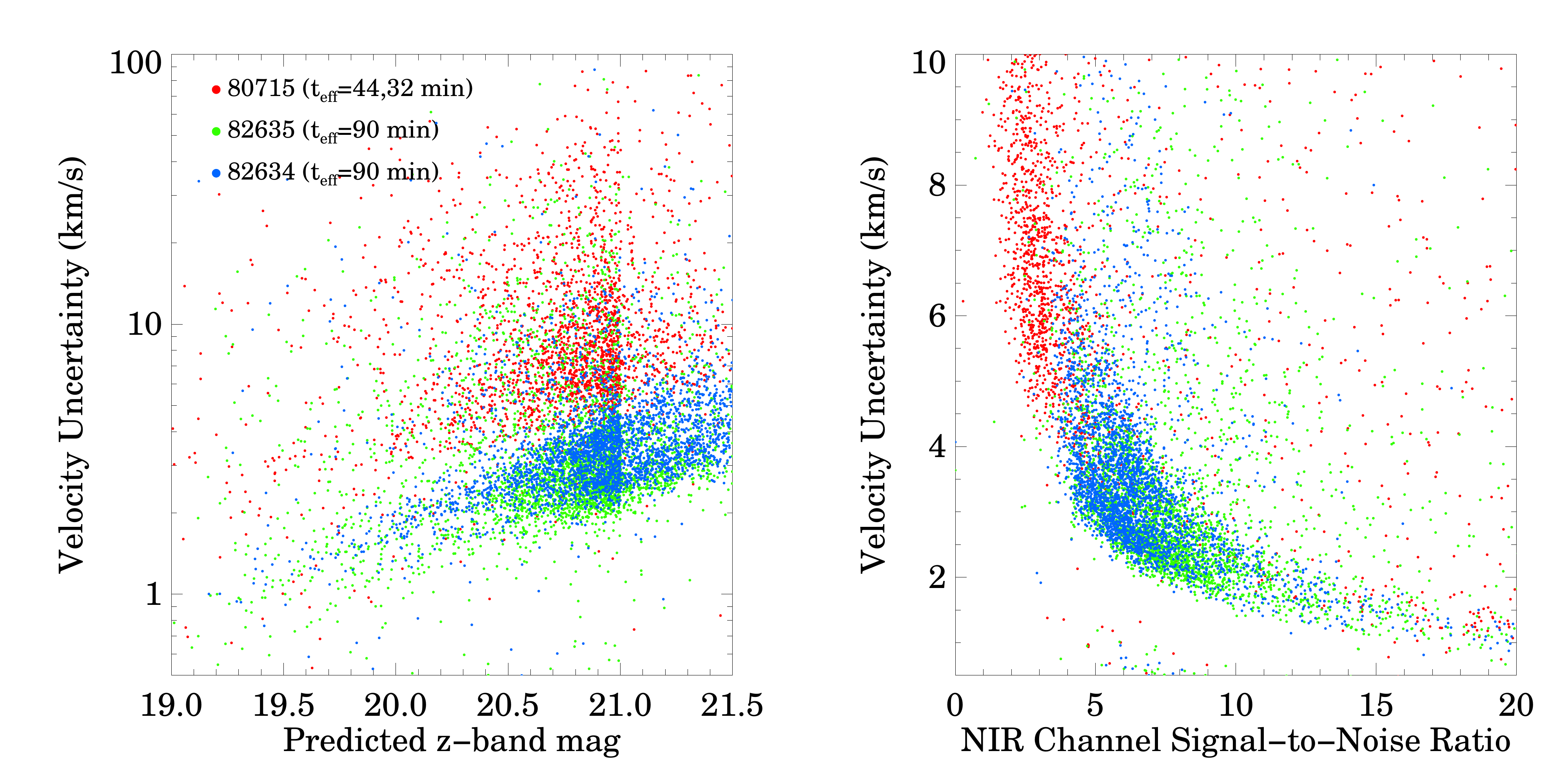}
    \caption{The distribution of measured radial velocity 
    uncertainty in the DESI spectra as a function of predicted $z$-band magnitude (left panel) and the mean signal-to-noise ratio per pixel in the NIR channel of the DESI spectrographs (right panel) in the fields centered on the disk (red points), GSS (blue points) and Northeast shelf (green points). For the nominal effective exposure time of 90 min (achieved for the halo tiles 82634 and 82635), the majority of the $z<21.5$~mag stars have velocity uncertainties $\sigma(V_{\rm los}) < 5$~\kms. 
    }
    \label{fig:velocityaccuracy}
\end{figure}

For each DESI target, we therefore have two velocity estimates, one from Redrock and the other from the RVS pipeline. For stars, the two pipelines agree extremely well: the median radial velocity difference is 0.05~\kms\ and the RMS scatter is 3~\kms. 
The 
accuracy of the stellar parameter determination by the RVS pipeline is discussed in the MWS overview paper, although the M31 data, especially the observation of the outer halo, represent a very different regime than most of the 
main MWS, as the majority of the M31 targets are very faint cool giants, where the dominant spectral information comes from the molecular absorption bands (the stellar atmosphere grid used by the RVS pipeline extends to effective temperatures of 2300\,K and works well for these sources; \citet{Cooper2022}). We found that the surface gravity estimates are particularly useful to identify M31 members and separate them from the Milky Way contaminants. For the nominal effective exposure time of 90 min (achieved for tiles 82634 and 82635), the majority of the $z<21.5$~mag stars have velocity uncertainties $\sigma_V < 5$~\kms\ (Figure~\ref{fig:velocityaccuracy}). We expect that the estimates of  [Fe/H] should be accurate to $\sim 0.2$ dex except for the faintest objects \citep{Cooper2022}. The estimates of [$\alpha$/Fe] are more uncertain and require better calibration datasets for comparison; the discussion of the [$\alpha$/Fe] measurements is therefore postponed to a future study.


Four of the authors (GM, JJZ, JN, AD) visually inspected 
\NVI\ of the spectra using the ``Prospect'' spectral inspection (VI) tool. We inspected all spectra for which Redrock  returned a SPECTYPE of GALAXY or QSO, found a redshift of $z>0.001$, or where the target was selected to be a QSO. In addition, we visually inspected the spectra of all targets selected from previous catalogs (i.e., the globular cluster, planetary nebulae, and variable star candidates). 
Further, we also visually inspected all the well-measured 
\citep[i.e., RVS\_WARN=0; see][for details]{Cooper2022} sources for which the Redrock- and RV-measured velocities differed by more than 50~\kms.
Spectra were visually classified according to three broad types (STAR, GALAXY, and QSO) and assigned a quality flag (varying from 0 = `No useful data' to 4 = `robust redshift and spectral type') based on the reliability of the redshift estimate. 



To create a final catalog, we retained only sources for which a velocity could be determined, i.e., sources with quality flags of 3 or 4 (which only excludes 6.3\% of the VI-ed sources).
The catalog reports a ``best'' velocity, selected from among the velocity measured by VI and those reported by the analysis pipelines. 
If the VI velocity was within 100~\kms\ of either the corresponding RVS or Redrock values, both of which 
were determined with greater precision 
than the VI value, we selected the value closer to the VI value. Conversely, if the VI velocity was $>100$~\kms\ away from the RVS and Redrock values, we selected the VI velocity.

\section{Spectroscopic Results}
\label{sec:results}

\subsection{Measurements}
\label{sec:measurements}


\begin{figure}
    \centering
    \includegraphics{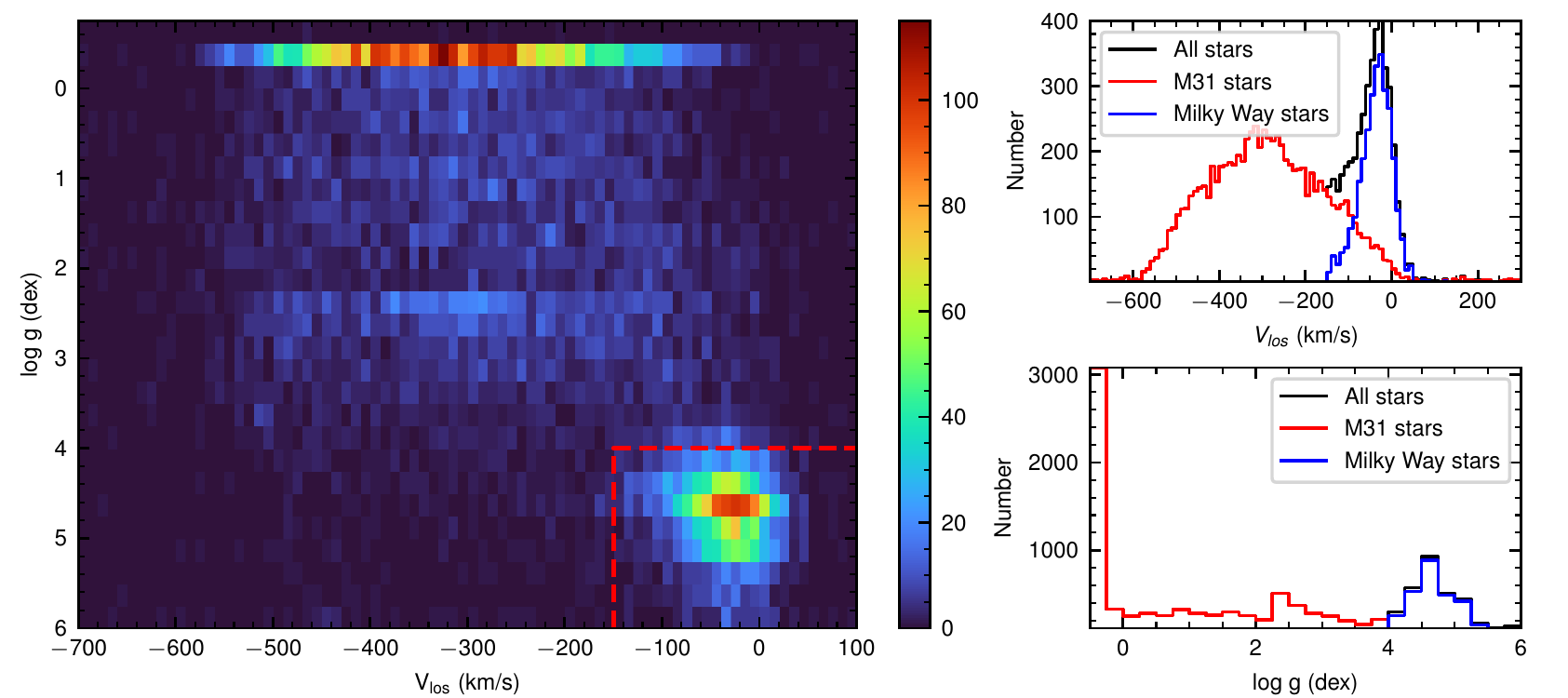}
    \caption{Isolating a robust sample of M31 sources. {\it Left panel:} distribution of stars with RVS\_WARN=0 in the space of radial velocity and surface gravity. The color-bar indicates the number of stars per bin. Milky Way stars are mostly nearby disk dwarfs with $V_{\rm los}\approx-50$~\kms\ and high surface gravities (log (g)), and are seen as the clump shown in the lower right. Because the M31 stars 
    are giants with low log(g), 
    we can exclude Milky Way stars by requiring log(g)$\le$4 or 
    heliocentric line-of-sight velocity $V_{\rm los}\le-150$~\kms 
    (red dashed line). The log(g) estimates for stars in M31 are not evenly distributed, but exhibit ``gridding''; i.e., they tend to values that form the grid of PHOENIX spectral models \citep[see][for details]{Cooper2022}. {\it Right panels:}  resulting radial velocity (top) and surface gravity (bottom) selection histograms of M31 stars (red) and Milky Way stars (blue). 
}
    \label{fig:logg_RV_selection}
\end{figure}

The DESI observations resulted in spectra of \Ntargets\ unique astronomical targets. Of these, \Ngalaxies\ are confirmed as galaxies and \NQSO\ as QSOs (see Appendix A). \Nstars\ of these are sources within M31 or foreground stars in the Milky Way. As shown in Figure~\ref{fig:logg_RV_selection} we can effectively isolate a robust sample of the M31 sources using the following combined criteria: 

$$ {\rm RVS\_WARN} = 0 $$ 
$$ \sigma(V_{\rm los}) \le 20 {\ \rm km\,s^{-1}}$$
$$ \log g \le 4\quad {\rm or}\quad  V_{\rm los} \le -150 {\ \rm km\,s^{-1}} $$


For the subset of sources that were visually inspected, we excluded those sources with VI\_SPECTYPE = GALAXY or QSO or 0 $\le$ VI\_QUALITY $\le$ 2. We note that this is not a 100\% complete selection, as there a few objects ($\sim$ 100)  that seem to belong to M31 based on the radial velocity but have a measured ${\log g}>4$.

These criteria result in a final sample of 
\NAndStars\ stars, \NAndHIIPN\ HII regions or planetary nebulae, and \Nclusters\ open or globular clusters. Of the 
9266 targets selected using the Random Forest algorithm, 
8416 have reliable radial velocity measurements (i.e., no processing errors and $\sigma(V_{\rm los})\le 10$~\kms), and of these 
6768 (73\% of all targeted) are M31 stars. This high success fraction demonstrates the efficiency of the Random Forest selection. 
For the backup selection, 
213 of the 562 targets (38\%) are M31 stars. 

\begin{figure}
    \centering
    \includegraphics[width=0.9\textwidth]{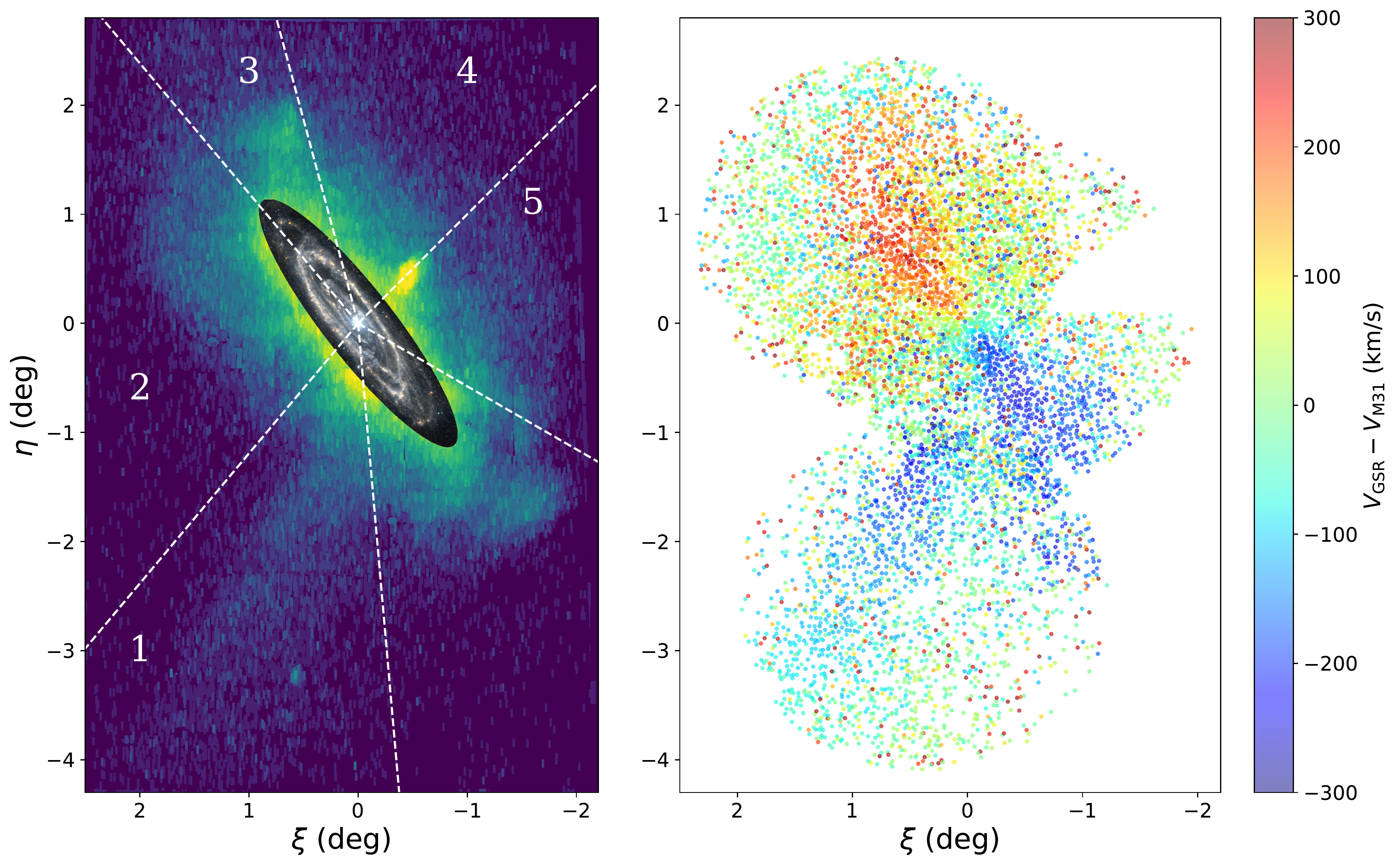}
    \caption{{\it Left:} Spatial density distribution of sources in the inner halo of M31 from the PAndAS catalog (green-purple color scale; see the text for details) with the unWISE coadded W3/W4 color image superposed for the central galaxy \citep[greyscale;][]{unWISEcoadd_2014}. Dashed lines indicate the five azimuthal zones in which the observed position-velocity structure is analyzed.    
    {\it Right:} Spatial distribution of the subset of DESI targets 
    selected according to the criteria described in \S~\ref{sec:measurements}. The points are color-coded by radial velocity relative to M31's recession velocity (all in the GSR frame). Highly redshifted and blueshifted stars extend far from the disk. The GSS appears as a stream of blueshifted stars  approaching the disk from the south. 
    }
    \label{fig:spatialveldistribution}
\end{figure}

In this paper, we present results based on the stars in the M31 halo, i.e., the region outside the ellipse encompassing the disk (e.g., see Figure~\ref{fig:spatialveldistribution}). The spectra of M31 disk sources will be discussed in a separate publication. The measured velocities and positions of the 6,436 confirmed M31 stellar sources are presented in Table~\ref{tab:stars}. The list of spectroscopically confirmed cluster, HII region and planetary nebula candidates is presented in Table~\ref{tab:HIIPNGC}. We also publish a FITS data table containing the measurements resulting from the analysis of the DESI spectra (see Appendix~\ref{appendix:datamodel}). Digital versions of all these tables are available online at \url{https://doi.org/10.5281/zenodo.6977494}.


\begin{rotatetable}
\begin{deluxetable}{ccccccccccccccl}
\movetabledown=3mm
\movetableright=0.1mm
\tablecaption{M31 Stars\tablenotemark{a}}
\tablecomments{Table 3 is published in its entirety in the machine-readable format.
      A portion is shown here for guidance regarding its form and content.}
\tablehead{
\colhead{\bf ID} & 
\colhead{\bf RA ($^\circ$)} & 
\colhead{\bf Dec ($^\circ$)} & 
\colhead{${\bf V_{\rm los}}$} &  
\colhead{${\bf \sigma(V_{\rm los})}$} &
\colhead{\bf [Fe/H]} &  
\colhead{\bf $\sigma$([Fe/H])} &
\colhead{${\bf T_{\rm eff}}$} &  
\colhead{${\bf \sigma(T_{\rm eff})}$} &
\colhead{\bf log(g)} &  
\colhead{\bf $\sigma$(log(g))} &
\colhead{\bf $G_{\rm DR2}$} & 
\colhead{\bf $g_{\rm PAndAS}$} & 
\colhead{\bf $i_{\rm PAndAS}$} &
\colhead{\bf Alternate Name} 
}
\startdata
   1 &  9.7371301 & 40.2178233 &  -472.5 &     6.7 & -1.85 &  0.06 &  3643.0 &     5.0 &  2.46 &  0.28 &   NaN & 24.90 & 21.89 &  PANDAS  170264                \\
   2 &  9.9989009 & 40.3012761 &  -520.7 &     5.5 &  0.44 &  0.19 &  4389.0 &    84.4 &  2.99 &  0.02 &   NaN & 25.48 & 22.20 &  PANDAS  96604                 \\
   3 & 10.1620093 & 40.4871788 &  -479.8 &    10.2 & -0.05 &  0.05 &  4162.9 &    20.9 &  3.04 &  0.02 &   NaN & 22.42 & 19.74 &  PANDAS  5541                  \\
   4 & 10.2713676 & 40.3688261 &  -521.1 &     4.4 &  0.42 &  0.06 &  3870.8 &     7.2 &  2.12 &  0.01 &   NaN & 25.53 & 21.77 &  PANDAS  176484                \\
   5 &  9.9539718 & 40.3950344 &  -509.8 &     5.0 & -0.70 &  0.05 &  4058.6 &    15.9 &  1.94 &  0.01 &   NaN & 25.85 & 22.20 &  PANDAS  96676                 \\
   6 &  9.7809551 & 40.2289816 &  -489.6 &     4.6 &  0.21 &  0.12 &  4055.0 &    69.2 &  3.21 &  0.01 &   NaN & 25.40 & 22.09 &  PANDAS  54369                 \\
   7 &  9.8708426 & 40.2325316 &  -413.9 &     6.0 &  0.16 &  0.09 &  4067.7 &    39.8 &  2.82 &  0.28 &   NaN & 25.86 & 21.92 &  PANDAS  171130                \\
   8 & 10.1081926 & 40.4273066 &  -503.8 &     4.6 & -0.62 &  0.04 &  4130.8 &    19.5 &  2.47 &  0.19 &   NaN & 24.74 & 21.76 &  PANDAS  176049                \\
   9 & 10.0735259 & 40.3537205 &  -454.4 &     4.9 & -1.08 &  0.07 &  3779.1 &     5.8 &  0.74 &  0.00 &   NaN & 25.24 & 21.78 &  PANDAS  96643                 \\
  10 &  9.9964951 & 40.4045608 &  -524.8 &    11.5 &  0.50 &  0.07 &  9800.0 &     6.2 &  1.51 &  0.00 & 19.92 &   NaN &   NaN &  Gaia DR2  381120004684395520  \\
  11 &  9.9106488 & 40.2671799 &  -158.2 &     1.8 & -1.12 &  0.07 &  4576.7 &    11.9 &  4.24 &  0.01 & 20.34 &   NaN &   NaN &  Gaia DR2  369107977589805824  \\
  12 &  9.7966509 & 40.2791149 &  -395.6 &    11.1 &  0.88 &  0.00 &  4139.0 &    45.8 &  4.00 &  0.01 &   NaN & 25.03 & 21.69 &  PANDAS  96533                 \\
  13 & 10.0611343 & 40.2721011 &  -452.8 &     5.2 & -0.88 &  0.07 &  3788.7 &     5.9 &  0.21 &  0.00 &   NaN & 24.92 & 21.91 &  PANDAS  174999                \\
  14 & 10.2674759 & 40.5024538 &  -461.4 &     9.7 & -0.44 &  0.04 &  4067.8 &    15.3 &  1.84 &  0.01 &   NaN & 24.69 & 21.85 &  PANDAS  179680                \\
  15 & 10.2065926 & 40.4144844 &  -501.4 &     5.7 & -2.12 &  0.04 &  3371.2 &     3.7 &  0.51 &  0.00 &   NaN & 24.90 & 21.82 &  PANDAS  176513                \\
  16 &  9.7446926 & 40.3032594 &  -485.0 &     4.2 &  0.07 &  0.11 &  5758.0 &   103.1 &  3.00 &  0.01 &   NaN & 23.08 & 21.59 &  PANDAS  54401                 \\
  17 &  9.8513801 & 40.3184038 &  -493.3 &     4.0 & -0.82 &  0.05 &  3945.2 &     7.2 &  1.69 &  0.01 &   NaN & 24.29 & 21.72 &  PANDAS  174596                \\
  18 & 10.1946551 & 40.3319622 &  -482.1 &     3.9 &  0.46 &  0.06 &  3759.3 &     5.8 &  0.97 &  0.01 &   NaN & 25.71 & 22.13 &  PANDAS  176249                \\
  19 & 10.0397343 & 40.4225761 &  -499.6 &     4.7 &  0.09 &  0.03 &  4230.9 &    18.6 &  2.60 &  0.13 &   NaN & 25.76 & 21.83 &  PANDAS  96749                 \\
  20 &  9.9798933 & 40.4267977 &  -506.4 &     0.4 & -0.39 &  0.02 &  3851.9 &     3.2 & -0.50 &  0.00 & 20.08 & 21.62 & 19.35 &  PANDAS  5270                  \\
  21 & 10.0378426 & 40.3165594 &  -493.0 &     5.7 &  0.41 &  0.07 &  4143.4 &    35.1 &  2.78 &  0.22 &   NaN & 24.67 & 21.81 &  PANDAS  175068                \\
  22 & 10.1195551 & 40.3864344 &  -436.8 &     9.9 & -0.37 &  0.09 &  4044.9 &    32.2 &  2.84 &  0.26 &   NaN & 25.02 & 21.90 &  PANDAS  175857                \\
  23 &  9.9955093 & 40.3262261 &  -478.3 &     3.5 &  0.36 &  0.07 &  4153.6 &    32.8 &  2.54 &  0.20 &   NaN & 24.80 & 21.87 &  PANDAS  175062                \\
  24 &  9.9753384 & 40.2584122 &  -482.7 &     6.0 &  0.07 &  0.09 &  4270.8 &    48.8 &  3.67 &  0.01 &   NaN & 25.13 & 21.93 &  PANDAS  174942                \\
\enddata
 \tablenotetext{a}{See Online Version for complete Table. The columns are: (1) a running index; (2,3) RA and Dec in J2000; (4,5) $V_{\rm los}$, the line-of-sight heliocentric velocity and its formal uncertainty; (6,7) the spectroscopic estimate of [Fe/H] and its formal uncertainty; (8,9) $T_{\rm eff}$, the effective temperature and its formal uncertainty; (10,11) the surface gravity (log g) and its formal uncertainty; (12) the {\it Gaia} DR2 G-band flux (NaN if not available); (13,14) the PAndAS $g$ and $i$ magnitude (NaN if not available); (15) an alternate name for the target (i.e., from {\it Gaia} DR2 or PAndAS).}
 
    \label{tab:stars}
\end{deluxetable}
\end{rotatetable}

\begin{deluxetable}{cccccccccl}
\tablecaption{M31 HII, PNe, and GC Targets\tablenotemark{a}}
\tablecomments{Table 4 is published in its entirety in the machine-readable format.
      A portion is shown here for guidance regarding its form and content.}
\tablehead{
\colhead{ID} & \colhead{RA ($^\circ$)} & \colhead{Dec ($^\circ$)} &
        \colhead{$V_{\rm los}$} &  \colhead{$\sigma(V_{\rm los})$} & \colhead{\bf Target Class} &
        \colhead{\bf $G_{\rm DR2}$} & \colhead{\bf $g_{\rm PAndAS}$} & \colhead{\bf $i_{\rm PAndAS}$} &
        \colhead{\bf Alternate Name}
}
\startdata
  35 & 10.5797000 & 40.9525000 &  -442.1 &     5.1 & H2PN       &   NaN &   NaN &   NaN &                          PN168 \\
  36 & 10.6240000 & 41.0584000 &  -438.9 &     7.0 & H2PN       &   NaN & 20.94 & 22.10 &                          PN184 \\
  37 & 10.3057000 & 41.1931000 &  -400.6 &    11.5 & H2PN       &   NaN & 22.73 & 25.33 &                          PN080 \\
  38 & 10.4541000 & 41.0738000 &  -513.6 &     1.4 & H2PN       &   NaN &   NaN &   NaN &                          PN129 \\
  39 & 10.4348000 & 41.0364000 &  -489.7 &     3.3 & H2PN       &   NaN & 21.42 & 22.63 &                          PN118 \\
  40 & 10.6293000 & 40.8846000 &  -408.9 &     0.2 & H2PN       &   NaN &   NaN &   NaN &                          PN185 \\
  41 & 10.5604000 & 40.8729000 &  -471.8 &     0.1 & H2PN       &   NaN &   NaN &   NaN &                          PN165 \\
  42 & 10.4922000 & 41.1361000 &  -470.6 &     2.9 & H2PN       &   NaN & 21.47 & 26.18 &                          PN144 \\
  43 & 10.3843000 & 41.0033000 &  -486.6 &     3.8 & H2PN       &   NaN & 21.80 & 22.02 &                          PN102 \\
  44 & 10.1449766 & 40.4436903 &  -496.4 &    15.2 &    cluster & 20.46 &   NaN &   NaN & GC2196,B196D,B196D-SH08    \\
  45 &  9.8293724 & 40.3661097 &  -553.9 &     0.1 &    cluster &   NaN &   NaN &   NaN & GC3106,SH06,SH06          \\
  46 &  9.9358057 & 40.2355292 &  -475.7 &     1.0 &    cluster &   NaN &   NaN &   NaN & GC308,B314,B314-G037      \\
  47 &  9.8905182 & 40.5207430 &  -507.2 &     0.2 &    cluster & 18.17 &   NaN &   NaN & GC305,B311,B311-G033      \\
  48 & 10.1869141 & 40.8855375 &  -536.7 &     2.0 &    cluster & 21.15 &   NaN &   NaN & GC3710,BH10,BH10          \\
  49 & 10.2529516 & 39.9317236 &  -235.2 &     0.3 &    cluster & 19.74 &   NaN &   NaN & GC332,B339,B339-G077      \\
  50 & 10.2205474 & 40.5888014 &  -552.0 &     1.1 &    cluster &   NaN &   NaN &   NaN & GC9074,KHM31-74,KHM31-74      \\
\enddata
 \tablenotetext{a}{See Online Version for complete Table. The columns are: (1) a running index; (2,3) RA and Dec in J2000; (4,5) $V_{\rm los}$, the line-of-sight heliocentric velocity and its formal uncertainty; (6) the reason the source was targeted, i.e., whether it was a potential emission line (``H2PN'') or star cluster (``cluster'') candidate; (7) the {\it Gaia} DR2 G-band flux (NaN if not available); (8,9) the PAndAS $g$ and $i$ magnitude (NaN if not available); and (10) alternate name(s) for the target from \citet{Sanders2012,galleti2007,galleti2014cat}. Note: `GC' stands for ``Galactic Cluster" candidate and does not distinguish between young and old clusters.}
    \label{tab:HIIPNGC}
\end{deluxetable}

\subsubsection{Comparison to Previous Work}
\label{sec:previouswork}




M31 has been the target of several spectroscopic campaigns over many decades. A search of the SIMBAD database \citep{SIMBAD2000} resulted in a total of 139,078 entries (for 35,374 sources with unique names) within 5$^\circ$ of M31, of which 73,090 (representing 14,617 unique sources) have reported radial velocities. In addition, the \href{https://oirsa.cfa.harvard.edu/signature_program/}{CFA Optical/Infrared Science Archive} \citep[][and references therein]{Sanders2012, Caldwell2016, Bhattacharya2019}
consolidates the many years of MMT/Hectospec and Hectochelle campaigns in M31. Of the 10,322 sources in the Archive that are within 5$^\circ$ of M31, 5064 have measured radial velocities and 2,099 are also included in the SIMBAD list. In summary, there are 17,582 sources with published radial velocities in this region. 

The bulk of the literature radial velocities are foreground (Milky Way) stars, and only 6,939 sources have radial velocities typical of M31 ($<-100$~\kms). The bulk of these stars lie within the projected area of the M31 main disk and, unlike the DESI data, do not sample the M31 inner halo well. 

Thus, the new DESI radial velocities presented here only have a small overlap with the published radial velocity measurements. Only 145 DESI targets have matches (within 1\arcsec) to sources with radial velocities in the literature. Where there is overlap, the DESI radial velocities agree well: they have a median offset of $\approx 2.8$~\kms\ and an rms scatter of $\approx 14$~\kms. For the matched sources, the median velocity uncertainty of the measurement quoted in the literature is $\sim15$~\kms. The DESI data provide more precise radial velocities, with $\approx88$\% of the sources having velocity uncertainties $\le 10$~\kms. 

While most of the spectroscopy to date of individual stars in M31 has been carried out with 6.5-m to 10-m class telescopes \citep[e.g.,][and references therein]{Ibata2004, Bhattacharya2019, Caldwell2016, Guhathakurta2006, Kalirai2006a,Kalirai2006b, Gilbert2007, Gilbert2009a, Dorman2012, Gilbert2020,Escala2019,Escala2020a,Escala2020b,Escala2021}, the present results illustrate the science potential of highly multiplexed spectrometers on smaller aperture telescopes. 
%
%
A caveat here is that although several campaigns by different groups have targeted the fainter M31 halo populations, the data have not been published along with the papers reporting the results. 
These campaigns (primarily with Keck/DEIMOS) have targeted primarily giants and horizontal branch stars in M31 in a number of pencil beams scattered across the region.  These prior studies typically reach targets 
fainter than our DESI observations, 
and include more metal poor RGB targets \citep[e.g.,][]{Ibata2004,Kalirai2006a,Gilbert2020,Kirby2020,Escala2020a,Escala2020b,Escala2022}.
However, the advantage DESI offers is the ability to (approximately) uniformly sample large spatial regions of the M31 halo both quickly and efficiently: a total DESI on-sky exposure time of $\approx 3.75$~hr yielded \Ngoodspec\ velocities, {\NAnd} of which are M31 sources with well-measured line-of-sight velocities with uncertainties $\sigma_V\le 10$~\kms. 

\subsection{Position-Velocity Diagrams}
\label{sec:PVdiagram}

The left panel of Figure~\ref{fig:spatialveldistribution} shows the density distribution of sources in the 
inner halo of M31 selected from the PAndAS catalog \citep[]{McConnachie2018} in the region covered by the DESI spectroscopy, with the unWISE coadded W3/W4 image superposed on the central galaxy  \citep{unWISEcoadd_2014}.
The distribution of inner halo sources shows the previously identified morphological features: the GSS to the SSE; the Southeast and Northeast Shelves; and the Western Shelf \citep[see][for details]{Ferguson2016}. 
To create the image of the inner halo, we selected catalog sources from the $i$ vs $(g-i)$ color-magnitude diagram that lie within the polygon defined by 
[$(g-i)$,$i$]=[[0.9, 1.8, 5.0, 5.0, 2.2, 2.0], [23, 21, 22, 22.5, 22.5, 23]] and used a Gaussian kernel density estimator to adaptively smooth the spatial point distribution of sources. 
The 
ellipse separating the inner halo and central galaxy 
(with semi-major axis $a_e=1.5^\circ$, semi-minor axis $b_e=0.337^\circ$, and PA=38$^\circ$ and centered at (RA, DEC)=(10.6847$^\circ$, +41.26875$^\circ$))
denotes the disk of M31 and roughly traces the ring of star formation so clearly visible in young stars and mid-infrared observations of the galaxy \citep[e.g.,][]{barmby2006,lewis2015}. 

In the left panel, radial dashed lines demarcate the zones in which we explore the position-velocity distributions of the observed sources. 
The zonal boundaries (see Table~\ref{tab:kinematicfeatures} for details) are chosen to overlap known overdensities and to distinguish these from each other and the M31 disk.
Zone 1 is dominated by the GSS; Zone 2 contains the Southeast Shelf (a portion of which begins in Zone 1) and more than half of the Northeast Shelf; Zone 3 includes the Northeast Shelf and the blobby feature located at $(\xi,\eta)\approx(0.8,1.8)$; Zone 4 contains the inner halo region just north of the M31 disk; and Zone 5 is dominated by the Western Shelf. 
The remaining range of azimuth 
does not contain much DESI spectroscopy beyond the boundary of the disk. Several of these zones include Andromeda dwarf galaxies: Andromeda 1 and M32 lie in Zone 1; Andromeda 9 lies in Zone 2; and NGC205 straddles the boundary between Zones 4 and 5. We neither explicitly targeted nor excluded stars that may be associated with these companions.

\begin{figure}
    \centering
    \includegraphics[width=0.85\textwidth]{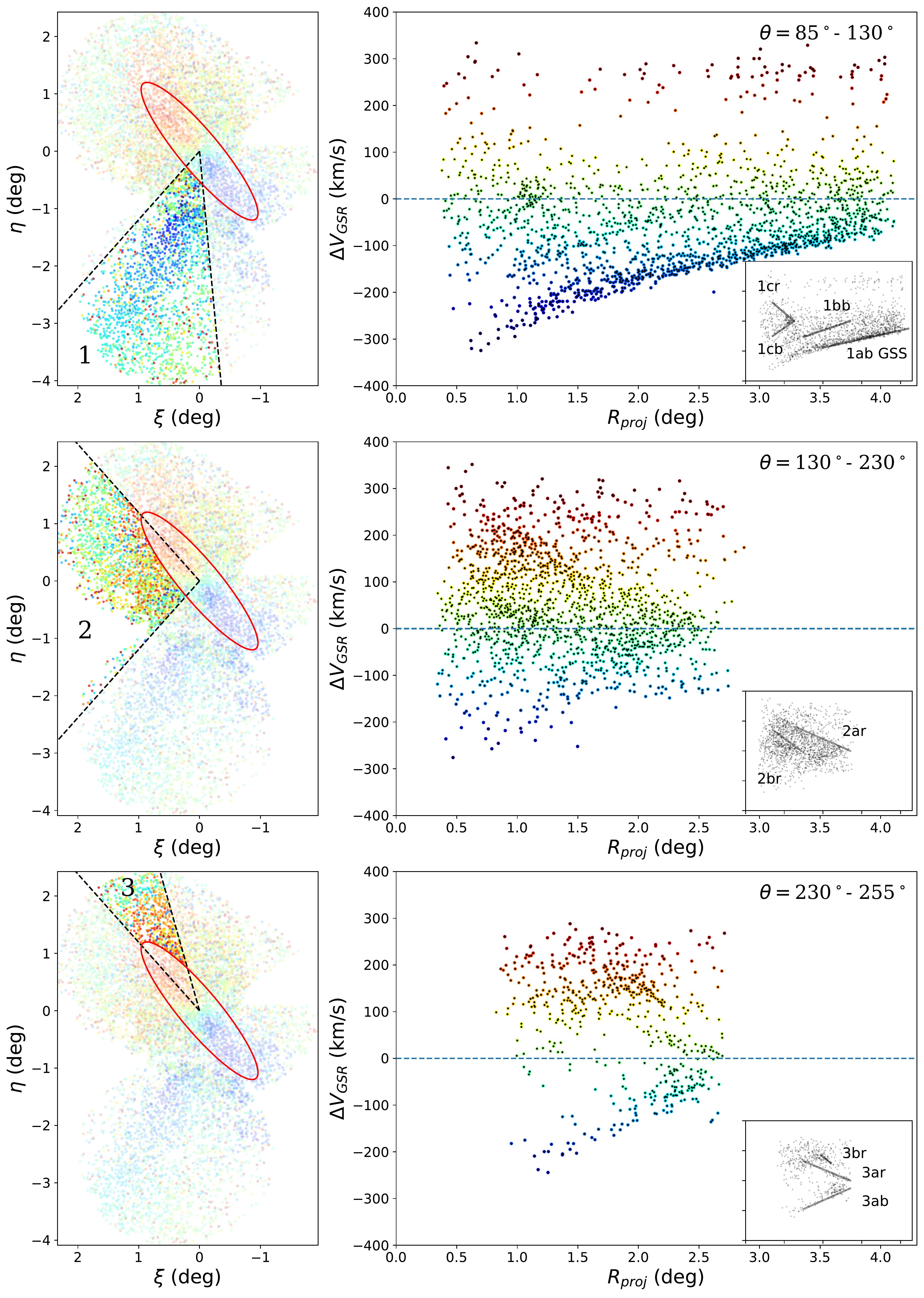}
    \caption{Line-of-sight position-velocity diagrams (right hand panels) for angular zones of interest (left panels). Stars lying within the ellipse centered on M31 (red line) are excluded from the position-velocity diagram. Velocities are relative to the M31 central velocity of $V_{\rm los}({\rm M31})=-300$~\kms\ (i.e., $V_{GSR}({\rm M31})=113.6$~\kms). Insets highlight linear features apparent in each region, which are shown as grey lines. Zone 1 contains the Giant Stellar Stream. Zones 2 and 3 are part of the Eastern Shelf.  }
    \label{fig:zoneposvela}
\end{figure}

\begin{figure}
    \centering
    \includegraphics[width=0.85\textwidth]{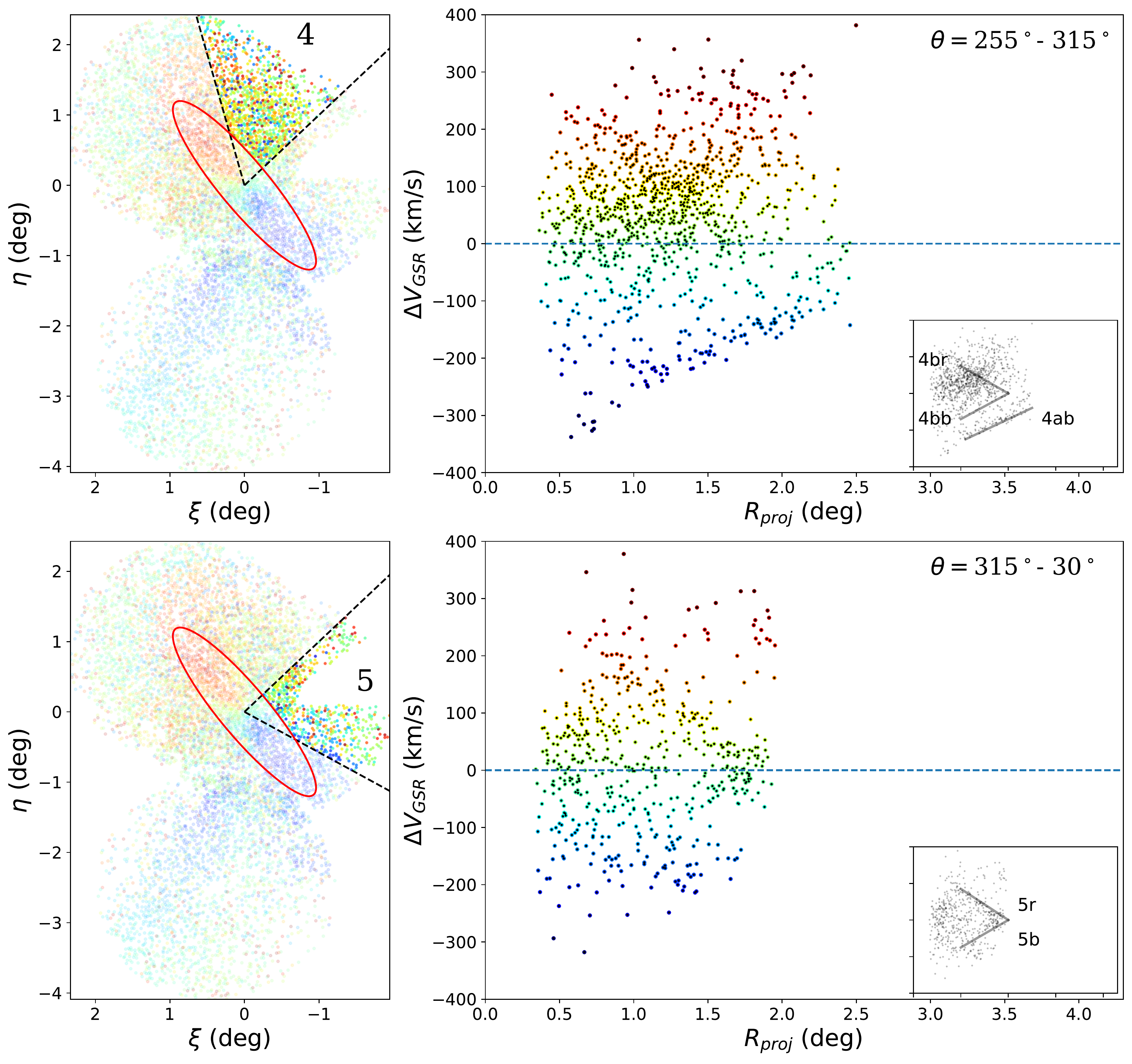}
    \caption{As in the previous figure, for two additional zones. Zone 5 includes the Western Shelf.}
    \label{fig:zoneposvelb}
\end{figure}

\begin{figure}
    \centering
    \includegraphics[width=0.85\textwidth]{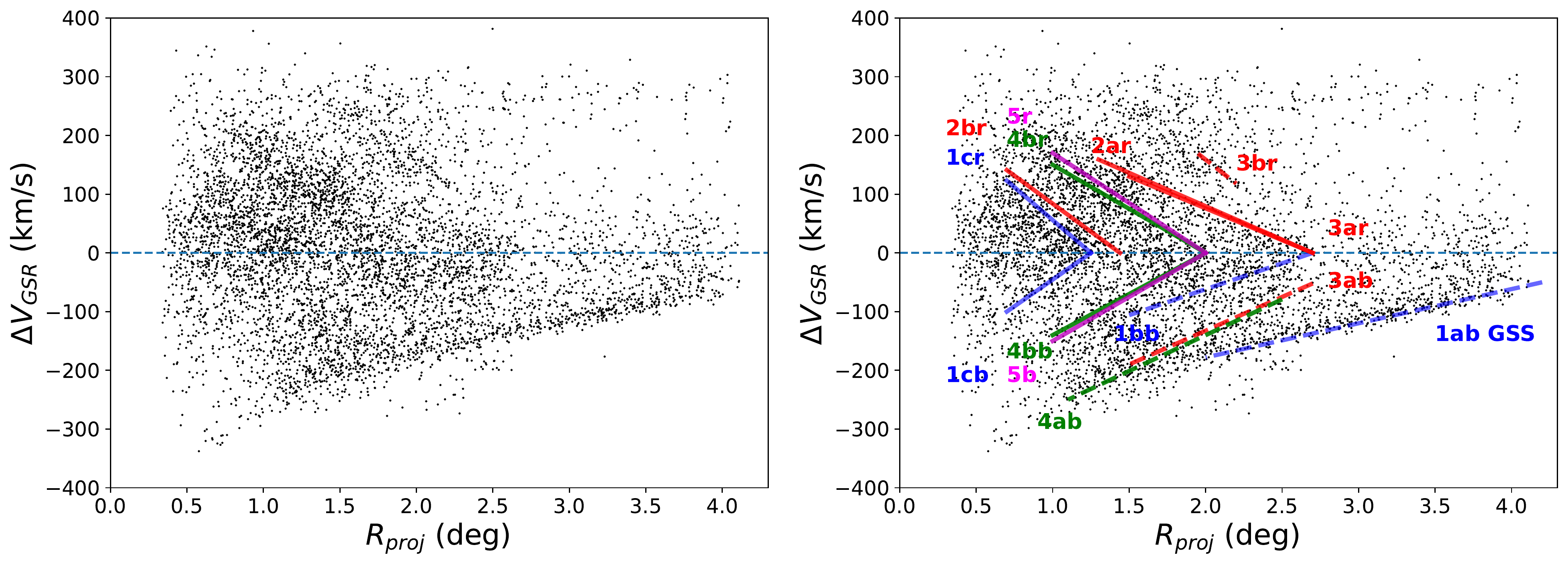}
    \caption{Combined line-of-sight positions and velocities for stars in all zones (black dots, left and right panels). The right panel also shows the linear features highlighted in the insets from Figures~\ref{fig:zoneposvela} and \ref{fig:zoneposvelb} and described in Table~\ref{tab:kinematicfeatures}. 
    }
    \label{fig:allfeaturesannotated}
\end{figure}

The right panel of Figure~\ref{fig:spatialveldistribution} shows the positions of the measured M31 sources color coded by line-of-sight velocity. 
There is a clear red-blue asymmetry along the major axis of the galaxy, with an apparent strong flaring and/or a warp near [$\xi,\eta$] $\approx$ [$+0.7^\circ,+1.8^\circ$], also observed in the stellar density distribution. 
We can examine the kinematics of each zone 
by plotting the line-of-sight radial velocity in the Galactic Standard of Rest relative to M31 ($\Delta V_{\rm GSR}$) as a function of projected distance from the center of M31 ($R_{\rm proj}$) for the sources in each sector, as shown in 
Figures \ref{fig:zoneposvela} and \ref{fig:zoneposvelb}.
In each panel, stars at velocities $\sim 300$\,\kms relative to M31 are primarily foreground Milky Way stars. 
Figure~\ref{fig:allfeaturesannotated} shows the line-of-sight positions and velocities for stars in all zones and summarizes the linear features in position-velocity space detected in each zone. 

The kinematic features are also tabulated in Table~\ref{tab:kinematicfeatures} 
and identified by the following convention: a number for the zone,  a letter index
 to distinguish multiple features in the same zone,
and a ``b'' or ``r'' based on whether the feature is blue- or redshifted relative to the M31 systemic velocity. For example, the GSS feature is labelled as ``1ab'', meaning that it is feature ``a'' in Zone 1, and is blue-shifted relative to the M31 systemic velocity. 
We refer to the GSS and other linear features in Figures~\ref{fig:zoneposvela} and \ref{fig:zoneposvelb} 
as ``streams” based on the previous use of the term in naming the GSS. These ``streams'' are only-redshifted or only-blueshifted, mostly linear features, in contrast to the features we refer to as ``shells”, which have the morphology of chevrons or wedges and typically have both red- and blue-shifted components in 
Figures~\ref{fig:zoneposvela} and \ref{fig:zoneposvelb}.
In contrast to true narrow Galactic streams such as GD1 \citep[e.g.,][]{Koposov2010}, the structures we call ``streams" may be more accurately described as ``one-sided shells”,
with their member stars possibly spanning a range of total energies.



\begin{figure}
    \centering
    \includegraphics[width=1.0\textwidth]{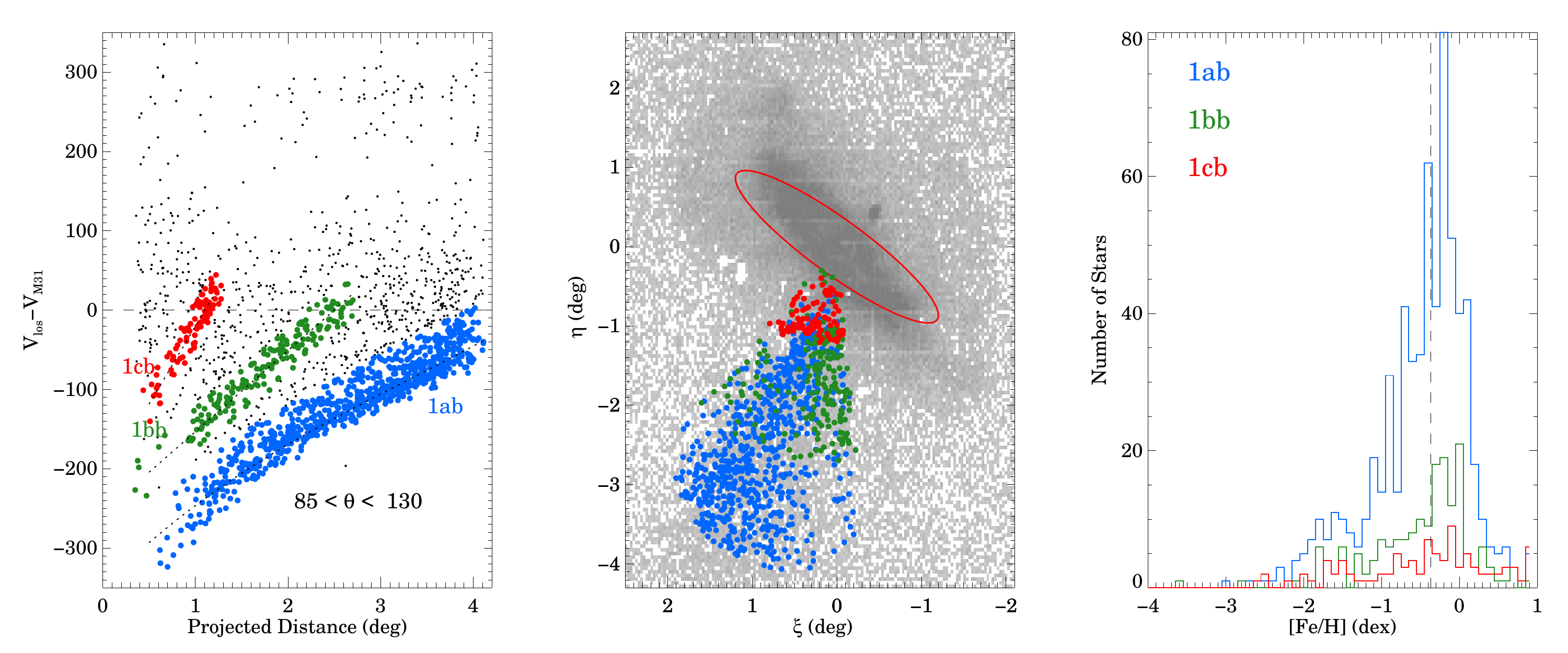}
    \caption{The left panel shows the line-of-sight position-velocity diagram for Zone 1 with the selection of stars in the three main kinematic features as described in the text. The center panel shows the spatial distribution of all observed stars lying within these selection windows. 
    The right panel shows the spectroscopic metallicity distribution of these points and demonstrates that the features 1ab (GSS) and 1bb have similar broad distributions with median [Fe/H]$\approx-0.37$~dex (vertical dashed line). The metallicity distribution of 1cb is flatter and broader. 
    }
    \label{fig:zone2feh}
\end{figure}

The position-velocity diagram for {\bf Zone 1} reveals at least three main features. Most prominent is the GSS (labeled `1ab' in Figure \ref{fig:zoneposvela}), which appears as a tight band of blueshifted stars whose average velocity varies smoothly with distance from $\sim -300$~\kms at 0.5$^\circ$ to $\sim -50$~\kms at 4$^\circ$ separation from M31. These stars are also highlighted in blue in Figure 
\ref{fig:zone2feh}. Our data cover most, but not all, of the entire visible extent of the GSS. Extrapolating linearly, we expect the kinematic structure to cross the zero velocity line at a projected distance of 
$\approx 5.0^\circ$ 
from M31. 
%
%

\begin{figure}
    \centering
    \includegraphics{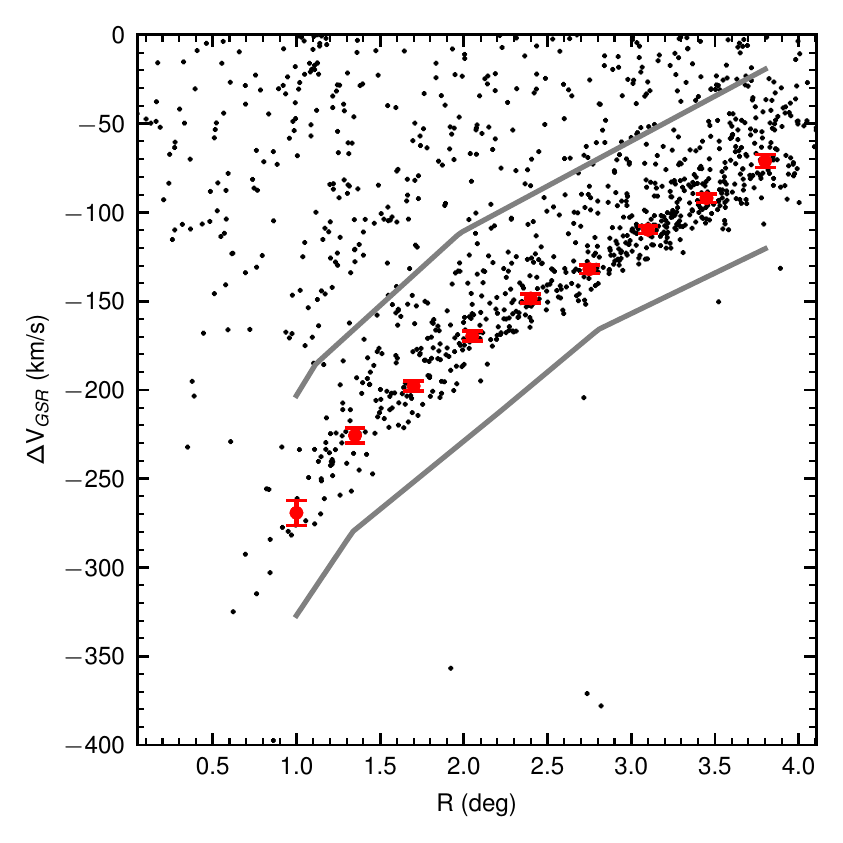}
    \includegraphics{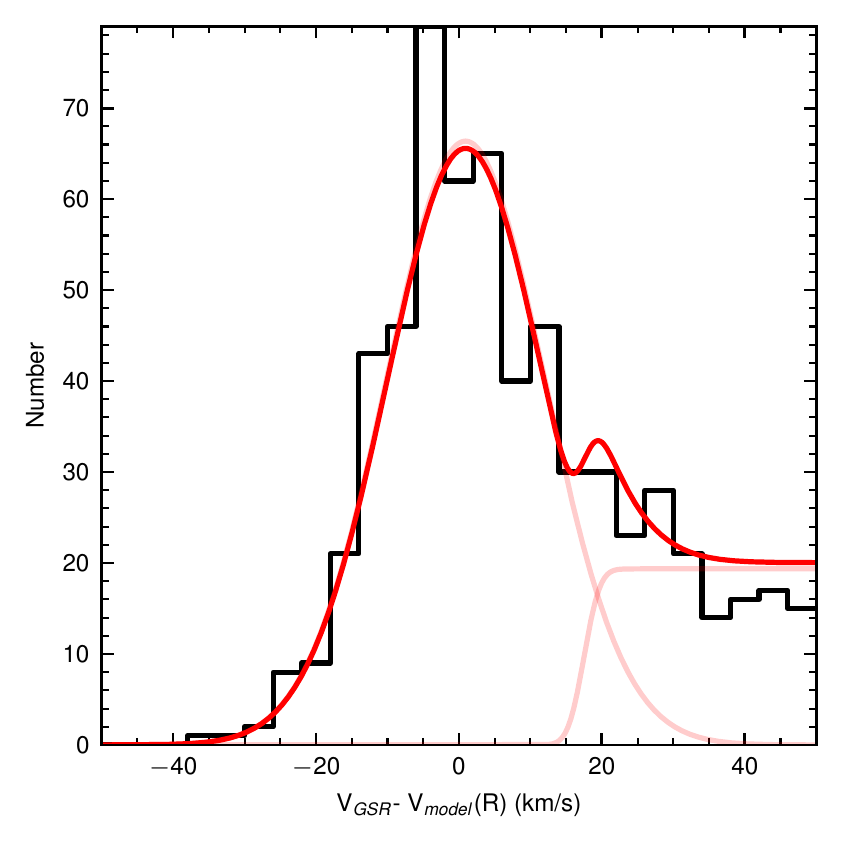}
    \caption{{\it Left:} Extracted projected radial velocities as a function of projected distance used to define the locus of the GSS. 
    Black points show individual stars in the direction of the GSS. Grey lines bracket the sample of stars used in the fit. Red points with error bars show the extracted 
    locus of the GSS in distance bins. {\it Right:} Histogram of radial velocity residuals relative to the best-fit locus for the GSS. The model for the radial velocity distribution is overplotted in red. See the text for details. 
    }
    \label{fig:Sergey_GSS_RV}
\end{figure}

To determine the velocity dispersion of the GSS, we fit the observations with a two-component model: 
$$P(V|R) = f N(V|V_{\rm str}(R),\sigma) + (1-f)  S_{\rm bg}(V-V_{\rm str}(R)),$$
where $N(V)$ is a Gaussian density in projected radial velocity that represents the GSS and 
$S_{\rm bg}(x),$ which represents the foreground component, is 
an appropriately normalized piece-wise linear function of the form $\min(\max(x,0),1)$. The quantity $f$ is the mixing fraction between the 
stream and the 
more smoothly distributed M31 halo stars, $V_{\rm str}(R)$ is the radial velocity of the 
stream as a function of distance parameterized by a cubic spline with nine knots, and $\sigma$ is the velocity dispersion of the GSS. The model has 14 parameters in total and was fitted to the sample of stars between the grey lines shown in the left panel of Figure~\ref{fig:Sergey_GSS_RV}. We use in the fit only stars at projected distances between 1$^\circ$ and 3.8$^\circ$ from M31 and at position angles between 147$^\circ$ and 175$^\circ$. The posterior of the parameters is sampled using the {\tt dynesty} nested sampling code \citep{dynesty,speagle2020}, and the fitted velocity dispersion is determined to be $10.80 \pm 0.75$ \kms. 
This velocity dispersion is lower than most of the measurements by \citet{Gilbert2009b} of the primary GSS component in the GSS core and envelope region, but is more consistent with their measurement in the ``m4" field of $11.4^{+5.2}_{-4.1}$~\kms\ centered on Stream C. This may be due to the better velocity resolution and better spatial sampling of the DESI study, which results in the ability to cleanly isolate the primary GSS kinematic structure (1ab) from the 
other components. Given the multiple structures that make up this region of the halo, measuring a single velocity dispersion for all the components together is not physically meaningful. The fitted position-velocity locus of the GSS 
is 
provided in Table~\ref{tab:gss_rv_track} and 
used in the mass estimate analysis presented in Section~\ref{sec:gssmassest}. 

%
%

Zone 1 also includes another band of blueshifted stars that runs parallel to the GSS in the position-velocity diagram. Less blueshifted by $\sim 100$~\kms\ than the GSS, this kinematically cold component (labeled `1bb' in Figure \ref{fig:zoneposvela}) has a velocity dispersion similar to that of the GSS (see also green points in Figure \ref{fig:zone2feh}),
and is more limited in length, extending to $\sim 2.7^\circ$ from the center of M31.
As shown in Figure \ref{fig:zone2feh}, feature `1bb' is also spatially offset from the GSS, extending outward from the center of M31 at a different mean angle 
than the GSS stars. 
Feature 1bb was previously identified in spectroscopy carried out in pencil-beam surveys of discrete portions of the M31 halo 
\citep{Kalirai2006a,Gilbert2009b}.
Our results are consistent with the velocities previously reported and illustrate, for the first time, the spatially continuous nature of the structure and its spatial offset from the GSS. 

We also see in Zone 1 a hint of a more compact feature: a 
{\bf chevron} pattern, i.e., a concentration of stars along a triangular-shaped edge
(its blue- and redshifted edges labeled `1cb' and `1cr' in Figure~\ref{fig:zoneposvela}), similar to the general shape expected for radial shells \citep{Merrifield1998}.
The chevron extends to $\sim 1.3^\circ$ in projected distance and reaches an apex at a velocity 
within $\sim 30$~\kms\ of M31 (Figure~\ref{fig:zoneposvela}; red points in Figure~\ref{fig:zone2feh}).
Higher density sampling is needed to confirm this feature and define its kinematic structure. 

\begin{table}[h]
    \centering
    \caption{Approximate Parameters of Kinematic Features}
    \begin{tabular}{ccllrcl}
    \hline
{\bf Zone} & {\bf Angular Range\tablenotemark{a}} &  {\bf Feature} & {\bf $R_{\rm max}$} & {\bf d$V_{\rm los}$/d$R$} & {\bf $M_{\rm enc}$}\tablenotemark{c} & Type\\
           &                     &                & (deg)      & (\kms/deg)                & ($10^{11} M_\odot$) & \\
\hline
        1 & $85^\circ-130^\circ$ & 
  1ab (GSS) & 5.057\tablenotemark{b}    & 58      &         & Stream \\  
&& 1bb      & 2.70      & 88       &        & Stream? \\    
&& 1cb      & 1.25      & 182      & 2.0    & Shell \\    
&& 1cr      & 1.25      & -224      & 3.1    & Shell \\      
2 &   $130^\circ-230^\circ$& 
   2ar      & 2.70      & -114      & 8.0    & Shell \\      
&& 2br      & 1.44      & -191      & 3.4    & Shell, related to 1cr \\      
3 &  $230^\circ-255^\circ$ & 
   3ar       & 2.70      & -108      & 7.3    & Shell, related to 2ar \\      
&& 3br      & --         & -202      &         & Short linear feature \\   
&& 3ab       & 3.15\tablenotemark{b}      & 115      &         & Stream? \\      
   
4 & $255^\circ-315^\circ$ & 
   4ab      & 3.15\tablenotemark{b}      & 122      &         & Stream? related to 3ab \\  
&& 4bb      & 2.00      & 140      & 5.0    & Shell \\        
&& 4br      & 2.00      & -150      & 5.7    & Shell \\        
5 & $315^\circ-30^\circ$ & 
   5b       & 2.00      & 150      & 5.7    & Shell, related to 4bb \\        
&& 5r       & 2.00      & -170      & 7.3    & Shell, related to 4br \\        
          \hline
    \end{tabular}
    \tablenotetext{a}{Angle is measured clockwise from the $\eta=0$ axis; i.e., $\theta=270^\circ-$PA}
    \tablenotetext{b}{$R_{\rm max}$ for features 1ab (GSS), 3ab, and 4ab are determined from the linear extrapolation to $V_{\rm los}=0$.}
    \tablenotemark{c}{Enclosed masses $M_{\rm enc}$ for shells estimated from Merrifield \& Kuijken (1998).}
    \label{tab:kinematicfeatures}
\end{table}

{\bf Zones 2, 3, and 4} include the Northeast Shelf, which extends out to $\sim 2.5^\circ$ from M31 (Figure \ref{fig:spatialveldistribution}) and has a
shell-like morphology. 
The position-velocity diagram for Zone 2, which samples the portion of the Northeast Shelf south of the M31 disk,  
shows a large, prominent triangular {\bf ``wedge'' shape (a filled chevron)}, with an apex at $\sim 0$~\kms\ relative to M31 at a distance of $\sim 2.5^\circ$ and extending to $\pm 300$~\kms\ at $\sim 0.5^\circ$.
The redshifted edge of the wedge (labeled `2ar' in Figure~\ref{fig:zoneposvela}) is better defined than the blueshifted edge, and the interior of the wedge is more populated at redshifted velocities. 
Within this feature, a smaller wedge-shaped feature also appears to be present, with an apex at $\sim 1.5^\circ$ distance and extending to $\sim 150$~\kms\ at 0.5$^\circ$ distance (feature `2br' in Figure~\ref{fig:zoneposvela}). This feature (2br) may be the continuation of the wedge defined by 1cb and 1cr seen in Zone~1.
\citet{Escala2022} also recently reported a wedge-shaped distribution in Zone 2 based on five pencil-beams at distances between $1-2^\circ$ from M31.

The position-velocity plots for Zones 3 and 4, which are radially opposite from Zone 1, show a narrow
blueshifted feature with kinematics similar to that of the GSS in Zone 1 (feature `3ab' in Figure~\ref{fig:zoneposvela}
and `4ab' in 
Figure~\ref{fig:zoneposvelb}. The stars comprising the feature are widely distributed spatially across both zones. 
Perhaps these are stars that were once in the GSS and have passed back through M31 to the northern side of the galaxy. 
Such features do appear in merger simulations (e.g., that discussed in Section~\ref{sec:cosmo}). 
Zone 3 also includes a hint of a narrow redshifted feature (feature `3ar' in Figure~\ref{fig:zoneposvela}), which is likely related to feature 2ar in Zone 2. Unlike the wedge associated with 2ar, the wedge associated with 3ar is mostly empty. A striking feature of Zone 3 is the presence of a group of about 40 stars that define a short ``stub'' in the position-velocity diagram, defined as 3br in Figure~\ref{fig:zoneposvela}. These stars appear to be at the northeastern and southwestern edges, respectively, of the overdensities defined as the Northern Spur and the Northeast Clump by \citet{mackey2019}. The 3br feature is notable for its small velocity dispersion of $4.6\pm1.5$~\kms\  despite its stars covering the width of the zone.
Apart from features 3ab, 3ar and 3br, 
the rest of the stars in Zones 3 are preferentially redshifted, as in Zone 2, and scattered across position velocity space. 
Zone 4 also shows a preference for redshifted stars. The major feature in Zone 4 is a more completely filled wedge bordered by 4br and 4bb.


Finally, the position-velocity plot for {\bf Zone 5} shows a chevron pattern (i.e., the outline of a wedge-like shape; labeled 5b and 5r),  
similar to the shape expected for radial shells \citep{Merrifield1998}.
A similar feature was reported for the Western Shelf region by \citet[see their Figure 8]{Fardal2012} based on spectroscopy of stars in a narrow strip along the minor axis of the M31.
Here, the stars that make up the chevron pattern are broadly distributed across the Western Shelf feature in Zone 5. 
The stars that make up the red- and blue-shifted edges (5b and 5r) spatially overlap each other as expected for an umbrella-like fan viewed tangentially \citep{Merrifield1998}.  The 5b/5r chevron pattern overlaps the edge of the filled wedge bordered by 4bb/4br (Figure~\ref{fig:allfeaturesannotated}).


\begin{figure}[th]
    \centering
    \includegraphics[width=0.65\textwidth]{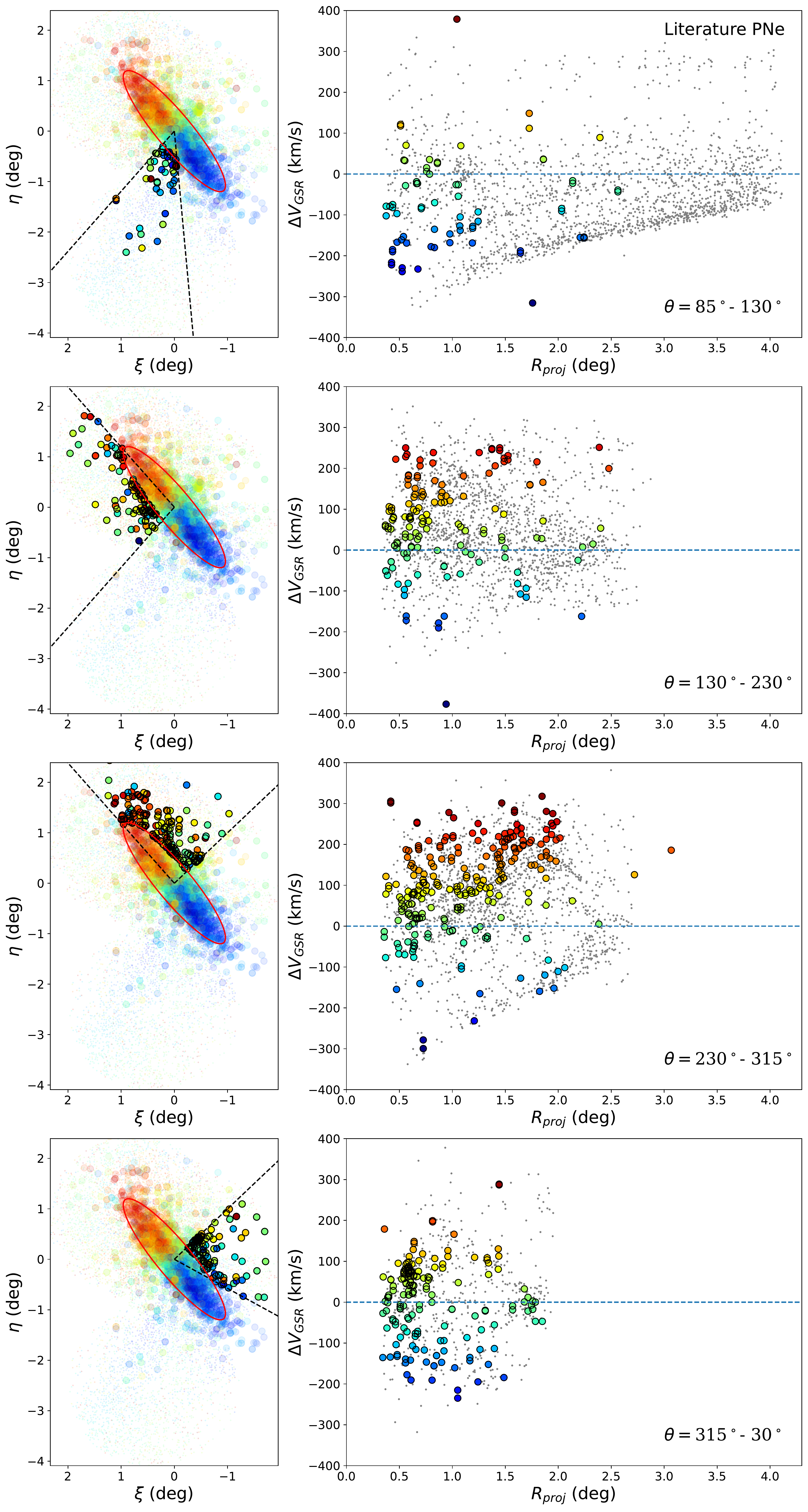}
    \caption{The distribution of known PNe from the literature (large colored circles; see the text) overlaid on M31 stars (smaller points).}
    \label{fig:pnlit}
\end{figure}

\subsection{Comparison to Planetary Nebulae from the Literature}

Figure~\ref{fig:pnlit} compares the spatial and velocity distributions of M31 stars with those of planetary nebulae (PNe) reported in the literature. The angular zones shown are the same as those shown in Figures~\ref{fig:zoneposvela} and \ref{fig:zoneposvelb}, with the exception that Zones 3 and 4 are combined. The PNe shown were identified using SIMBAD and the MMT/Hectospec archive, and are the result of a large body of work by many authors 
\citep[see][and references therein]{Merrett2006,Yuan2010,Sanders2012,Bhattacharya2019}.
The comparison shows that the known PNe that lie beyond the main disk of the galaxy trace the same kinematic structures visible in the DESI data. The similarity is apparent in all spatial regions of M31, but most strikingly in the regions shown in the bottom two panels. In the angular range $230^\circ < \theta \le 315^\circ$ (zones 3+4 in Figures~\ref{fig:zoneposvela} and \ref{fig:zoneposvelb} covering the northern portion of the Northeast shelf), the PNe are preferentially redshifted, echoing the distribution of the stars, and roughly demarcate the two wedges visible in the stellar data. At $315^\circ < \theta \le 30^\circ$ in the Western Shelf (bottom panel, zone 5 in Figure~\ref{fig:zoneposvelb}), the PNe trace the red- and blue-shifted edges of the chevron. \citet{Fardal2007}
previously pointed out how the PNe in the Western Shelf preferentially fall near the boundary of a triangular region in position-velocity space. The present comparison shows how the PNe distribution echoes the more densely sampled stellar distribution over much of the inner halo, as expected. We compare to the distribution of star clusters and dwarf galaxies in \S~\ref{sec:NatureofProgenitor}.

\subsection{Metallicities}
\label{sec:metallicities}


Photometric studies have demonstrated that the M31 halo shows a wide range of stellar metallicities with much of the substructure being metal rich \citep{Ibata2001,Brown2006,Ibata2007,Gilbert2009a,Gilbert2009b,Ibata2014,Conn2016}. Spectroscopy from Keck/DEIMOS has not only found evidence for a low-metallicity halo component that is detectable both in the inner regions and at large distances, but also confirmed that the stars associated with some of the kinematic substructure are metal rich \citep[e.g.,][]{Guhathakurta2006,Kalirai2006a,Kalirai2006b,Gilbert2020,Escala2020a,Escala2020b,Escala2021}. 
The metallicity of the Western Shelf (measured photometrically) is the same as that of the GSS 
\citep{Fardal2012, Tanaka2010},
with a typical metallicity of [Fe/H] = --0.7 for the satellite debris and --1.2 for the spheroid component of M31. 
Since the selection of targets for the DESI observations presented here is biased toward redder colors (and thus higher-metallicity populations) and does not sample the metal-poor RGB populations, we cannot use the DESI data to infer directly the metallicity distributions in the different kinematic components.
However, we do find significant numbers of metal-rich stars across all regions surveyed. 
For the stars in the region of the GSS, Figure~\ref{fig:zone2feh} shows that we measure similar median metallicities in the three different kinematic components (the median metallicities in 1ab, 1bb, and 1cb are $-0.33$, $-0.26$ and $-0.32$, respectively, with all the observed stars in this zone -- represented by the dashed line -- showing a median metallicity of $-0.37$). 

\begin{figure}[ht]
    \centering
    \includegraphics[width=0.49\textwidth]{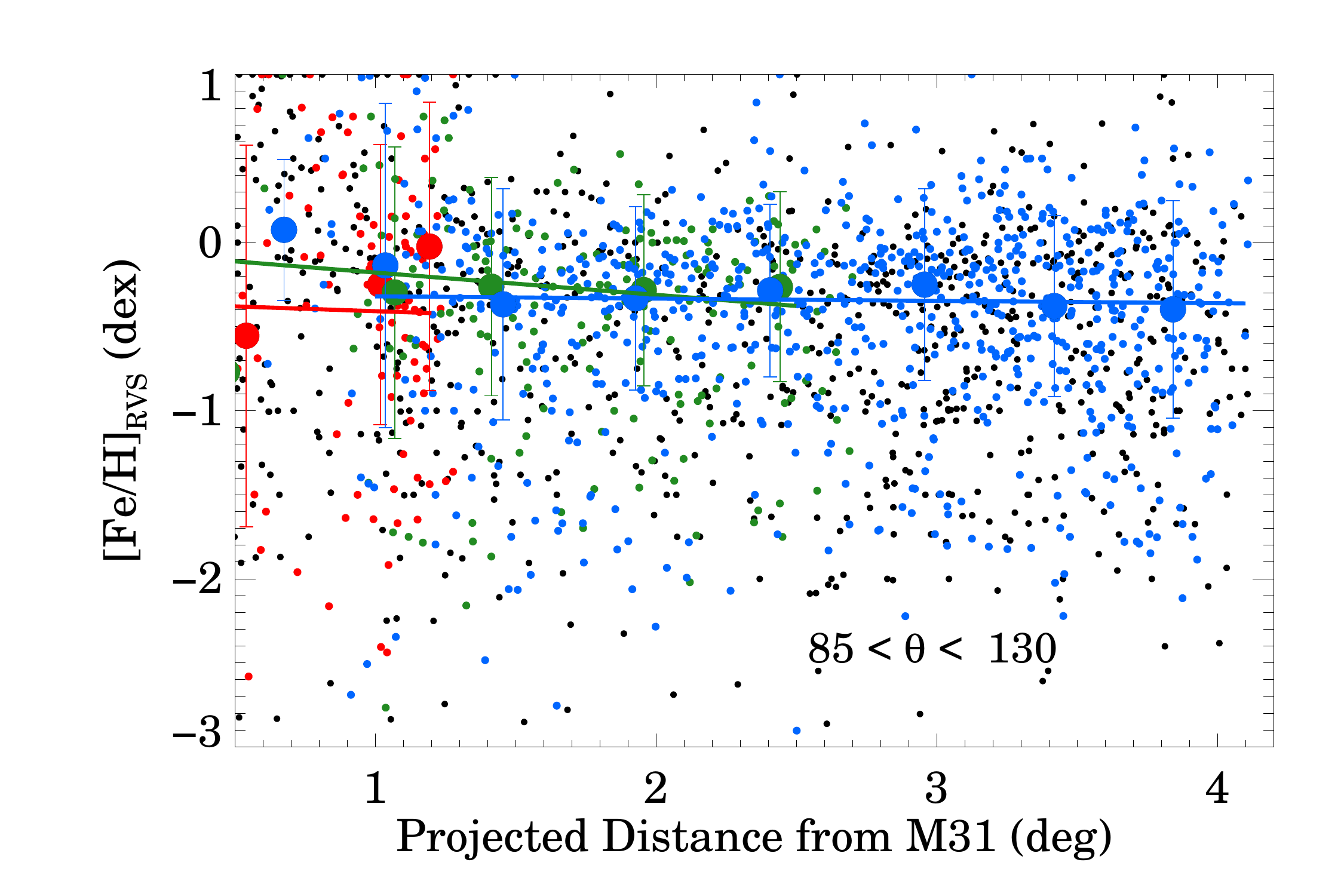}
    \includegraphics[width=0.49\textwidth]{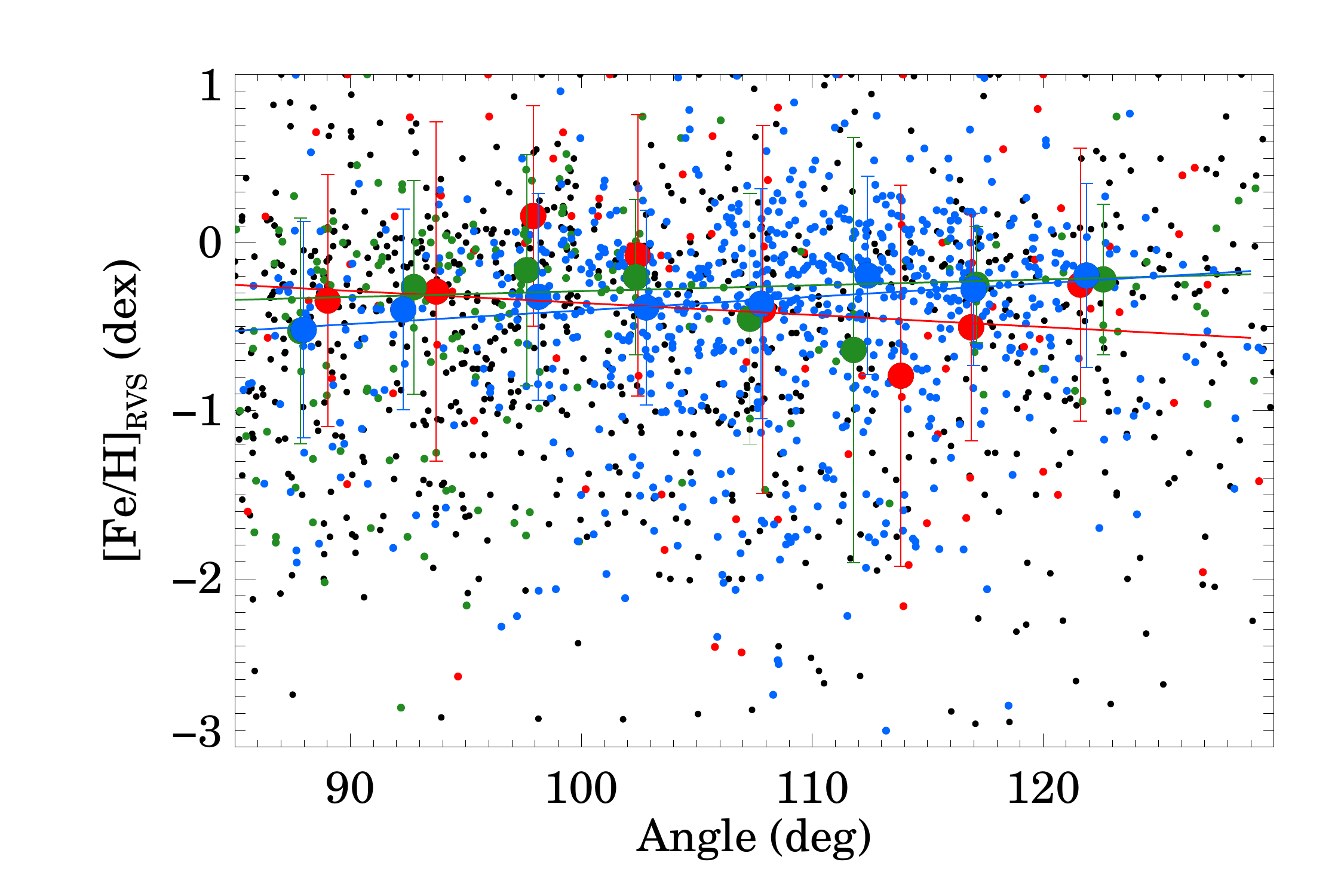}
    \caption{Variation of [Fe/H] as a function of position in the region of the GSS
    as a function of projected radial distance from M31 (left) and angle around M31 (right). The blue, green and red dots respectively represent stars associated with the three kinematic structures 1ab, 1bb, and 1cb shown in the left panel of Figure~\ref{fig:zone2feh}; the black dots represent the remaining stars in Zone 1. The large solid dots with error bars represent the median values and the 1$\sigma$ scatter, respectively, in the [Fe/H] values in equal bins of projected distance or angle. The solid lines show least absolute deviation fits to each subset. There is no significant variation in the mean metallicity 
    in either direction within the DESI sample.}
    \label{fig:zone2fehvsposn}
\end{figure}

The overall distribution of metallicities is remarkably similar to that presented by \citet[][see their Figure~11]{Fardal2012}, showing a skewed distribution with a tail to lower metallicities. 
This similarity is surprising given that our target selection is biased toward the high-metallicity regions of the color-magnitude diagram. The presence of lower-metallicity stars in our sample may result from photometric scatter in the PAndAS data (i.e., with the more metal-poor stars scattering into our selection region). Nevertheless, assuming that the DESI data only sample the high-metallicity tail of the distribution, the measurements suggest [Fe/H]$\lesssim-0.4$ is a strong upper limit to the median metallicity in these regions. There is weak evidence that the metallicity distribution in the 
compact wedge component in Zone 1 (1cb; shown by the red points in Figure~\ref{fig:zone2feh}) is flatter (i.e., stretching to higher metallicities) than the main 1ab (GSS) and 1bb components. However, this component may also be contaminated by stars from the 
M31 disk and bulge.

We see no significant variation in the metallicities in the region of the GSS (Figure~\ref{fig:zone2fehvsposn}) either along the radial direction (left panel) or with azimuthal angle (right panel). Previous photometric studies have reported spatial variations of the metallicity: \citet{Conn2016} found that the metallicity in the GSS region increases from [Fe/H]~$\approx-0.7$ at $R_{\rm proj}\approx1^\circ$ to about $-0.2$ near $R_{\rm proj}\approx2.8^\circ$, and then decreases steadily to [Fe/H]~$\approx -1$ at $R_{\rm proj}\approx5.9^\circ$. The pencil-beam spectroscopic metallicity estimates by \citet{Escala2021} found a gradient of $-0.25$ dex/degree, even stronger than those reported by \citet{Conn2016}. 
While the DESI measurements in Figure~\ref{fig:zone2fehvsposn} show a high mean value of the metallicity, they also show large scatter with no statistically significant systematic trends.   However, we caution that these results may be due to our biased selection of targets, and a more comprehensive study of the metallicity variations would require a more complete sampling of the low-metallicity portions of the RGB (by future observations) and a careful accounting of the selection function. 






\section{Comparison to Simulations} 
\label{sec: simulations}


\begin{figure}
    \centering
    \includegraphics[width=1.\textwidth]{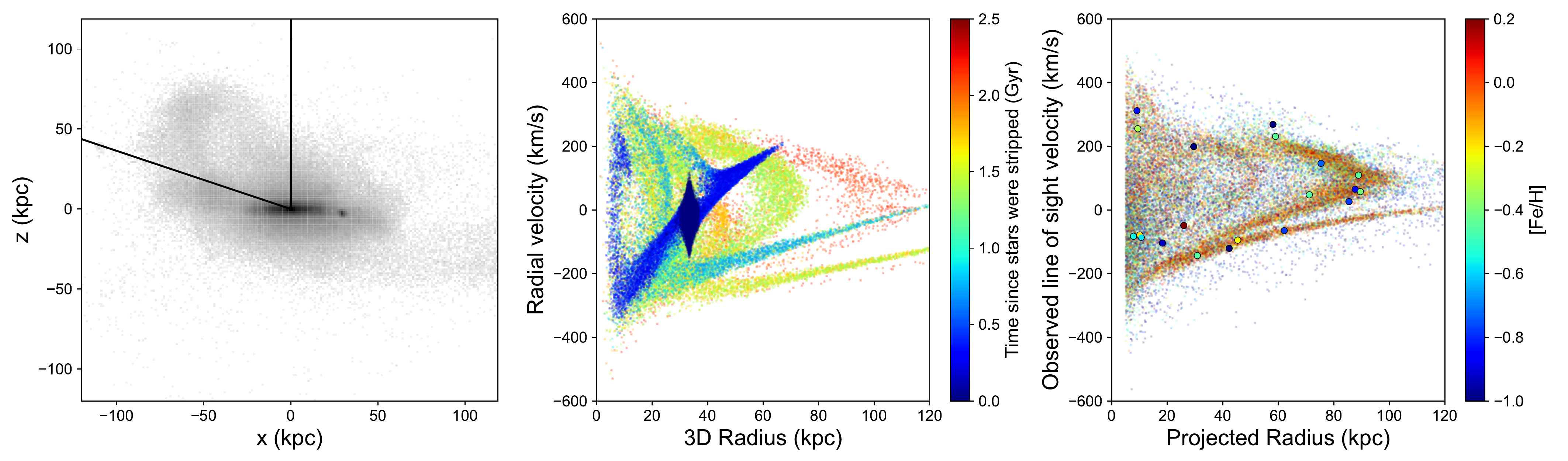}
    \caption{{\it Left:} Distribution of stellar particles in an M31 analog from the TNG-50 simulation. The system is matched in stellar mass to M31 and, like M31, has a prominent disk and  large stellar halo, and experienced a recent encounter with a massive satellite (see the text for details). 
    This particular analog shows a giant stream, numerous shells, and a compact core from the still-bound portion of the 
    large accreted satellite at (x,z)$\approx$(30,$-$2.5)~kpc. 
    {\it Center:} Particles from the merging satellite in 3D radius vs.\ radial velocity, color-coded by when they were stripped from the satellite. The distribution shows complex shell structure from a single large progenitor.  {\it Right:} The distribution of accreted stellar particles (from all satellites)  
    in the upper left wedge indicated in the left-hand panel, color-coded by metallicity. The larger solid points show 9\,Gyr old to 12\,Gyr old particles selected as possible (poor) proxies for globular clusters. In contrast to the globular cluster-like particles, which show relatively little kinematic substructure, the most metal rich stars display rich substructure, which results entirely from the recent large merger. 
    }
    \label{fig:tng50}
\end{figure}

\subsection{Comparison to Galaxy Formation Simulations in a cosmological context} \label{sec:cosmo}

Simulations of galaxy formation in a cosmological context illustrate how 
mergers can generate complex, organized structure similar to that observed in M31. 
To illustrate this point, we show in Figure 13 an example 
of a system like M31 which experienced a fairly massive merger in the last few Gyr. 
This  example is not meant to replicate M31 in any detail, but is provided only to illustrate how streams and shells emerge naturally in cosmological simulations. 
The example is taken from the TNG-50 simulation \citep{Pillepich2018,Pillepich2019}, which
%
simulates a large cosmological volume (51.7 Mpc on a side) with high
resolution (300\,pc softening length for the collisionless particles), enabling an analysis of the detailed kinematics of merger
debris. 

To identify this system within the simulation, 
we began by selecting systems 
with properties similar to that of M31 \citep{Ibata2014,dsouza2018,dsouza2021}, i.e., systems with stellar masses 5$\times 10^{10}\,M_{\odot}$ to 15$\times 10^{10}\,M_{\odot}$. We selected galaxies that have a prominent disk by requiring that more than 40\% of stars are on orbits that have a circularity $\epsilon = J_z/J(E) > 0.7$, where $J_z$ is the specific angular momentum of a particle around the angular momentum axis of the stellar body of a galaxy, and $J(E)$ is the maximum angular momentum of the 100 particles with the most similar total binding energies (see also \citealt{genel2015}). In addition, we required that the galaxy have 
a total accreted stellar mass of at least $3 \times 10^9\,M_{\odot}$ and have had 
an encounter with a massive satellite (with stellar mass $M_{\rm sat} > 10^{10}\,M_{\odot}$) that
fell into the system 2 Gyr to 8 Gyr ago. We then examined recent
snapshots of these systems for visual analogs of M31's giant stream and shells. The best match
is subhalo ID 482155, which has a present-day dark halo mass of 2.2$\times 10^{12}\,M_{\odot}$ and a present-day stellar mass of 1.2$\times 10^{11}\,M_{\odot}$, and is in the process of accreting a large satellite (stellar mass
$10^{10}\,M_{\odot}$) that experienced first infall 6.7 Gyr ago. 
As the merger is still underway, the dissipating satellite retains 
a compact core of stellar mass $1.8 \times 10^9\,M_{\odot}$ located $\approx$30\,kpc from the center of the
primary galaxy (see Figure \ref{fig:tng50}, left panel); the median metallicity of all the particles from this
massive satellite is nearly solar, with ${\rm [Fe/H]} = -0.07$. The
satellite is no longer star-forming, but underwent star formation as
recently as 2.5\,Gyr ago.

Fig.\ \ref{fig:tng50}, which shows three different projections of this
system,  
illustrates how the merger of a single progenitor galaxy can generate a stream, multiple shells, and nested wedges in phase space, similar to those seen in M31.
The left panel shows the projected stellar mass density in greyscale with logarithmic scaling. The giant stream analog
is clearly visible, as are shell structures and the compact core of
the satellite (seen as the dark dot near x$\approx$30, z$\approx-$2.5). The center panel shows the overall kinematic structure of
the stellar particles from the 
infalling satellite
using the conventional (simulation) visualization of  radial velocity (centered in the frame of the M31 analog) as a function of radius (in 3D rather than projected coordinates), in direct analogy to e.g., Fig.~10a from \citet{Fardal2007} or \citet{Pop2018}.
Particles are color-coded by the time when they
were last part of the satellite's subhalo (prior to their tidal
stripping), showing a clear progression in which the outermost tidal
debris arises from earlier episodes of stripping (e.g., red and green points), and material near the still-bound core of the satellite is the most recently stripped (darker blue). 

Finally,
the right panel shows the line-of-sight velocity as a function of
projected radius for stars in the angle wedge containing the giant stream
analog, color-coded by metallicity. 
Although the contributions from all merged satellites are shown in this panel (not just that of the most recent merger as in the center panel), the earlier accreted satellites are all low mass and they merged long enough ago that they no longer contribute fine-scale kinematic structure \citep[e.g.,][]{Beraldo_e_Silva2019}. As a result, the most recent merger completely dominates the properties of the inner halo. Indeed, {\it all} of the substructure in this particular halo is metal rich, and arises from the
stripping of this most massive satellite. 
We discuss this topic further in Section~\ref{sec:discussion}.

\subsection{Comparison to an N-Body Model}
\label{sec:nbody_model}


The DESI observations can be compared in greater detail to simulations that have been customized to replicate the structure of M31. 
In order to understand whether a single encounter could account for much of kinematic structure observed in our data, we constructed a simple model informed by the results of previous studies. 
Previous modeling efforts in this field are described in greater detail in Section~\ref{sec:discussion}.
We also publish all our good velocity measurements to aid future modeling attempts. It is our hope that these observations, insights from the cosmological models, and the comparisons presented here can inform future modelers in their efforts to reproduce more of the density and phase-space structure of M31's halo. 

%
The model consists of a single component Plummer sphere \citep{Plummer1911} describing the progenitor of the GSS, 
which has a total mass of $\sim 2 \times 10^8\,M_\odot$, a half-mass radius of 1 kpc, and is represented by 300,000 particles. The model does not distinguish between dark matter and stellar particles. M31 is represented by a static analytic potential that consists of the disk, halo, and a bulge, where we use parameters similar to those employed in previous modeling efforts  \citep{Fardal2006,Kirihara2017a}. The bulge is a Hernquist bulge \citep{Hernquist1990} with a mass of $3.2\times 10^{10}\,M_\odot$ and size $a=0.5$\, kpc. The dark matter halo is described by an NFW profile \citep{NavarroFrenkWhite1997} with $V_{\rm max}=215$\,\kms\ and a scale radius of $r=7.63$\,kpc. The disk is assumed to have  
an exponential scale length of $r_d=5.4 $\,kpc,  vertical height $h=0.6$\,kpc, and a total mass of $M_{\rm disk} = 3.7\times 10^{10} M_\odot$; it is represented as a linear combination of 3 Miyamoto-Nagai disks, following the prescription of \citet{smith2015}. The progenitor falls in on an approximately radial orbit.
Further details regarding the simulation are provided in Appendix B.

\begin{figure}
    \centering
    \includegraphics[width=0.8\textwidth]{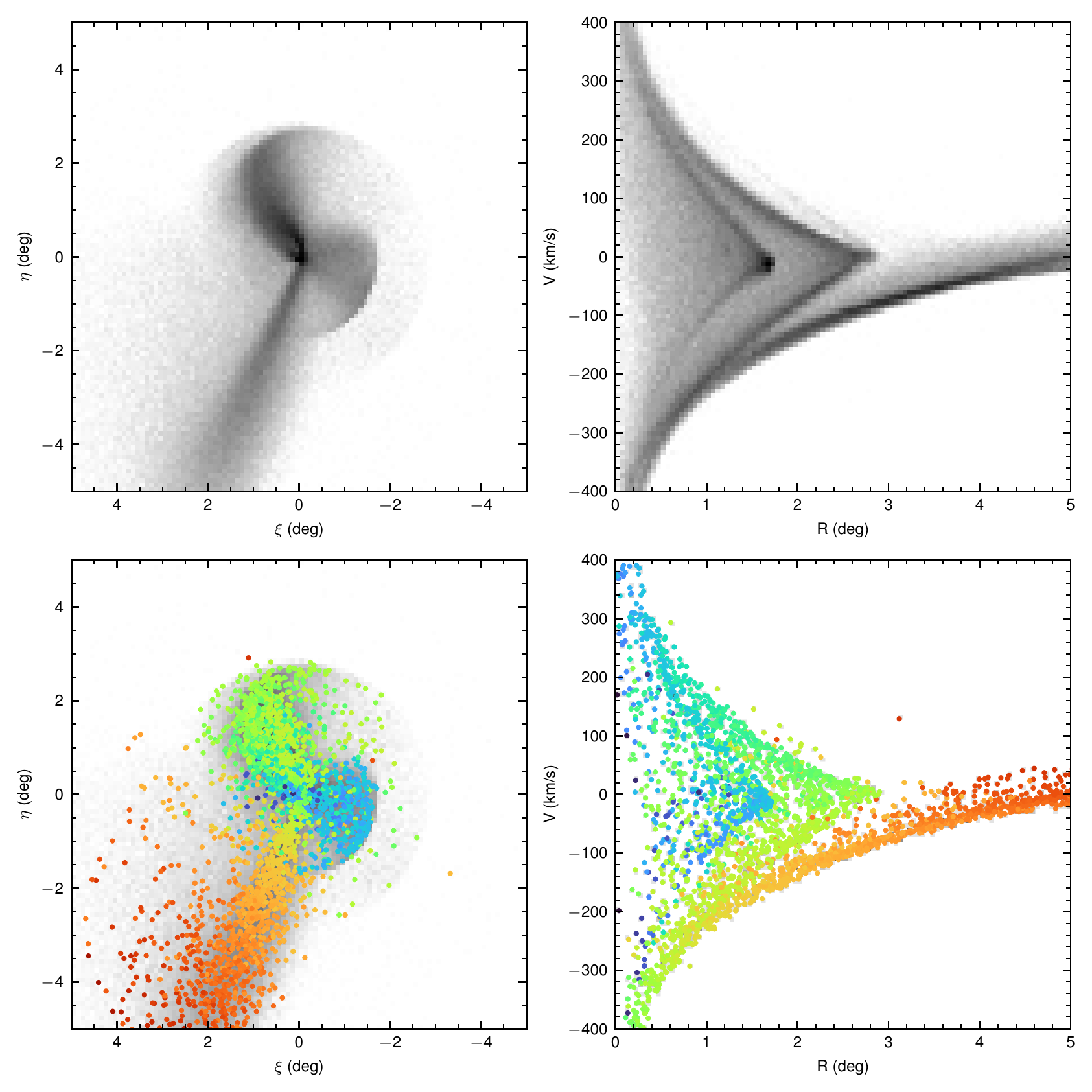}
    \caption{N-body model of the interaction between a progenitor and M31. Top panels show the spatial distribution
      (left) and projected position-velocity diagram (right) of particles 791 Myr after the start of the simulation (or 585 Myr after the pericentric passage of the GSS progenitor).
    In the bottom panels, 
    a random 0.1\% of the simulation particles are color-coded by their total energy (kinetic + potential) in the host potential. 
    The progenitor is fully disrupted on the first encounter. As a result, the particles in the shell system are simply arranged by energy \citep[][as  energy is directly related to orbital period]{DongPaez2022}, and 
    the southern stream (orange points) and  shells (green, cyan, and navy points) are cleanly separated in energy. The similarity to the M31 observations suggests that the observed structures could result from a single encounter, with the nested structures being subsequent wraps (i.e., different pericentric passages) of stars from the same progenitor.}
    \label{fig:Sergey_model}
\end{figure}
Figure~\ref{fig:Sergey_model} shows results 791 Myr after the start of the simulation. The initial pericentric passage of the GSS progenitor occurred at 188 Myr. While our simulation has not been tuned to match the data perfectly, it does provide a heuristic interpretive guide to the complex kinematic structures in M31. 
In the bottom panels of Figure~\ref{fig:Sergey_model}, a random selection of 0.1\% of particles are color-coded by their total energy (i.e., kinetic + potential), with the color range extending from red representing particles with the least negative total energy, to blue representing the most negative (i.e., most tightly bound) particles.

Unlike the higher-mass merger in  Section~\ref{sec:cosmo} that retains a bound remnant to the present day, this progenitor
is fully disrupted on the first apocentric passage; thus the resulting set of shells can be understood as the debris from one 
disruption event, arranged according to energy \citep{DongPaez2022}.  
The GSS-like southern stream (orange-red points) and the nested shells (green, cyan, and navy points) are cleanly separated in energy. Specifically the stars with the least negative energy have not yet had a second pericentric passage after being stripped, 
while the particles with the most negative energies and therefore much shorter orbital period (the cyan points), have already had multiple pericentric passages after the initial stripping episode.

Comparing Figures~\ref{fig:Sergey_model} and \ref{fig:allfeaturesannotated}, we see that the main part of the GSS (feature 1ab) is similar to the orange-red points in the simulation; and that the shells denoted by the structures 2a, 4b+5 and 1c+2b (i.e., the Northeast Shelf, Western Shelf, and Southeast Shelf) are similar to the simulation points shown in green, cyan, and navy, respectively, in Figure~\ref{fig:Sergey_model}. Stars with the most negative energies (navy colored points), located in the southeastern sector of M31, are the stars from the leading part of the debris. 

\begin{figure}
    \centering
    \includegraphics[width=0.4\textwidth]{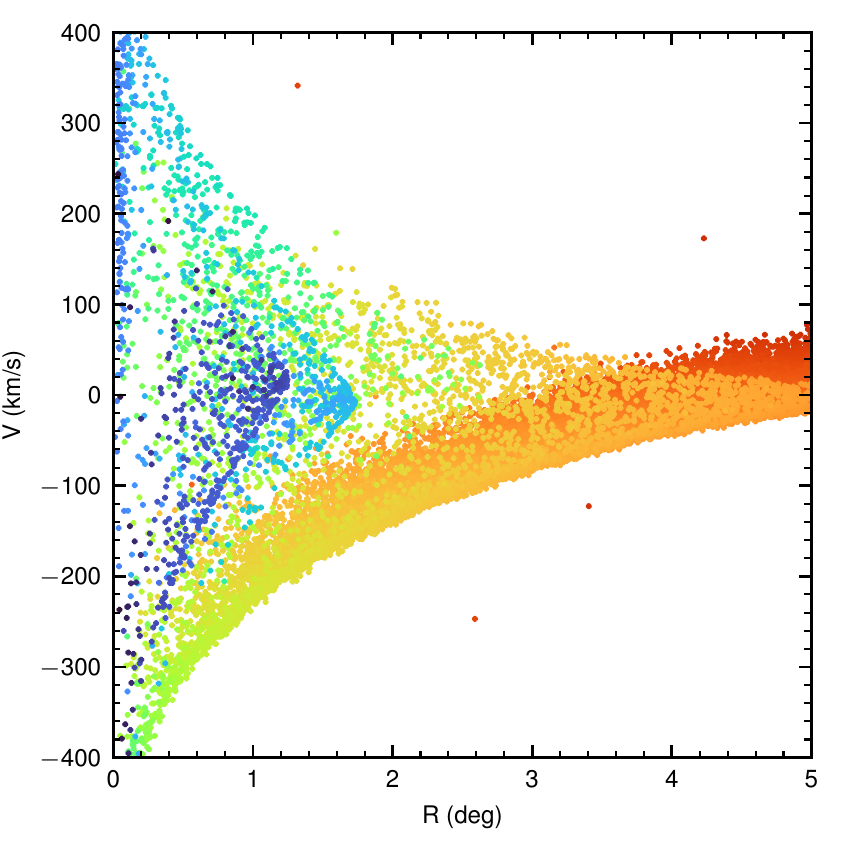}
    \caption{Velocity vs.\ projected distance of particles 
    in the southeastern sector of the simulation at 
    770 Myr (a slightly earlier time than that shown in  Figure~\ref{fig:Sergey_model}).
      The navy coloured particles (those with the most negative energies)  do not form a full chevron, only the bottom half of it, as the stars in this structure have not yet reached their apocenter.
      The resulting feature is reminiscent of feature 1bb in M31, despite not quite matching its range in projected distance.
    }
    \label{fig:kcc_rvr}
\end{figure}

Figure~\ref{fig:kcc_rvr} shows only the radial velocities of particles in the southeast sector, using the same energy color-mapping scheme as in Figure~\ref{fig:Sergey_model}. Interestingly, the leading particles in position-velocity space do not occupy a full chevron, but instead primarily trace out a locus at negative velocities, because many of these stars have not yet experienced a turnaround at apocenter. The 1bb feature in M31 may have a similar origin. Similar to the situation shown in Figure~\ref{fig:kcc_rvr}, 1bb appears in the same sector as the GSS (equivalent to the orange points in the Figure), is blueshifted, and has no companion redshifted feature that would create a chevron-like pattern.


The results suggest that the multiple structures observed in M31 could arise from a single encounter, with the various nested structures produced by subsequent wraps (i.e., different pericentric passages) of stripped stars from the same progenitor. Importantly, in our simulation, the progenitor is fully disrupted in the encounter, so all of the shells that result are essentially a single set of stars wrapping around the galaxy. 
If the progenitor instead preserves
some mass for a second pericentric encounter, 
as in the cosmological model discussed in Section~\ref{sec:cosmo},
an additional set of shells would be created;
these are unaccounted for in our current simulation. Simulating the shell system from a possible second pericentric encounter would be somewhat more challenging, as the dynamical friction of the progenitor likely would need to be taken into account. This second shell system is probably needed to explain some of the
smaller chevrons observed in our data.
%
%
%
Future observations that more densely sample the structures in the inner halo will provide a unique opportunity to constrain the mass and orbit of the progenitor. 


\smallskip

\section{Constraints on the Mass of M31} 
\label{sec:MassConstraints}

\subsection{Shell Kinematics} 
\label{sec:ShellKinematics}

The shell-shaped tidal signatures of galaxy mergers that we observe also offer the opportunity to measure the gravitational potential of the host galaxy, with nested shells probing the gravitational potential as a function of galactocentric distance
\citep{Merrifield1998, Sanderson2013}.
Our current sample measures radial velocities for stars in multiple shells spanning a range of distances and thereby offers a rare opportunity to constrain the dynamical mass of the galaxy as a function of galactocentric distance using this technique. 

As described by \citet{Merrifield1998}, for a shell oriented in the plane of the sky, the projected velocity of the shell has a distinctive triangular shape as a function of projected distance (a filled `wedge' or empty `chevron' shape), and the slope of the projected velocity near the outer edge of the shell can be used to infer the gravitational potential. That is, for a spherical shell of radius $r_s$ with a projected velocity $v_{\rm los}$ that increases with decreasing projected distance $R,$ the velocity gradient   
$d v_{\rm los}/dR = -\Omega,$
where $\Omega$ is the circular frequency at $r_s.$
\citet{Sanderson2013} derived a related expression that includes the effect of the outward velocity of the shell (their eq.~23) and argued that the simpler \citet{Merrifield1998} method will tend to overestimate the enclosed mass.

The multiple shell structures observed in M31 allow us to explore these ideas. 
To explore how well the simple \citet{Merrifield1998} prescription recovers the expected gravitational potential of M31, we measured (by eye) the velocity gradient of the red- and blue-shifted edges of the wedges and chevrons seen in the Northeast Shelf, the Western Shelf, and in the region of the GSS. These features have approximate projected extents of $\sim 1.3^\circ$ (Features 1cb/1cr and 2br), $\sim 2^\circ$ (Features 2ar and 3ar), and  $\sim 2.7^\circ$ (Features 4bb/4br and 5b/5r), which correspond to  projected distances of $\sim 19$ kpc, $\sim 28$ kpc and $\sim 38$\,kpc. 

The lines shown in the insets of Figures~\ref{fig:zoneposvela} and \ref{fig:zoneposvelb} show the regions used to estimate the velocity gradients. Slopes are better defined for features that are densely populated (e.g., 2ar). Feature crowding, the possibility that features overlap each other or are embedded in a distributed background halo, can make it difficult to define the slope of a feature (e.g., 2br resides within 2ar). Higher density spectroscopy of the M31 halo can potentially mitigate these challenges. 

\begin{figure}
    \centering
    \epsscale{1.15}
    \plottwo{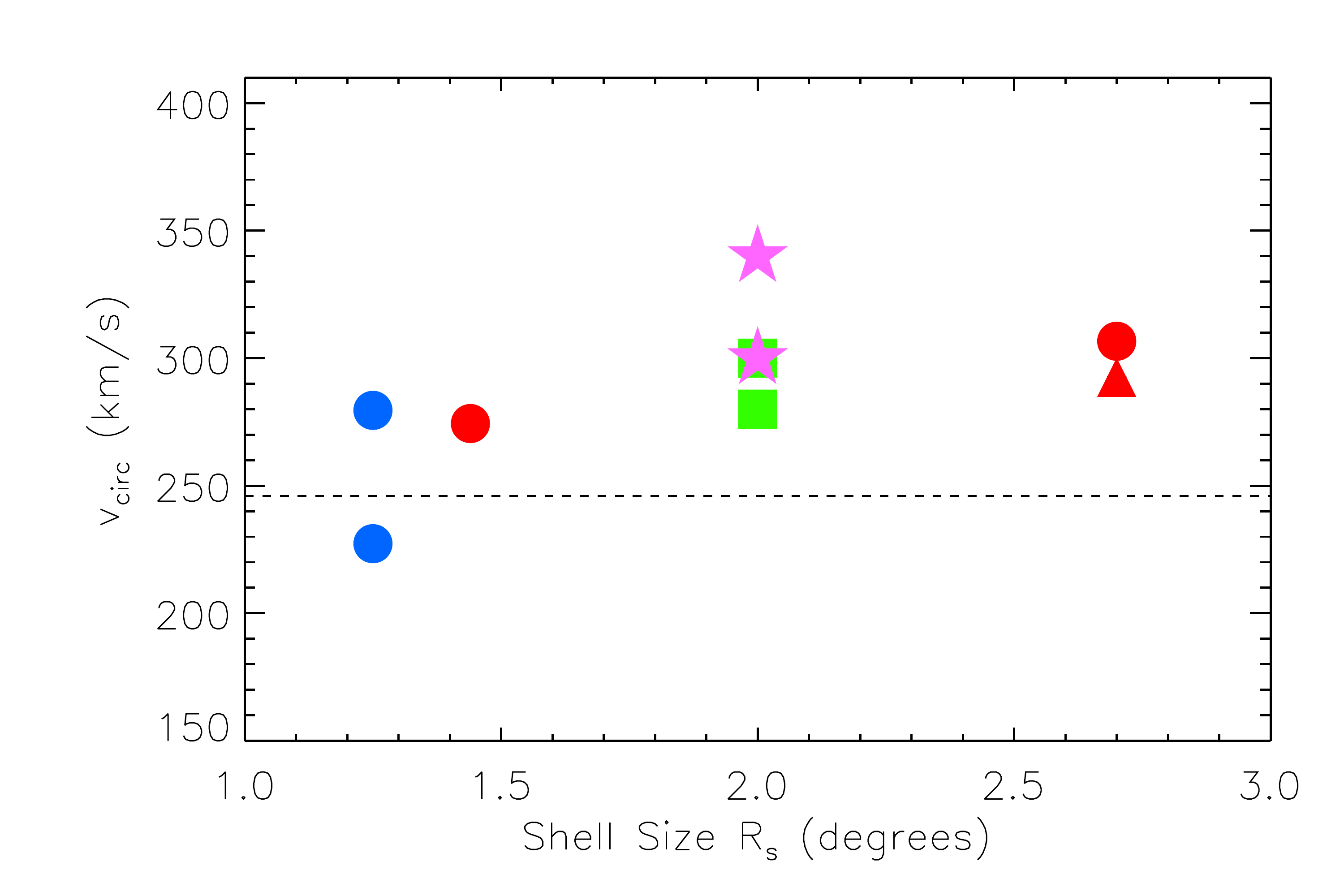}{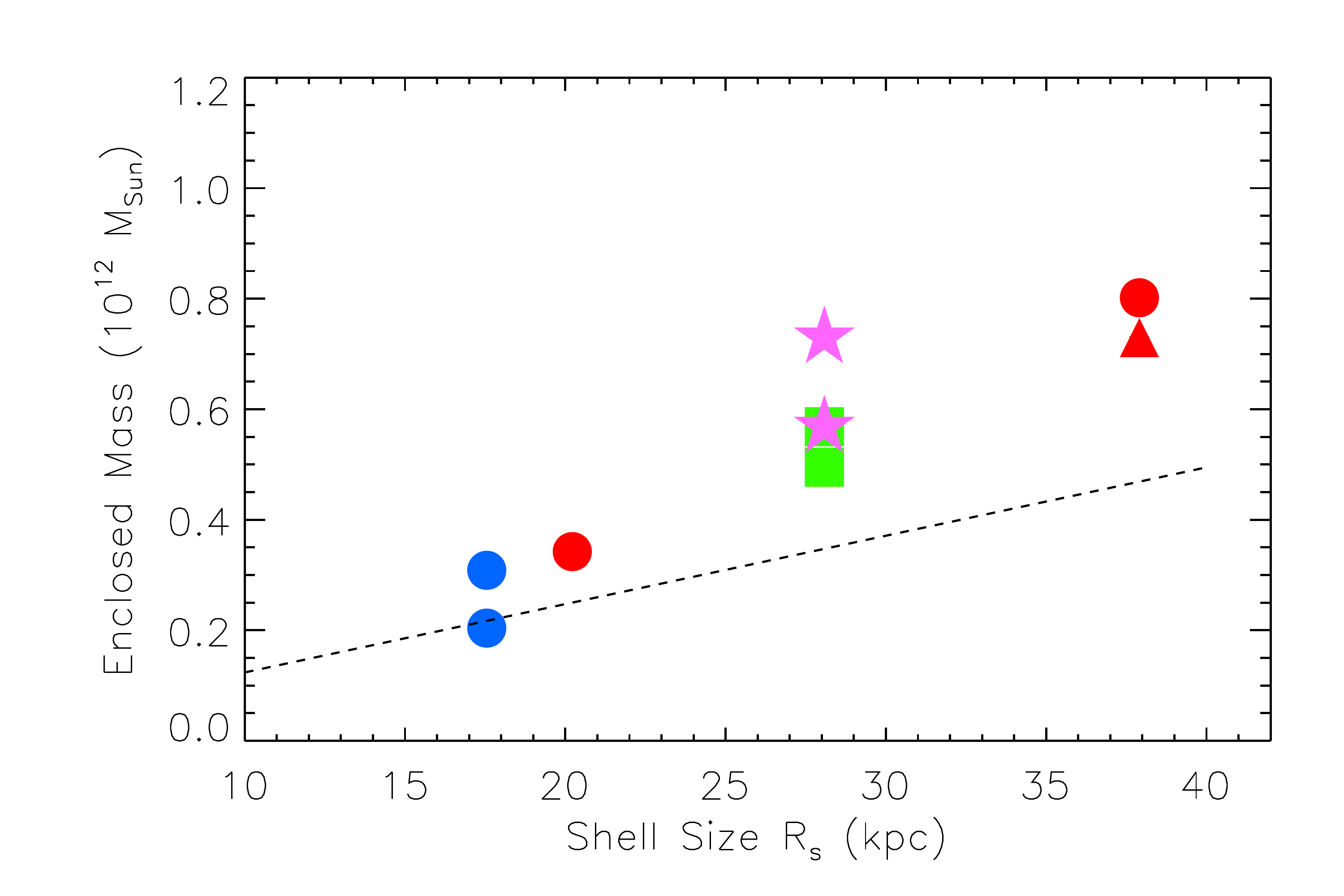}
    \caption{Circular velocities (left) and enclosed masses (right) inferred from the velocity gradients ($dV_{\rm los}/dr$) of linear features in line-of-sight velocity and projected distance in Zone 1 (blue dots), 2 (red dots), 3(red triangles), 4(green squares), 5 (pink stars). Dashed lines show the typical circular velocity of the flat portion of the HI rotation curve of M31 \citep{Chemin2009} and the corresponding enclosed mass for a flat rotation curve out to $\sim 40$~kpc.}
    \label{fig:vcirc_Me}
\end{figure}

The measured slopes correspond to 
circular velocities of 230~\kms\ to 340~\kms\ at the shell radius (i.e., the apex of the wedge or chevron) and imply enclosed masses of $2\times 10^{11}\,M_\odot$ to
$8\times 10^{11}\,M_\odot$
over this range of distances (Figure~\ref{fig:vcirc_Me}).
The circular velocities are 
similar to, or larger than, the velocity of the \ion{H}{1} rotation curve of M31 measured over the same range of radii  (horizontal line in left panel of Figure~\ref{fig:vcirc_Me}). The rotation curve, which is roughly flat at $\sim 250$~\kms\ from 10 kpc to 40 kpc, 
implies an enclosed mass that is $4.7\times 10^{11}\,M_\odot$ within 38 kpc \citep{Chemin2009}
and declines toward smaller radii as $1/r$ (dashed line, right panel).  

In comparison, the circular velocity and enclosed mass derived from the properties of the smallest shell (radial extent $\sim 1.3^\circ$) are close to the values inferred from the \ion{H}{1} rotation curve at the same distance. The values for the larger shells (radial extents of 2.0$^\circ$ and 2.7$^\circ$) are larger than the corresponding values from the \ion{H}{1} rotation curve.  
%
These results are consistent with the findings of \citet{Sanderson2013}, that the \citet{Merrifield1998} prescription can correctly recover the enclosed mass in some cases, but that it often overestimates enclosed mass by a factor of 2 to 3. A similar result was reported by \citet{Escala2022} in their analysis of the kinematics of stars in a portion of the Northeast Shelf. 

In summary, mass estimates from the 
observed velocity gradients of shells rely on the assumption of shells of stars  oriented in the plane of the sky and do not account for complexities introduced by geometry, angular momentum, or the details of the interaction. Consequently, while the overall idea of using shells to estimate the mass of M31 is potentially useful, it is clear that more sophisticated modeling and more extensive spectroscopic samples will be necessary to reach an interesting level of accuracy. 


\subsection{Kinematics of the GSS}
\label{sec:gssmassest}



As a complementary approach, stellar streams like the GSS also probe the galactic potential and can plausibly be interpreted using a more detailed dynamical model that is driven by a few simple assumptions. 
The shells from a single pericentric passage represent a group of stars on a  sequence of orbits ordered by energy \citep[see a detailed exposition of shell formation in][]{DongPaez2022}. Stars with 
the most negative energies have 
the shortest orbital periods, while 
stars with less negative energies have longer orbital periods. Thus the shell system will have an energy gradient. Although the energy gradient makes the analysis of the shells more cumbersome, as we cannot rely on the constant energy assumption that approximately works for thin tidal streams \citep{Koposov2010}, we can still  effectively use the assumption that the energy changes monotonically along the structure due to energy sorting in the shell. 
The strength of the energy gradient in the shell is itself limited by the total energy spread in it, which in turn is determined by the energy spread of stars at the pericentric passage of the progenitor i.e $\delta E \sim \frac{1}{2} (V_{\rm peri}+\sigma)^2-(V_{\rm peri} -\sigma)^2 = V_{\rm peri} \sigma$, where $\sigma$ is the velocity dispersion of the progenitor and $V_{\rm peri}$ is the velocity of the progenitor at the pericenter \citep[see][for more details]{DongPaez2022}.
It turns out that these basic principles, together with a few assumptions about the GSS geometry, can help us model the radial velocity vs.\ distance behavior observed in the GSS and constrain the M31 gravitational potential.

To define the model we begin by defining a coordinate system $x,y,z$ in which the $z$-direction is oriented along the line connecting the Sun and M31 pointing away from the Sun, the $x$-direction is aligned with the East, and the y-direction points North. 
Projected on the sky, the GSS forms an essentially linear structure, with position angle $\phi_{GSS}\sim $155$^\circ$. We assume that the GSS is also a linear structure in 3D, defined by the unit vector ${\bf \hat k} = [k_x,k_y,k_z]$ and that the line defined by this vector intersects the projected center of M31 (i.e., $x,y=(0,0)$) at a (small) distance $z_{\rm off}$ from  the center of M31.  We then assume that stars in the GSS move along this vector $ {\bf \hat k}$. Thus the GSS stars are assumed to be on nearly radial orbits.

As discussed earlier, we expect an energy gradient along the GSS and therefore we assume that the total energy (potential and kinetic) of stars in the GSS 
can be approximated by a linear gradient along the stream (see, e.g., Appendix B and Figure~\ref{fig:sergey_energy_plot}), i.e., 
$ E(R) = E_0 + \frac{dE}{dR}(R-R_0)$
where $R$ is a projected distance along the stream and $E_0$ is the energy at $R_0,$ the projected distance at the same point (i.e., the middle of the stream). We assume that the energy gradient is positive and limited by the maximum range of energies along the stream $\delta E_{\rm max}$  $0<\frac{dE}{dR} (R_2-R_1)<\delta E_{\rm max}$, where $R_1$ and $R_2$ are the projected distances that limit the observed portion of the GSS. The reason for the assumption of the positive energy gradient is that this is exactly what we expect for the trailing part of the shell. The GSS shell stars are currently falling back to M31 (from the first pericentric encounter of the GSS progenitor) and the most distant stars have the least negative energies (and therefore longest orbital periods). See for example the bottom left panel of Figure~\ref{fig:Sergey_model} showing the positive energy gradient in the GSS.

The adopted upper bound on the energy spread is $\delta E_{\rm max}$ = 30  $\times$ 500 (\kms)$^2$, the energy spread resulting from a progenitor with an initial pericentric velocity of 500~\kms\ and velocity dispersion of 30~\kms. Because the GSS is only the very end of the shell system in M31, and the middle and leading part of the shell system are likely responsible for the Northeast Shelf and Western Shelf respectively, we expect that the actual energy spread for stars in the GSS is much smaller than $\delta E_{\rm max}$. The final assumption is that energies $E(R)$ are always negative along the stream, i.e., all the GSS stars are bound.

While we make several assumptions here (e.g., that the energy gradient is linear as a function of projected radius and that the stars within the GSS are moving on primarily radial orbits), 
we have verified that in the fiducial N-body model of the disruption of the GSS progenitor presented in Section~\ref{sec:nbody_model} and Appendix~B that these assumptions are satisfied.

The geometric assumptions that the stream is linear and the stars move along it tell us that the 3D velocity should be changing as a function of projected distance $R$ along the stream as

\begin{equation}
{\bf V}(R) = a(R) {\bf \hat k}
\label{eqn:v_ar}
\end{equation}
where $a(R)$ is an unknown function. Since the (line-of-sight) radial velocity is simply a projection of the 3D velocity along the $z$-axis, 
\begin{equation}
    V_{\rm los}(R) = a(R) k_z.
    \label{eqn:vrad}
\end{equation}
Under the assumption of a linear change of energy with radius along the GSS we can write
\begin{equation}
 \frac{V^2(R)}{2}+ \Phi({\bf X}) = E_0 + (R-R_0) \frac {dE}{dR},
\label{eqn:energy}
\end{equation}
where ${\bf X}$ is the 3D position along the stream
\begin{equation}
    {\bf X}=(R {k_x},\ R {k_y},\ z_{\rm off}+R k_z/\sqrt{1-k_z^2} )
\label{eqn:definitionX}
\end{equation}
corresponding to a projected distance $R$ along the stream and
 $\Phi({\bf X})$ is the gravitational potential.
 
Combining Eq.~\ref{eqn:v_ar} and \ref{eqn:energy} allows us to write an expression for $a(R)$:
\begin{equation}
 a(R)  =\sqrt{2 \left( E_0 +\left[\frac{dE}{dR} (R-R_0) \right] - \Phi( {\bf X})\right)}
\label{eqn:ar}
\end{equation}
which gives us the expression for $V_{\rm los}(R)$ through Eq.~\ref{eqn:vrad} if we know $E_0$, $\frac{dE}{dR},{\bf \hat k},$ and the gravitational potential.
Essentially we now can write the likelihood 
for the radial velocity as a function of projected distance $P(V_{\rm los}|R,{\bf \hat k},\Phi(R),E_0, \frac{dE}{dR}) $ that we can fit to the velocity track of the GSS. 

\begin{figure}[ht]
    \centering
    \includegraphics[width=1.0\textwidth]{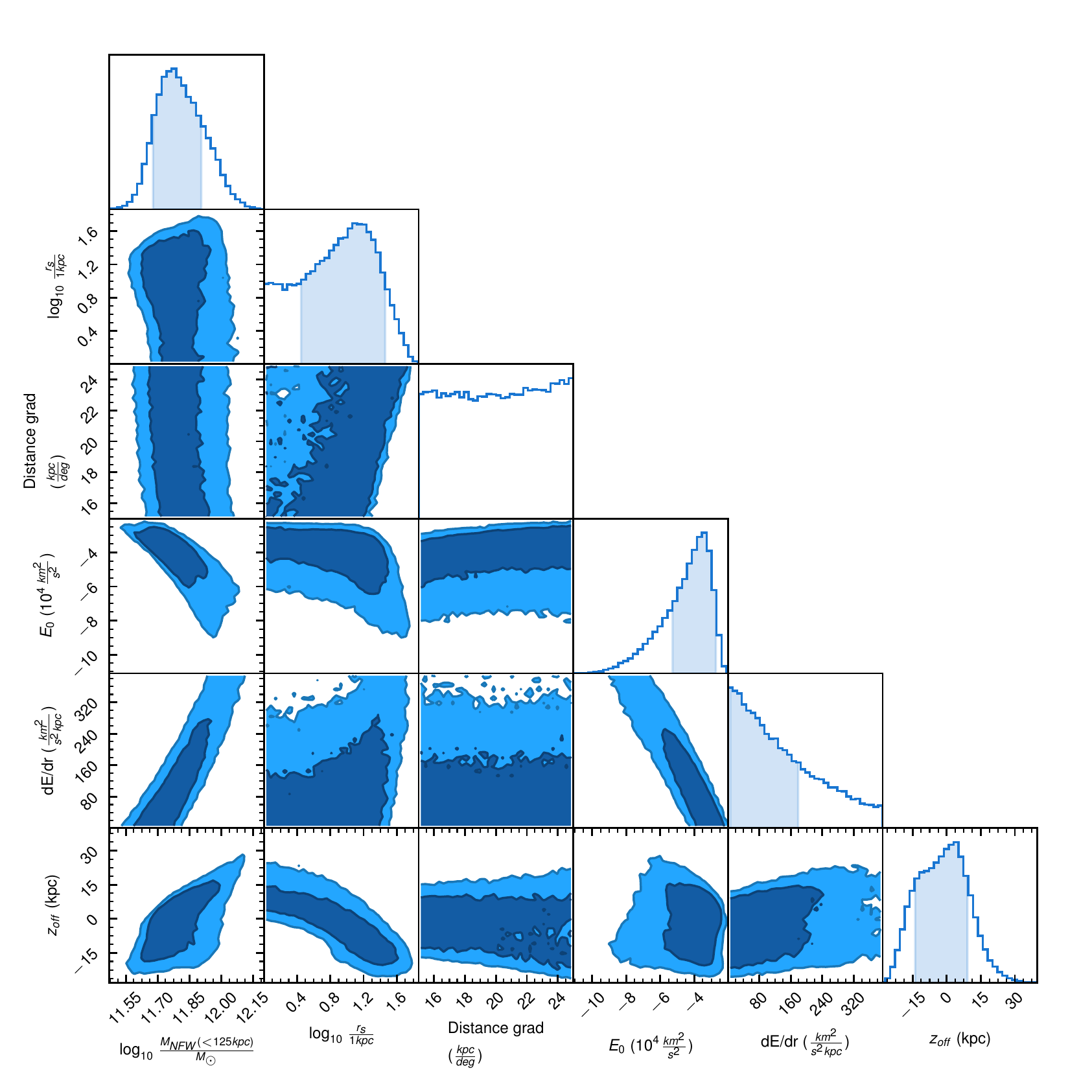}
    \caption{Corner plot showing the posterior probability distributions resulting from the simple dynamical mass estimates derived by modeling the GSS as a linear structure with a linear energy gradient along its extent (see the text in \S~\ref{sec:gssmassest} for details).}
    \label{fig:sergeymassest}
\end{figure}

While the number of parameters is potentially quite large, we can adopt informative priors on many of them. We have previously described our constraints on energy and energy gradients $E_0$ and $\frac{dE}{dR}$. 
Furthermore, the GSS orientation parametrized by ${\bf \hat k}$ is
well constrained by its projected orientation on the sky and 
the measured distance gradient of 
20\,kpc\,deg$^{-1}$ along its 6$^\circ$ extent \citep{Conn2016}. 
We therefore adopt a uniform prior for the distance gradient to be between 
15\,kpc\,deg$^{-1}$
and 25\,kpc\,deg$^{-1}$.
A simple algebraic equation for the distance gradient provides a prior on $k_z$.
For the gravitational potential we adopt a typical bulge/disk/halo decomposition with the bulge and disk models to be Hernquist and  Miyamoto-Nagai  models respectively with fixed parameters from \citet{Kirihara2017a}. We assume a disk inclination angle of 77$^\circ$ and a position angle of the major axis of 38$^\circ$. The dark matter halo component is modeled as an NFW \citep{NavarroFrenkWhite1997}, where the halo mass $M_{\rm halo}$ and scale length $r_s$ are to be determined. We adopt a log-uniform prior on the mass 
$10^8\,M_\odot <M_{\rm halo}<10^{14}\,M_\odot$ 
and scale length 
$1\,{\rm kpc}<r_s < 100\,{\rm kpc}$. 
This completes the definition of our model likelihood and parameter priors.



For the locus of the GSS, we used the result of the two-component fit described in Section~\ref{sec:PVdiagram}. 
The radial velocity measurements in 9 positions together with their uncertainties along the GSS 
were then fit by the $V(R)$ model as described in Eq.~\ref{eqn:v_ar} and \ref{eqn:ar}. The posterior was sampled with the {\tt dynesty} nested sampler. The model had 6 parameters in total: the halo mass and scale length, the distance gradient, energy and its gradient $E_0$ and $dE/dR$, and the offset of the stream from pointing directly at the M31 center $z_{\rm off}$.  The posterior on these parameters is shown in  Figure~\ref{fig:sergeymassest}. To avoid the typical mass-size degeneracy, we show the posterior for the mass inside 125 kpc rather than the total halo mass. Multiple parameters are unconstrained (such as the distance gradient, where we are purely driven by the prior), which is not very surprising given the limited data available. We also note that the offset of the GSS from pointing directly at the M31 center ($z_{\rm off}$) 
is consistent with zero, confirming that the orbits are very close to radial. We also see that the energy gradient prefers significantly lower values than our threshold, which is reasonable, given that we expect the GSS to be only a small (trailing) part of the shell.

We find the halo mass within 125 kpc to be $\log_{10}\, M_{\rm NFW}(<125\,{\rm kpc})/M_\odot = 11.80_{-0.10}^{+0.12}$ or if we include the disk and the bulge $\log_{10}\,M_{\rm total}(<125\,{\rm kpc})/M_\odot = 11.84_{-0.10}^{+0.12}$. 
As the method we employed 
makes
significant assumptions, we have also applied exactly the same fitting procedure to the sample of stars from the simulation presented in Section~\ref{sec:nbody_model} and obtained the halo mass with the bias of $\log_{10} M_{\rm halo,fit} - \log_{10} M_{\rm halo,true}\approx 0.1$ which is within our uncertainty.


Our mass estimate of 
$\log_{10}\, M_{\rm NFW}(<125\,{\rm kpc})/M_\odot = 11.80_{-0.10}^{+0.12}$
is consistent with estimates from the literature of the enclosed mass at this distance \citep[e.g., graphical summary in][]{Kafle2018}. 
In particular, our result is similar to that of \citet{Ibata2004}, who carried out the first kinematic study of the GSS, measuring the velocities of 184 stream stars and using
the velocity gradient along the stream to estimate a halo mass of
$M_{125}=7.6\pm1.2\times 10^{11} M_\odot$ for a logarithmic halo and  $M_{125}=6.4\pm1.3\times 10^{11} M_\odot$  for an NFW halo.





\section{Discussion}
\label{sec:discussion}

As described in the previous sections, DESI spectroscopy reveals 
intricate, coherent spatial-velocity structure in the inner halo of M31, including nested chevrons and wedge-shaped structures  (Figures~\ref{fig:zoneposvela}, \ref{fig:zoneposvelb}),
with a spatial and kinematic clarity never-before observed in an extragalactic source (Section~\ref{sec:results}).
The DESI results affirm earlier ``pencil beam'' spectroscopy carried out in restricted portions of the inner halo. 
The observed structures are consistent with the expected kinematic signatures of shells and streams produced in galaxy mergers (\S~\ref{sec:cosmo}, \ref{sec:nbody_model}) 
and suggest that most, if not all, of the structure observed in M31 arises from a single merger event (\S~\ref{sec:nbody_model}). We illustrated how the kinematics of the structure induced by the merger---the shells and the GSS---can dynamically probe the mass distribution of M31 as a function of galactocentric distance  (\S~\ref{sec:ShellKinematics}, \ref{sec:gssmassest}).
In this section we situate our results in the context of 
prior work and turn to the question of the nature of the progenitor that produced the observed substructure. 


\subsection{Comparison to Previous M31 Merger Models}
\label{sec:PrevObsandModels}


Many previous studies have explored and advanced 
a picture in which much of the inner halo substructures of M31 
are tidal debris from 
a single companion galaxy that encountered M31 on a nearly radial orbit
\citep[e.g.,][]{Ibata2004, font2006, Fardal2006, Fardal2007, Fardal2008, Fardal2012, Fardal2013, Mori2008, Sadoun2014, Kirihara2017a,Milosevic2022}. These simulations have explored a wide range of parameters, and found that 
a wide range of progenitor stellar masses can reproduce the observed morphologies. Several studies have suggested that the visible debris is the result of a minor merger ($\sim$ 1:10 to 1:5), with the stellar mass of the companion in the range 1--5$\times 10^9\,M_\odot$ \citep[e.g.,][]{Fardal2013, Kirihara2017a,Sadoun2014}. 
In contrast, a few studies have suggested that the observational data suggest a major merger (i.e., $\sim$1:4--5) with a progenitor of stellar mass 
$>10^{10}\,M_\odot$ \citep{dsouza2018,Hammer2018,Bhattacharya2019}. 
\citet{dsouza2018} advocated for a major merger based on the mass, metallicity, and star formation history (SFH) of the halo. They also hypothesized that M32, M31's compact satellite, could be the core of the disrupting satellite based on metallicity and star-formation history. M32 is located within the debris field close to where the GSS meets the M31 disk and has a very different velocity from the GSS, indicating that M32 would be at a very different phase of its orbit than the GSS material. Other studies predict that the progenitor lies elsewhere in the debris or may be completely disrupted. \citet{Hammer2018} additionally note that the 2$-$4~Gy-old star-formation episode and significant thick disk of M31 might both be the product of the interaction with a massive progenitor. 
Recent studies of M31 halo stars also report higher [Fe/H], a stronger [Fe/H] gradient, higher mean [$\alpha$/Fe], and a larger [$\alpha$/Fe] spread than observed in the Milky Way halo, suggesting that much of the M31 inner halo region and GSS may result from the assimilation of a fairly massive galaxy with a complex star-formation history  \citep{Gilbert2019,Escala2020a,Escala2020b,Escala2021}. This major-merger picture is also consistent with the observed steep age-velocity dispersion relation, large asymmetric drift, and other chemical signatures observed in the disk populations \citep{Dorman2015,Bhattacharya2019,Arnaboldi2022}.

Although the morphology of the debris appears to be insensitive to the mass of the progenitor \citep[e.g.,][]{Hammer2018,Boldrini2021}, the simulations reveal that it is sensitive to the orbital parameters of the encounter. 
Dynamical models that attempt to account for both the observed spatial distribution of the debris and the radial velocities available to date generally infer 
an initial pericentric passage within a few kpc of the center of M31 within the last 1--2 Gyr. 
%
%
%
In many of the models tailored to M31 
(including our own from \S~\ref{sec:nbody_model}), 
a companion galaxy plunges into M31 and its stars are pulled out on the far side of M31 to form the GSS following the first pericenter passage. The DESI observations do not detect any outward moving stars following the first pericentric passage, but they do detect the infalling stream of stars on their way back toward M31 after their first apocentric passage. 
The Northeast Shelf is produced as the second wrap of the orbit, 
and the Western Shelf constitutes the third wrap 
\citep{Fardal2007,Fardal2008,Fardal2012,Fardal2013}. 
The Southeast Shelf, tentatively identified by \citet{Gilbert2007}, may constitute the fourth wrap, the leading edge of the tidal debris.
In the models of \citet{Fardal2013} and \citet{Kirihara2017a},  
the core of the progenitor, if it has survived tidal disruption, is predicted to reside somewhere in the Northeast Shelf.

All of the models are successful in accounting for the general spatial morphology of the GSS, Northeast Shelf, and Western Shelf, as well as  spectroscopic observations of the GSS and the Western Shelf available to date.
The anticlockwise-rotating thick-disk model of \citet{Kirihara2017a}
better reproduces the edge-brightening observed on the eastern side of the GSS. 
In addition, some  models \citep[e.g.,][]{Milosevic2022,Kirihara2017a} also reproduce the metallicity variations observed in the GSS by \citet{Conn2016}.
%
Previous studies have compared their simulations with the velocities of either small numbers of PNe or larger numbers of RGB stars measured with Keck/DEIMOS in pencil beams located at a few radial positions within the inner halo \citep[e.g.,][]{Fardal2007,Fardal2013}.


In particular, the model of \citet{Fardal2007}, which was designed to replicate the observed substructure in photometric imaging studies of M31, is remarkable in capturing many of the observed features in the DESI radial velocity data. Since these model data were not available to us, we made simple comparisons of our observations plotted in the same way as the simulations,  
comparisons which corroborate many of the features predicted by \citet{Fardal2007}. 
In comparing to earlier data, \citet{Fardal2007} 
showed how 11 PNe from \citet{Merrett2003,Merrett2006} (which have kinematics classified as ``stream" or ``stream?'') trace out the predicted locus of the blue-shifted edge of the Northeast Shelf, the large wedge in the position-velocity diagram at $-500 \lesssim V_{\rm los} \lesssim -100$~\kms\ (see the right panel of 
Figure~3 of \citet{Fardal2007}). 
The DESI observations overlap with the PNe and chart out the wedge-like structure more completely and in greater detail on both the red- and blue-shifted edges 
and show that the structure extends to slightly larger projected distances than predicted by the \citet{Fardal2007} model (see also Figure~\ref{fig:pnlit}).

Earlier studies also compared the positions of PNe from \citet{Merrett2003,Merrett2006} and stars along the minor axis of M31 with the simulation predictions in the Western Shelf region  \citep[e.g.,][]{Fardal2007,Fardal2013}. The DESI data clearly trace out the shell-like nature of the kinematic structure in the Western Shelf over a large spatial extent, also showing only minor deviations, especially at projected radii $>1.5^\circ$.


While much of the structure observed with DESI is roughly consistent with the dynamical models published to date, some features remain unaccounted for.
Notably, the blueshifted feature 1bb in Zone 1 is not reproduced, as has been previously noted  \citep[e.g.,][]{vanderMarel2012}.
The DESI data also reveal new features.
For example, in the Zone 2 region of the Northeast Shelf, the DESI data show a second, smaller wedge extending out to 1.3$^\circ$ and highlighted in Figure~\ref{fig:zoneposvela}, which may be the continuation of features 1cb and 1cr into Zone 2. These Zone 1 features were previously identified by \citet{Gilbert2007} from pencil-beam spectroscopy along the southeast minor axis of M31. Associated with the Southeast Shelf, the Zone 1 and Zone 2 features may correspond to the fourth wrap predicted by the model of \citet{Fardal2013}.
In contrast,
the compact chevron in Zone 1 that is bounded by $|V_{\rm los}| \lesssim 150$~\kms\ does appear to be present in 
one simulation shown in \citet{Fardal2013}. Their Figure 6 shows a compact component in Zone 1 that extends to a similar distance from M31 as the observed structure. In the model, the component arises from Northeast Shelf stars on the near side of M31 that overlap the GSS. Finally, the 3br feature in Zone 3 is also not present in the models.

These initial DESI results represent a significant advance by covering large areas more uniformly and revealing the kinematic structures in unprecedented clarity. These data  inform future modeling efforts to understand the merger history responsible for the complex inner halo substructure of M31. 





\subsection{Clues to Nature of the Progenitor}
\label{sec:NatureofProgenitor}



As simulations have demonstrated that the morphology of the debris is relatively insensitive to the mass of the progenitor \citep[e.g.,][]{Boldrini2021}, other information is needed to constrain the nature of the progenitor. 
Previous studies have attempted to infer the nature of the progenitor using the metallicity (a wide range in metallicity, reaching more than 1/3 solar in the inner parts of the debris; \citealt{Gilbert2014,Ibata2014}) and star formation history (showing star formation
until around 2-3\,Gyr ago; \citealt{Brown2006}) implied by measurements of the inner
stellar halo and substructures \citep{dsouza2018}.
[Fe/H] and [$\alpha$/Fe] measurements of individual halo stars have also been used to argue that much of the M31 inner halo region and the GSS in particular may result from the assimilation of a fairly massive galaxy with a complex star-formation history  \citep{Gilbert2019,Escala2020a,Escala2020b,Escala2021}.
Here we contribute to this topic by commenting on 
(1) the metallicity of the stellar debris, (2) the number of  
dwarf galaxies and globular clusters potentially associated with the progenitor, and (3) whether there is any evidence for a surviving progenitor galaxy. 

%
%
%
Since our target selection introduces a bias towards
metal-rich RGB
stars (\S~\ref{sec:metallicities}), we cannot use the current DESI data to
reliably measure the metallicity distribution of the accreted stars. However, we do find that
significant numbers of metal rich stars are present across all regions
surveyed, suggesting that the progenitor responsible for these
structures is relatively high mass, high enough to have stars up to solar
metallicity. 
Future DESI observations that target stars more metal poor than those studied here can better characterize the metallicity distribution of the progenitor.

\begin{figure}[th]
    \centering
    \includegraphics[width=0.85\textwidth]{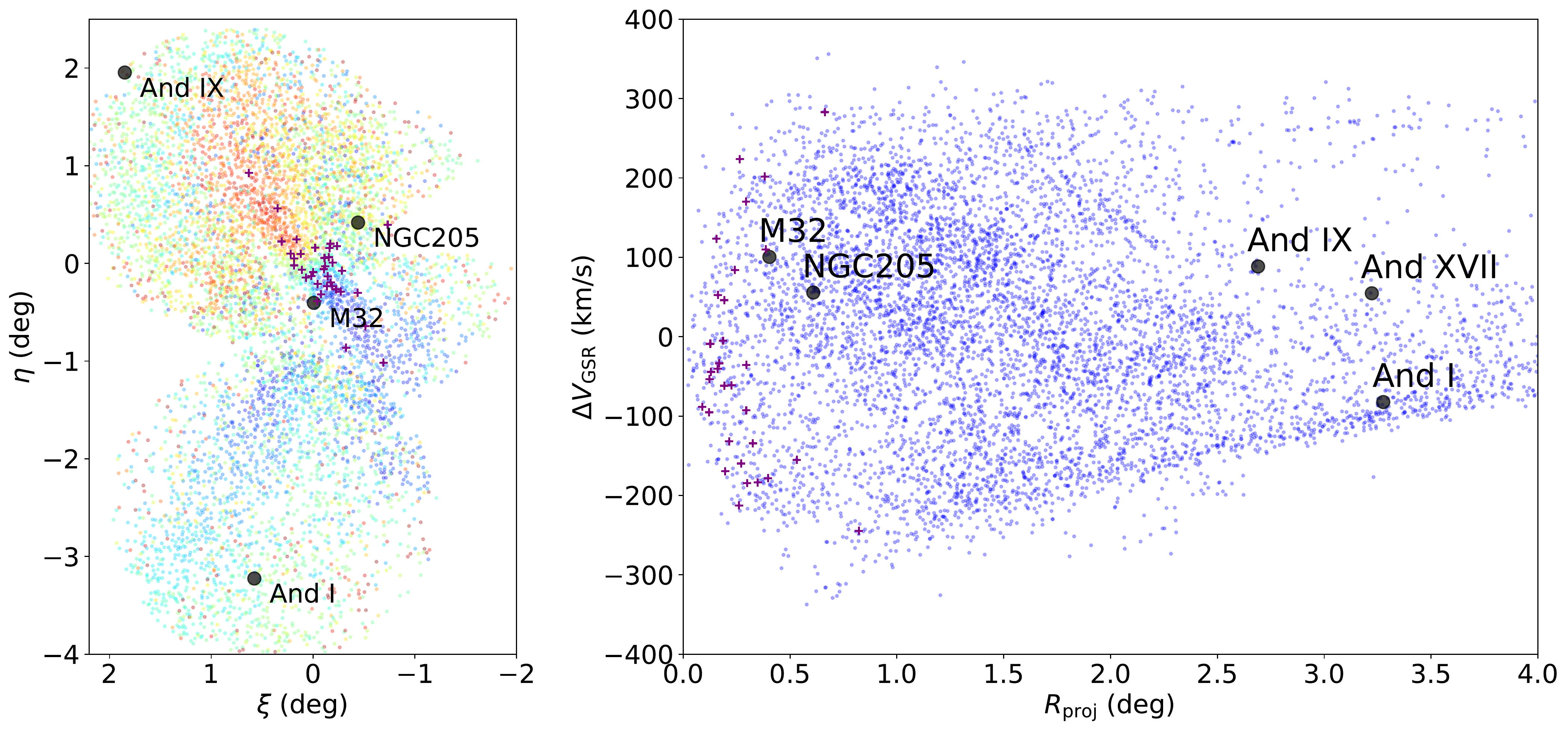}
    \caption{The distribution of the closest dwarf galaxy companions to M31 on the position-velocity diagram resulting from the DESI observations. The positions and velocities of the dwarf galaxies are taken from the updated (2021 January) version of the compilation by \citet{McConnachie2012}. The purple plus signs represent emission line sources (i.e., a combination of HII regions and PNe) among the DESI targets. 
    }
    \label{fig:dwarfgals}
\end{figure}
\begin{figure}
    \centering
    \includegraphics[width=0.65\textwidth]{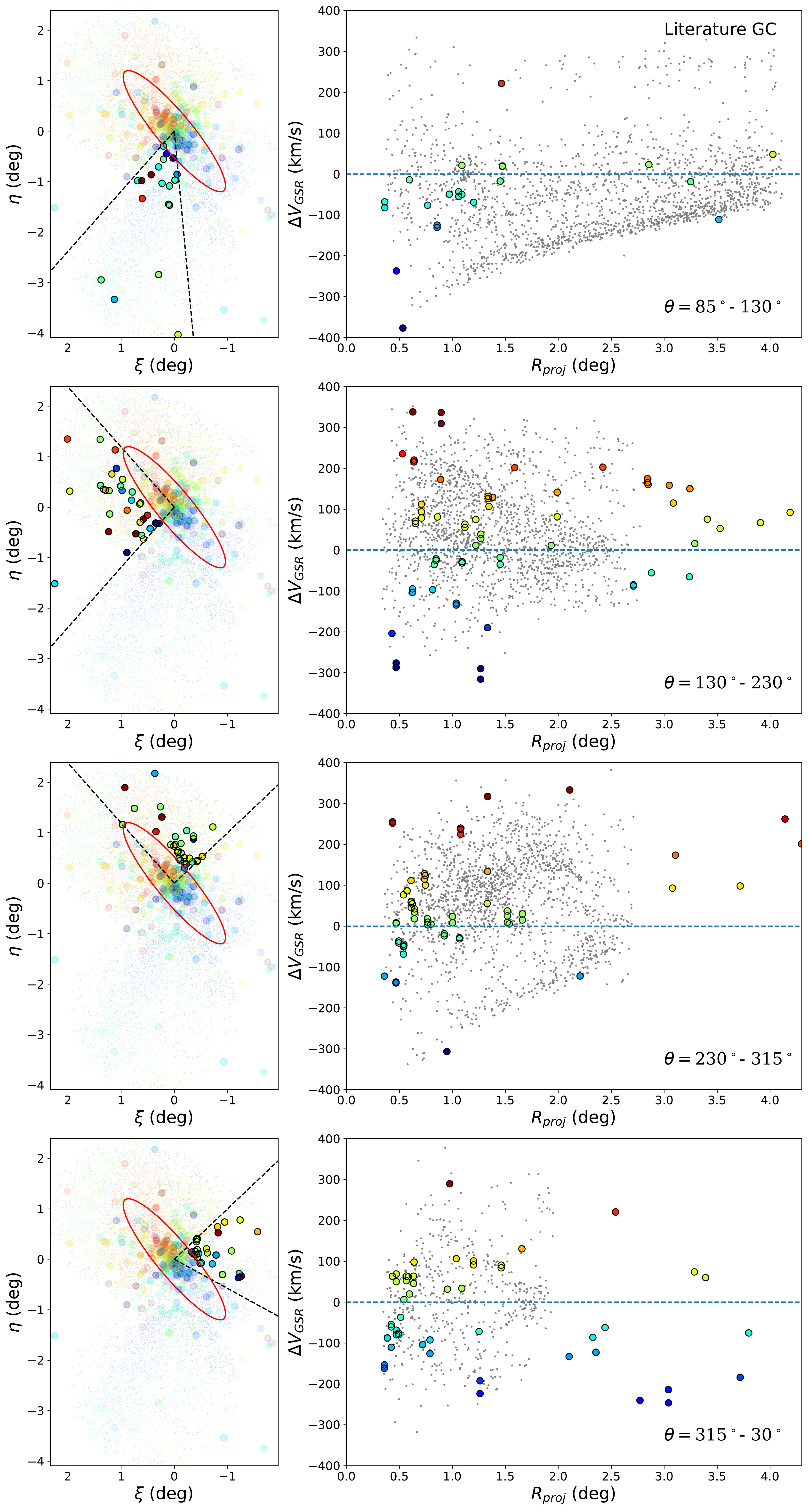}
    \caption{The distribution of old star clusters with ages $>2$~Gyr (large colored circles) overlaid on M31 stars (smaller points).
    }
    \label{fig:gclit}
\end{figure}

Given that the progenitor was probably massive (i.e., $> 10^9\,M_{\odot}$), it is possible that the merger
event will have delivered star clusters and dwarf galaxies to M31. 
Figures~\ref{fig:dwarfgals} and \ref{fig:gclit} show the distribution of dwarf galaxies and 
old ($>2$~Gyr) star clusters with measured ages and velocities from the literature compared to the stars measured with DESI. The dwarf galaxy measurements are from the compilation of \citet{McConnachie2012} and the star cluster measurements are primarily from LAMOST spectroscopic surveys \citep{Chen2015,Chen2016,Wang2021} and the Hectospec surveys \citep[e.g.,][]{Caldwell2016}. 

The dwarf galaxy Andromeda I closely overlaps the GSS (Figure~\ref{fig:dwarfgals}), as does at least one globular cluster at a distance of 3.5$^\circ$ (Figure~\ref{fig:gclit}), suggesting a possible physical association.
The association of this globular cluster, LAMOST-1, with the GSS has
been previously noted by \citet{Chen2015,Chen2016}.
LAMOST-1's
metallicity ([Fe/H] =$-0.4$) and age (9.2 Gyr) are consistent with an
association with the GSS progenitor. 
%
%
In Zones 2 through Zone 5, the distribution of globular clusters is similar to that of the stars in the inner halo. In particular, they populate the interior of the wedge in Zone 2 ($\theta$ = 130$^\circ$--230$^\circ$) and the small wedge in Zones 3+4 ($\theta$ = 230$^\circ$--315$^\circ$).
Thus, interestingly, many of the clusters are potentially associated with the wedge structures within 2$^\circ$ of M31, while relatively few clusters overlap the GSS. 

These results are roughly consistent with expectations from merger
simulations. For example, considering the representative galaxy merger
in the Illustris TNG-50 simulation (Fig.\ \ref{fig:tng50}, larger solid points in the right-hand panel), one could very crudely subsample particles from the dominant merger
companion that are `old' (9-12\,Gyr ago),  mirroring the epoch of globular cluster formation in the
Milky Way. 
These early-forming star particles tend to be more
centrally concentrated and kinematically hotter than the bulk of the
progenitor stars, and are not as clearly confined to
kinematically cold substructures. 
While more
detailed model predictions that follow globular cluster formation in
galaxies and their expected distribution among the tidal debris are
clearly needed to fully interpret the observations and obtain robust
constraints on the nature of the progenitor, this exercise tentatively suggests that it may be challenging to accurately attribute globular clusters to the progenitor solely on the basis of their clustering into kinematic substructures.
Future studies might explore the metallicities and orbits of M31 globular clusters to infer their association with the progenitor galaxy \citep[e.g.,][]{mackey2019,mackey2019b}.

Finally, with its ability to map out  stellar velocity structure over large areas, DESI offers the opportunity to locate the remnant core of the progenitor galaxy. 
\citet{Fardal2013} predicted that if the progenitor survives,  
stars from the core of the progenitor will populate a fairly compact structure in phase space located at a projected distance of $1^\circ \lesssim R_{\rm proj} \lesssim 2^\circ$ and a line-of-sight velocity of $\approx$ 0 to $-200$~\kms (i.e., in Zone 2 and blue-shifted relative to the M31 systemic velocity). 
\citet{Kirihara2017a} predicted that the stripped bulge of the progenitor lies in the eastern shell and in front of the disk of M31, compact in phase space and at a location of ($\xi,\eta,V_{\rm los})\approx (1.1^\circ,0.5^\circ,-200$~\kms).

No such structures are detected in the DESI data, although this may yet be due to the sparseness of our current sampling. We do find that the velocities in Zone 2 show a preferential {\it redshift}, rather than the blueshift predicted by \citet{Fardal2013}, and they show evidence of multiple shells rather than a component that is compact in phase space. 
Future DESI observations could 
place stronger constraints on (or possibly identify) a surviving progenitor galaxy.
More densely sampled spectroscopy will permit quantitative assessment of the possibility that M32 (D’Souza \& Bell 2018) or another existing galaxy is the progenitor, or perhaps identify a remnant or disrupting core in the inner halo of M31. 

M31 and the Milky Way show a remarkable parallel, in that the inner halos of both galaxies are dominated by debris from a single accretion event. The Milky Way's inner halo is dominated by the {\it Gaia}-Sausage-Enceladus structure, a radial accretion event of mass $>10^{10}~M_\odot$ nearly 8-11~Gyr ago \citep[e.g.,][]{Belokurov2018,Helmi2018}. The inner halo of M31 is also dominated by the single radial accretion event that produced the GSS and the intricate kinematic structures studied here, but which began only 1-2 Gyr ago. If the M31 shell system progenitor is indeed as massive as suggested based on its total stellar luminosity and stellar metallicities, M31 may provide a glimpse of what the Milky Way looked like several Gyr ago. Future spectroscopic surveys of the M31 inner halo will be able to explore this exciting possibility in greater detail.

\section{Summary and Conclusions}
\label{sec:conclusion}

We have obtained spectra, in three DESI pointings, of \Ntargets\ targets in the direction of M31. Using these observations, we have measured accurate radial velocities of 
\Nstars\ stellar sources, of which 
\NAnd\ are members of the M31 system. These include radial velocities for \NAndHIIPN\ HII regions and Planetary Nebulae, and \Nclusters\ M31 clusters. We have also identified 
\NQSO\ QSOs and \Ngalaxies\ galaxies behind M31, which can provide unique probes of the gas associated with the GSS progenitor and other circumgalactic and interstellar material associated with M31.

While most of the earlier spectroscopy of individual stars in M31 had been carried out with 6.5-m to 10-m class telescopes, a few hours of spectroscopy with DESI has added significantly to our knowledge of the stellar kinematics of the M31 halo. These data represent a $>$3-fold increase in the number of known M31 stars in the region outside the M31 disk, and provide a much more uniform sampling of the inner halo than any previous spectroscopic study. The rapid advance is due to (1) DESI's wide field of view, high multiplex, and high observing efficiency; (2) the use of selection criteria that efficiently select M31 stars with limited contribution from foreground Milky Way stars; (3) the strong molecular bands in the late-type spectra of the M31 sample, which enables reasonable radial velocity accuracy ($< 10$ \kms) on faint stars ($z$ = 21.5 AB mag); and (4) the good match of DESI's fiber density to the stellar target density of M31.

The DESI spectra reveal intricate coherent kinematic structure in the positions and velocities of individual stars in the inner halo of M31: streams, wedges, and chevrons that provide evidence of a recent merger, i.e., a galactic migration event. While hints of these structures have been glimpsed in earlier spectroscopic studies of M31, this is the first time wedges and chevrons have been
{\it mapped} with such detail and clarity in a galaxy beyond the Milky Way.  We find evidence for multiple coherent structures in the vicinity of the GSS and clear kinematic evidence for shell structures in the Western Shelf and Northeast Shelf regions. In particular, we identify 
750 stars in the largest kinematic component (feature 1ab) of the GSS and measure a narrow velocity dispersion of $10.80\pm0.75$~\kms.  The DESI data also reveal new structures not predicted by existing merger simulations. 
The kinematic structures seen in the stellar distribution of M31 halo stars are echoed in the position-velocity distribution of known M31 PNe.

Dynamical models from the literature that were constructed to explain the spatial morphology of the GSS and other inner halo features, as well as the models presented here,  
predict position-velocity structures that are remarkably similar to those observed. The results 
suggest that much of the substructure in the inner halo of M31 is produced by a single merger event with a companion galaxy a few Gyr ago. 
Taken together, the richness of the observed structure demonstrates that large spectroscopic samples can place valuable constraints on the recent merger history of M31 and that such samples are within the grasp of the Mayall/DESI system. 

We find significant numbers of metal-rich stars across all of the detected substructures, suggesting that the progenitor galaxy (or galaxies) had an extended star formation history, one perhaps more representative of more massive galaxies. Known populations of stellar clusters in the halo of M31 appear to be  more closely associated with the inner wedge structures (within 2$^\circ$ of M31) than the spatially extended GSS. The difference seems plausible if the clusters are predominantly older systems that originated in a kinematically hotter component in the progenitor galaxy.


The 
shell structures and the GSS also offer an opportunity to constrain the gravitational potential of M31 as a function of galactocentric distance. Using the simple prescription of \citet{Merrifield1998}, we obtained from the velocity gradients of the nested shell structures 
galaxy 
mass estimates ranging from $2\times 10^{11}M_\odot$ to $8\times 10^{11}\, M_\odot$ at projected distances between 17 and 38~kpc. These values exceed the enclosed mass estimates inferred from the \ion{H}{1} rotation curve at distances of $\sim 20$~kpc to $\sim 40$~kpc, but nevertheless are within a factor of 2 of those values.
A more detailed dynamical model fit to the GSS velocities implies a 
dark matter mass of $6.0^{+2.1}_{-1.2} \times 10^{11}\,M_\odot$ within 125 kpc, 
in good agreement with estimates from the literature \citep[e.g., ][]{Ibata2004}.

M31 is remarkably similar to the Milky Way in that the inner halos of both galaxies are dominated by stars from a single accretion event. Indeed, a recent study of the kinematics of Milky Way stars near the Sun reports chevron-shaped kinematic substructures 
\citep{Belokurov2022a} that are reminiscent of those reported here. If the progenitor of the M31 shell system studied here is $\gtrsim 10^{10}\,M_\odot,$ M31 may provide a close analog to what our own galaxy looked like several Gyr ago. 
%
More extensive DESI studies of M31 can explore this possibility by: 
(1) better characterizing kinematic substructures (shells, etc.) with higher sampling density;
(2) extending our study of the metal-rich halo population to a characterization of the metal-poor population;
(3) identifying the dwarf galaxies and globular clusters potentially associated with the progenitor; and 
(4) searching for evidence for a surviving progenitor galaxy. 

Although here we identified shells by eye---which was appropriate given the limited data available---with higher-density sampling, we can measure the shells more accurately. By combining these more precise measurements with a detailed dynamical model customized to M31, we can place better constraints on the orbit of the progenitor and the mass and shape of the gravitational potential of M31.
Characterizing the metal-poor population will allow us to better determine the metallicity of the progenitor and constrain its star formation history and total mass, as well as explore the more virialized (dynamically older) halo of M31. 
Extending over a large fraction of the galaxy’s volume,
the delicate chevrons we observe are also sensitive to the gravitational
perturbations from substructure within the M31 halo, such as satellites and dark
matter subhaloes. More refined mapping of the chevrons may be able to provide
constraints on the number of such substructures.
Finally, future work can also 
examine the structure and kinematics of the disk and the nature of the circumgalactic and interstellar media probed by the background QSOs and galaxies.  



The observations presented here, obtained in just three DESI pointings with effective exposure times of $\le90$~min, demonstrate the remarkable ability of DESI, on the Mayall 4-m telescope, to efficiently map out the large-scale kinematic structure of M31. 
Given DESI’s efficiency, 
we can extend these studies to a larger volume and probe the outer halo of M31 and its interaction with its galactic neighbors (M33 and others). Photometric imaging studies of this region show streams and other structures.
A future targeted survey could cover a significant fraction of M31's stellar halo with about 25 tiles. Such a survey would potentially increase the number of M31 halo stars by over an order of magnitude and reveal its structure and immigration history in unprecedented detail.

\begin{acknowledgments}

We thank Chien-Hsiu Lee, Monica Soraisam, Amanda Quirk, and Raja Guhathakurta for help in selecting filler targets for the original M31 DESI first light tile and for useful conversations regarding our remarkable neighbor. We also thank Nelson Caldwell for generously providing early access to the CFA Optical/Infrared Science Archive which records the many MMT/Hectospec spectroscopic campaigns on M31. We thank Vasily Belokurov and Risa Wechsler for stimulating discussions and detailed comments on the manuscript. 
We also thank the anonymous referee, Rosemary Wyse, Karrie Gilbert, Nelson Caldwell, and Magda Arnaboldi for constructive comments on the manuscript, Letizia Stanghellini for information regarding AGB stars, and Alyssa Goodman for help with the glue software package.

AD and JN's research activities are supported by the NSF's NOIRLab, which is managed by the Association of Universities for Research in Astronomy (AURA) under a cooperative agreement with the National Science Foundation. AD and JN's research is also supported in part by Fellowships from the John Simon Guggenheim Memorial Foundation and by the Institute for Theory and Computation at the Harvard-Smithsonian Center for Astrophysics. JN also acknowledges support from the Harvard Radcliffe Fellowship Program of the  Radcliffe Institute for Advanced Study at Harvard University.
JN, GM, and JJ-Z acknowledge support from the Harvard University's Radcliffe Research Partners Program of the Radcliffe Institute for Advanced Study. 
EFB is grateful for support from the National Science Foundation through grant NSF-AST 2007065. LBS acknowledges NASA-ATP award 80NSSC20K0509 and Science Foundation AAG grant AST-2009122. This research has made use of the SIMBAD database, operated at CDS, Strasbourg, France. 

This research is supported by the Director, Office of Science, Office of High Energy Physics of the U.S. Department of Energy under Contract No. DE–AC02–05CH11231, and by the National Energy Research Scientific Computing Center, a DOE Office of Science User Facility under the same contract; additional support for DESI is provided by the U.S. National Science Foundation, Division of Astronomical Sciences under Contract No. AST-0950945 to the NSF’s National Optical-Infrared Astronomy Research Laboratory; the Science and Technologies Facilities Council of the United Kingdom; the Gordon and Betty Moore Foundation; the Heising-Simons Foundation; the French Alternative Energies and Atomic Energy Commission (CEA); the National Council of Science and Technology of Mexico (CONACYT); the Ministry of Science and Innovation of Spain (MICINN), and by the DESI Member Institutions: \url{https://www.desi.lbl.gov/collaborating-institutions}.
The authors are honored to be permitted to conduct scientific research on Iolkam Du’ag (Kitt Peak), a mountain with particular significance to the Tohono O’odham Nation.

This work has made use of data from the European Space Agency (ESA) mission
{\it Gaia} (\url{https://www.cosmos.esa.int/gaia}), processed by the {\it Gaia}
Data Processing and Analysis Consortium (DPAC,
\url{https://www.cosmos.esa.int/web/gaia/dpac/consortium}). Funding for the DPAC
has been provided by national institutions, in particular the institutions
participating in the {\it Gaia} Multilateral Agreement.

This paper made use of the Whole Sky Database (wsdb) created by Sergey Koposov and maintained at the Institute of Astronomy, Cambridge by Sergey Koposov, Vasily Belokurov and Wyn Evans with financial support from the Science \& Technology Facilities Council (STFC) and the European Research Council (ERC).
This research also used the PAndAS data hosted by the facilities of the Canadian Astronomy Data Centre, which is operated by the National Research Council of Canada with the support of the Canadian Space Agency. 

For the purpose of open access, the author has applied a Creative Commons Attribution (CC BY) licence to any Author Accepted Manuscript version arising from this submission. 

This work was performed in part at Aspen Center for Physics, which is supported by National Science Foundation grant PHY-1607611. This work was partially supported by a grant from the Simons Foundation.


\facility{KPNO:Mayall (DESI), WISE, Gaia, CFHT:Megacam}
\software{{\tt pyyaml} (\url{https://github.com/desihub}), Astropy \citep{astropy:2013, astropy:2018}, sqlutilpy \citep{sqlutilpy}, dynesty \citep{speagle2020,dynesty}, Chainconsumer \citep{Hinton2016}, gala \citep{gala,adrian_price_whelan_2020_4159870}, Q3C \citep{Q3C}, gyrfalcon \citep{dehnen2014}, glue \citep{Beaumont2015,Robitaille2019}}

\end{acknowledgments}

\bibliographystyle{apj}
\bibliography{bibliography,bibliog_overview}

\appendix

\section{Quasars and Galaxies Behind M31}
\label{appendix:qso}


\begin{figure}[ht]
    \centering
    \includegraphics[width=\textwidth]{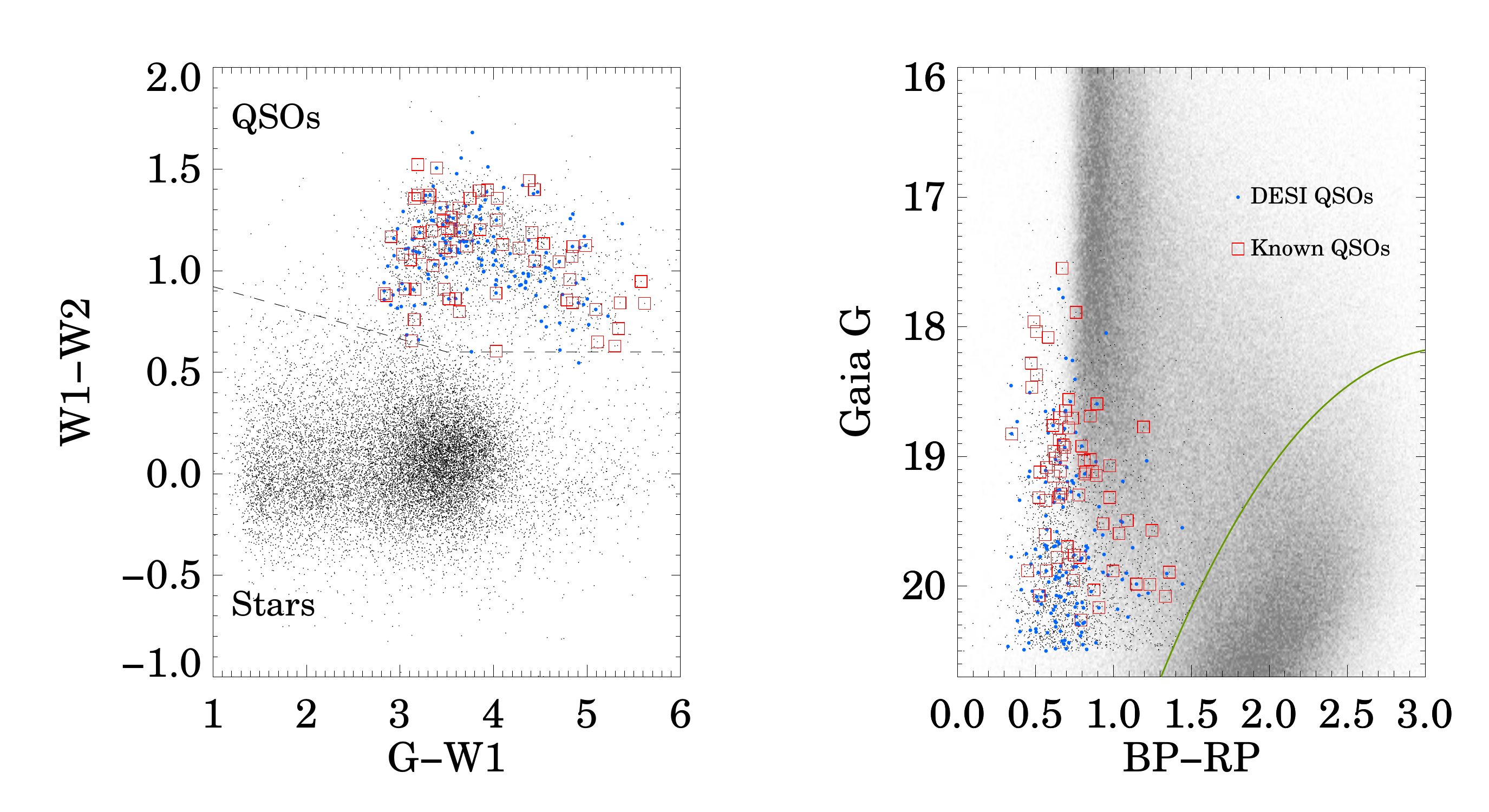}
    \caption{The selection of the DESI QSO targets. The left panel shows the criteria used (dashed line) to select QSO candidates. The known QSOs from \citet{massey2019} and \citet{Huo2010,Huo2013,Huo2015} are shown as red squares and the QSOs confirmed by DESI spectroscopy are shown as filled blue dots. The right panel shows the {\it Gaia} $G$ vs $B_p-R_p$ diagram as a greyscale for all {\it Gaia} stars within 5$^\circ$ of M31. The solid line shows the criterion used to exclude M31 sources. }
    \label{fig:QSOs}
\end{figure}

QSO candidates were selected using a combination of the {\it Gaia} DR2 \citep{GaiaMission2016,GaiaDR2summary} and the deep combined imaging data from the unWISE catalogs \citep{unWISEcoadd_2014,unWISE3_2017,unWISE1_2017,unWISE5_2019,unWISE5cat_2019} using the following criteria: 

\begin{itemize}
    \item $\pi-\sigma(\pi) \le 0.1$
    \item $\vert\mu_\alpha-2\sigma(\mu_\alpha)\vert \le 0.1$ and $\vert\mu_\delta-2\sigma(\mu_\delta)\vert \le 0.1$
    \item ($G<19$ and AEN$<10^{0.5}$) or ($G\ge 19$ and AEN$< 10^{0.5+0.2(G-19)})$
    \item $(W1-W2) > 0.5$
    \item $(W1-W2) > (1.0 - 0.125(G-W1))$
    \item $G \le 26.46 - 5.991(BP-RP) + 1.313(BP-RP)^2 - 0.07856(BP-RP)^3$
\end{itemize}
where $\pi$, $\mu_\alpha$, $\mu_\delta$ $\sigma(\pi)$, $\sigma(\mu_\alpha)$, $\sigma(\mu_\delta)$ are the parallax, proper motion, and associated uncertainties from the {\it Gaia} DR2 catalog; $G$, $BP$, $RP$ are the {\it Gaia} DR2 mean photometric magnitudes; $W1$, $W2$ are the {\it WISE} channel 1 and 2 magnitudes from the unWISE catalogs; and AEN is the Astrometric Excess Noise parameter from the {\it Gaia} DR2 catalog. 

The first three criteria are used to distinguish QSOs from Milky Way stars on the basis of parallaxes and proper motions consistent with zero in the {\it Gaia} DR2 catalog; these are generalized versions of the criteria used by \citet{vanderMarel2019} to select stars in M31 from the {\it Gaia} catalog. The AEN criterion is the same used by the DESI program to separate point sources from extended sources (i.e., galaxies) for {\it Gaia} DR2\footnote{See \url{https://github.com/desihub/desitarget/blob/2.5.0/py/desitarget/gaiamatch.py\#L207-L210}}. The $(W1-W2) > 0.5$ criterion is a more relaxed version of the {\it WISE} AGN selection discussed in \citet{SternWISEQSO2012}. The {\it Gaia} - {\it WISE} criteria were determined based on identifying the known spectroscopically confirmed QSOs \citep[from][]{massey2019,Huo2010,Huo2013,Huo2015} in  $G-W1-W2$ color-color space. Finally, the {\it Gaia} $G-BP-RP$ color criterion is an attempt to avoid stars from the M31 RGB in the QSO selection (see Figure~\ref{fig:QSOs}).

\begin{figure}[ht]
    \centering
    \includegraphics[width=0.6\textwidth]{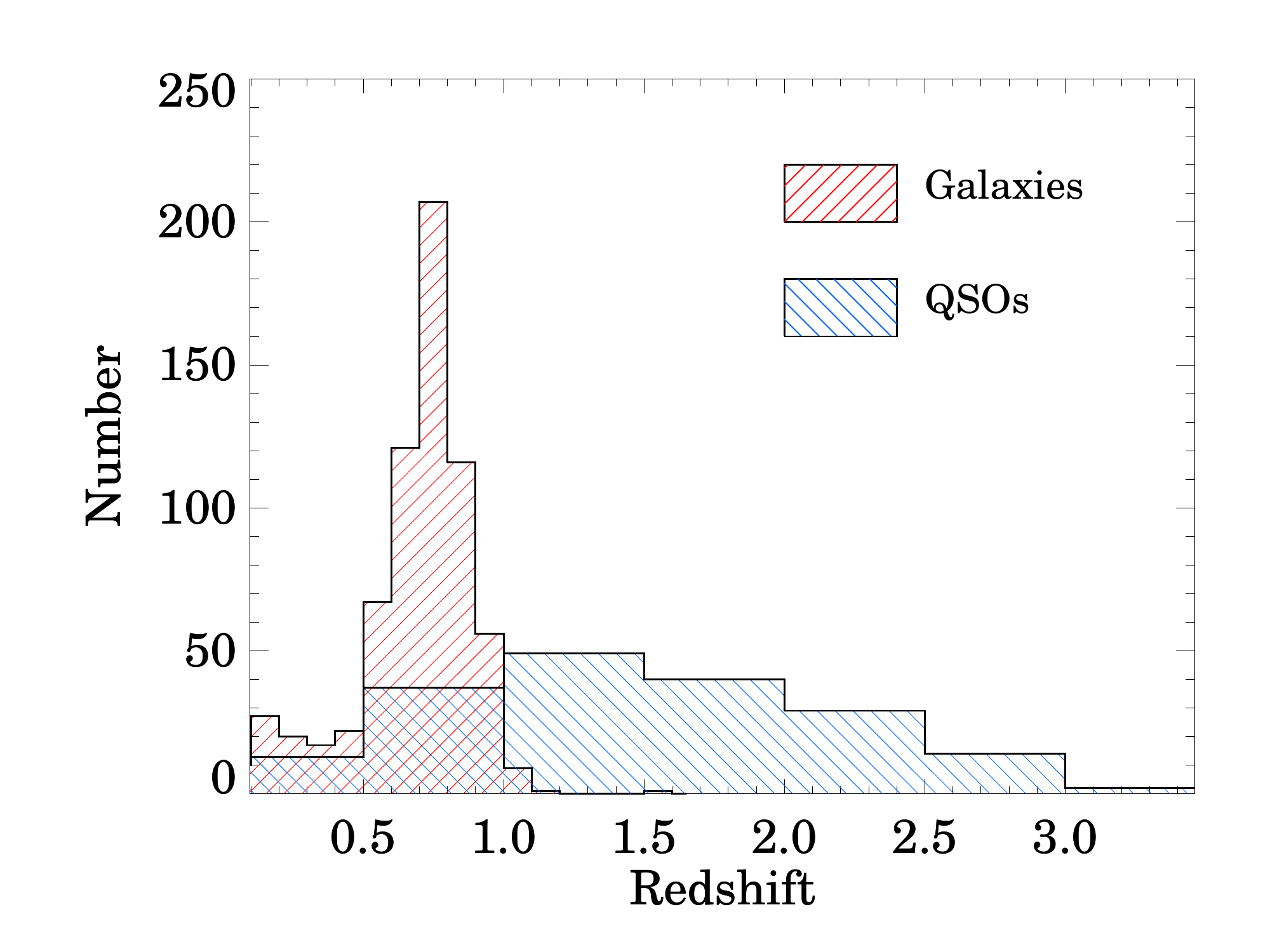}
    \includegraphics[width=0.35\textwidth]{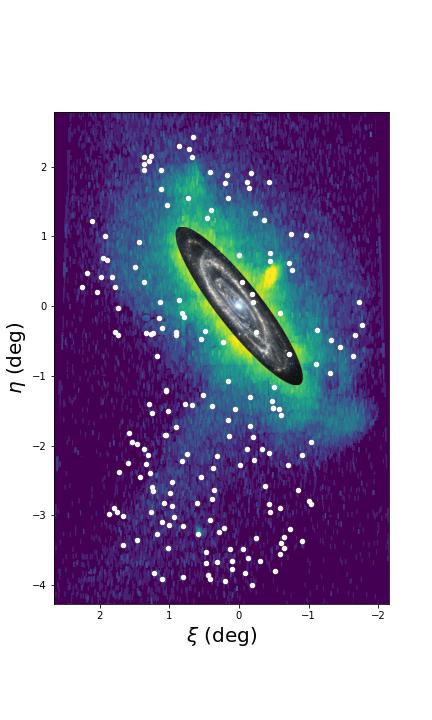}
    \caption{{\it Left:} Redshift distribution of the spectroscopically confirmed QSOs (blue hashed histogram) and galaxies (red hashed histogram). {\it Right:} The sky distribution of the spectroscopically-confirmed QSOs.}
    \label{fig:qsogalzs}
\end{figure}

183 QSO candidates were targeted successfully (i.e., without fiber positioning errors) on the three DESI tiles discussed in this paper, 172 of which were spectroscopically confirmed as QSOs. The remaining 11 includes 8 stars and 3 galaxies. This represents a $\approx$94\% success rate in the QSO selection criteria. In addition, 12 of our M31 stellar candidates turned out to be QSOs, and \Ngalaxies\ were background galaxies. The spectroscopically confirmed QSOs and galaxies are presented in Tables~\ref{tab:qsos} and \ref{tab:galaxies} respectively. Figure~\ref{fig:qsogalzs} shows the redshift distribution of the extragalactic sources and the sky distribution of the QSOs. 

These targets are useful probes of the interstellar and circumgalactic media of M31, and in particular provide a way of investigating any gas that may be associated with the various kinematic structures traced by the stellar debris \cite[e.g.,][]{Koch2015}. 

\begin{deluxetable}{cccccccl}
\tablecaption{QSOs Behind M31\tablenotemark{a}  \label{tab:qsos}}
\tablecomments{Table 5 is published in its entirety in the machine-readable format.
      A portion is shown here for guidance regarding its form and content.}
\tablehead{
\colhead{ID} & 
\colhead{RA ($^\circ$)} & 
\colhead{Dec ($^\circ$)} &
\colhead{Redshift} &  
\colhead{\bf $G_{\rm Gaia}$} & 
\colhead{\bf $g_{\rm PAndAS}$} & 
\colhead{\bf $i_{\rm PAndAS}$} &
\colhead{\bf Alternate Name}
}
\startdata
   1 & 10.0373918 & 40.1050291 & 2.196 & 19.88 & 19.68 & 19.46 &    Gaia DR3 369102106371697792 \\
   2 & 10.3592920 & 40.8907780 & 1.159 & 19.96 & 19.87 & 19.27 &    Gaia DR3 381161584267072896 \\
   3 & 10.0770974 & 39.8989696 & 0.284 & 19.76 & 19.90 & 19.54 &    Gaia DR3 368710061756016896 \\
   4 & 10.4851362 & 39.9701449 & 1.834 & 19.50 & 19.84 & 19.42 &    Gaia DR3 369044622529595520 \\
   5 & 10.0417886 & 39.7983967 & 0.675 & 20.06 &   NaN &   NaN &    Gaia DR3 368707720998989184 \\
   6 & 11.2855672 & 37.7439449 & 1.934 & 19.90 & 20.38 & 20.00 &    Gaia DR3 367495681227400192 \\
   7 & 10.8090715 & 37.6107886 & 2.489 & 19.61 & 19.98 & 19.86 &    Gaia DR3 367443454424848768 \\
   8 & 11.0832708 & 37.6063001 & 2.533 & 19.04 & 19.48 & 19.19 &    Gaia DR3 367490218028868352 \\
   9 & 11.2844517 & 37.5931928 & 1.287 & 20.28 & 20.74 & 20.23 &    Gaia DR3 367492142174203904 \\
  10 & 10.8512314 & 37.7872685 & 2.199 & 18.98 &   NaN &   NaN &    Gaia DR3 367543651717272064 \\
\enddata
 \tablenotetext{a}{See Online Version for complete Table. The columns are: (1) a running index; (2,3) J2000 RA and Dec in decimal degrees; (4) Redshift; (5) {\it Gaia} $G$-band magnitude from Gaia DR3 (NaN if not available); (6,7) the PAndAS $g$ and $i$ magnitude (NaN if not available); (8) {\it Gaia} DR3 identifier.}
\end{deluxetable}

\begin{deluxetable}{ccccccl}
\tablecaption{Galaxies Behind M31\tablenotemark{a} \label{tab:galaxies}}
\tablecomments{Table 6 is published in its entirety in the machine-readable format.
      A portion is shown here for guidance regarding its form and content.}
\tablehead{
\colhead{ID} & 
\colhead{RA ($^\circ$)} & 
\colhead{Dec ($^\circ$)} &
\colhead{Redshift} &  
\colhead{\bf $g_{\rm PAndAS}$} & 
\colhead{\bf $i_{\rm PAndAS}$} &
\colhead{\bf Alternate Name}
}
\startdata
   1 & 10.2211266 & 40.0121625 & 0.283 &   NaN &   NaN & GC7461,SK090C   ,SK090C        \\
   2 &  9.7017682 & 40.0505458 & 0.135 &   NaN &   NaN & GC7429,SK058C   ,SK058C        \\
   3 & 10.3558474 & 40.5148375 & 0.236 &   NaN &   NaN & GC7191,SK078B   ,SK078B        \\
   4 &  9.8651259 & 39.8292622 & 0.749 & 24.41 & 21.92 &  PANDAS  95102                 \\
   5 & 10.6379474 & 40.0817180 & 0.212 &   NaN &   NaN & GC7481,SK110C   ,SK110C        \\
   6 & 10.8841125 & 37.6855167 & 0.560 & 23.81 & 21.52 &  PSUPP 20989                   \\
   7 & 11.0291593 & 37.6711566 & 0.676 & 24.29 & 21.83 &  PANDAS  90320                 \\
   8 & 11.0145968 & 37.7888733 & 0.765 & 24.67 & 22.08 &  PANDAS  90419                 \\
   9 & 11.0623134 & 37.7701455 & 0.674 & 24.99 & 21.89 &  PANDAS  90457                 \\
  10 & 11.1578093 & 37.8240122 & 0.801 & 24.54 & 22.03 &  PANDAS  90472                 \\
\enddata
 \tablenotetext{a}{See Online Version for complete Table. The columns are: (1) a running index; (2,3) J2000 RA and Dec in decimal degrees; (4) Redshift; (5,6) the PAndAS $g$ and $i$ magnitude (NaN if not available); (7) Alternate name, where available.}
\end{deluxetable}

\section{N body simulation details}
\label{sec:nbodysimdetails}
For the simulations described in Section~\ref{sec:nbody_model}, we use a coordinate system that is aligned with the disk of M31, such that $x_{\rm M31}$ and $y_{\rm M31}$ are along the projected major and minor axes of the galaxy. The $x_{\rm M31}$ axis is oriented approximately towards the northeast in the plane of sky; the $y_{\rm M31}$ points out of the plane of the sky towards us but in a southeast direction; and the $z_{\rm M31}$ axis is perpendicular to the M31 disk plane, pointing in a northwesterly direction and tilted out of the plane of the sky slightly towards us.
The transformation between the M31 aligned coordinate system and a sky-oriented coordinate system in which $x$ is pointing east, $y$ is pointing north, and $z$ is pointing away from us along the line of sight can be done with the matrix $M$ (constructed assuming the position angle of the M31 line of nodes of 37$^\circ$ and and inclination of 77$^\circ$):
$$M=\begin{pmatrix}
 0.60181502 & -0.1796539 & -0.77816653 \\
  0.79863551 &  0.13537892 &  0.58639054\\
  0.     & -0.97437006 &   0.22495105
\end{pmatrix}$$
 
We start the simulation with  the progenitor at   $X_{\rm M31}=(-5.44,22.5,35.25)$\,kpc with velocity $V_{\rm M31}=(19.66,-28.79,-64.68)$ {\kms} (in the coordinate system aligned with the disk), where the initial coordinates and velocities are taken  from \citep{Kirihara2017a} and rotated using the matrix $M$. 

We run the model for 977 Myr using the gyrfalcON integrator \citep{dehnen2000,dehnen2014} from the NEMO software package \citep{Teuben1995} using the following command:
\begin{verbatim}
mkplum - 300000 r_s=1 seed=1 mass=9000 | snapshift rshift=-5.44,22.5,35.25
vshift=19.66,-28.79,-64.68  in=- out=- | gyrfalcON - out.snp
accname=nfw,miyamoto,miyamoto,miyamoto,hernquist 
accpars='0,7.63,215;10.68,.72,3.07e5;22.99,.72,-2e5;3.49,0.72,.329e5;0,1.39e5,.6' 
tstop=1 kmax=18 eps=0.1 step=0.001
\end{verbatim}
 
Some 1000 snapshots of the simulation  made are provided on zenodo at \url{https://doi.org/10.5281/zenodo.6977494}. After running the simulations, we convert the outputs back into the space of observables, i.e., the coordinate system aligned with the sky by applying the inverse rotation matrix and assuming that M31 is at a distance of 750 kpc.  We also compute the energies of each particle using the {\tt gala} package \citep{gala}.

\begin{figure}
    \centering
    \includegraphics[width=0.8\textwidth]{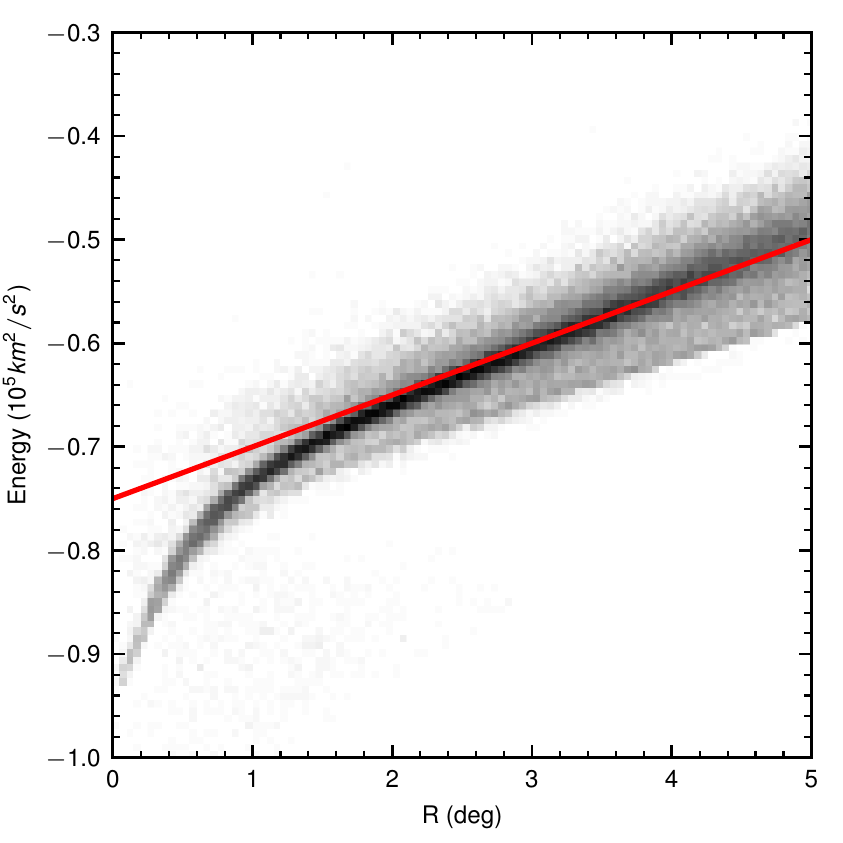}
    \caption{Total energy vs. projected galactocentric radius for the simulation discussed in Appendix B and \S~{\ref{sec: simulations}}. The solid red line represents a slope of $\approx-380~({\rm km\, s^{-1})^2\, kpc^{-1}}$. This shows that a linear relationship between energy and projected radius is  reasonable approximation for projected radii $>1.5^\circ$.}
    \label{fig:sergey_energy_plot}
\end{figure}

Figure~\ref{fig:sergey_energy_plot} shows the resulting energy as a function of projected distance for the particles in the simulation associated with the structure which matches M31's Giant Stellar Stream. In the range between 1$^\circ$ and 4$^\circ$, where we fit for the GSS in \S~\ref{sec:gssmassest}, the total energy is approximately linear with radius. 

\begin{table}[h]
    \centering
    \caption{GSS radial velocity measurements}
    \begin{tabular}{ccc}
    \hline
      { R } & { $V_{GSR}$ - $V_{GSR, M31}$} & { $\sigma_V$} \\
      { deg } & {km/s} & {km/s} \\
      \hline
1.00 & -269.23 & 7.01\\
1.35 & -225.67 & 4.20\\
1.70 & -197.83 & 2.86\\
2.05 & -169.69 & 2.78\\
2.40 & -148.52 & 2.57\\
2.75 & -132.03 & 2.49\\
3.10 & -109.78 & 2.17\\
3.45 & -92.01 & 2.29\\
3.80 & -71.10 & 3.67\\
      \hline
    \end{tabular}
    \label{tab:gss_rv_track}
\end{table}

\section{DESI M31 Measurement Catalog}
\label{appendix:datamodel}

The measurements resulting from our analysis of the DESI spectroscopic
observations are available as a FITS data table (DESI\_M31\_MEASUREMENT\_CATALOG.fits).  Table~\ref{tab:DESI_DataModel}
contains a description of the columns in the FITS data table. The data table
and other information are also available online \update{at
\url{https://doi.org/10.5281/zenodo.6977494}.}

\begin{deluxetable}{lcll}
\tablecaption{Data Model for the M31 Data FITS File}
\label{tab:DESI_DataModel}
\tablehead{
\colhead{Column Name}  &
\colhead{Data Type}  &
\colhead{Units}  &
\colhead{Column Description}
}
\startdata
TARGETID         & LONG64  & -           & DESI target identification \\
RA               & DOUBLE  & deg         & Right Ascention (J2000) in decimal degrees \\
DEC              & DOUBLE  & deg         & Declination (J2000) in decimal degrees \\
DUPLICATE        & LONG64  & -           & DESI TARGETID of the duplicate entry for this source if one exists; 0 otherwise \\
PRIMARY          & INT     & -           & 1 if this is the primary entry; 0 otherwise; only for duplicate entries \\
VRAD\_BEST        & DOUBLE  & ${\rm km\,s^{-1}}$ & "Best" heliocentric radial velocity in ${\rm km\,s^{-1}}$ \\
VRAD\_BEST\_ERR    & DOUBLE  & ${\rm km\,s^{-1}}$ & Uncertainty on "best" heliocentric radial velocity in ${\rm km\,s^{-1}}$ \\
VRAD             & DOUBLE  & ${\rm km\,s^{-1}}$ & Heliocentric radial velocity from DESI's RV Pipeline in ${\rm km\,s^{-1}}$ \\
VRAD\_ERR         & DOUBLE  & ${\rm km\,s^{-1}}$ & Uncertainty on heliocentric radial velocity from DESI's RV Pipeline in ${\rm km\,s^{-1}}$ \\
VRAD\_SKEW        & DOUBLE  & ${\rm km\,s^{-1}}$ & Skew on the heliocentric radial velocity from DESI's RV Pipeline in ${\rm km\,s^{-1}}$ \\
VRAD\_KURT        & DOUBLE  & ${\rm km\,s^{-1}}$ & Kurtosis on the heliocentric radial velocity from DESI's RV Pipeline in ${\rm km\,s^{-1}}$ \\
LOGG             & DOUBLE  & ${\rm log(cm\,s^{-2})}$ & Logarithm of the acceleration due to gravity from DESI's RV Pipeline in ${\rm log(cm\,s^{-2})}$ \\
TEFF             & DOUBLE  & K           & Effective temperature from DESI's RV Pipeline in degrees Kelvin \\
FEH              & DOUBLE  & dex         & [Fe/H] in units of solar metallicity \\
LOGG\_ERR         & DOUBLE  & ${\rm log(cm\,s^{-2})}$ & Uncertainty on log(g) \\
TEFF\_ERR         & DOUBLE  & K           & Uncertainty on Teff (K) \\
FEH\_ERR          & DOUBLE  & dex         & Uncertainty on [Fe/H] (dex) \\
CLASS            & STRING  & -           & Selection class(es) for the target \\
DESIGNATION      & STRING  & -           & Alternate designation of the target \\
SPECTYPE         & STRING  & -           & Spectroscopic type: STAR, GALAXY or QSO; blank if unknown \\
GDR3\_ID          & LONG64  & -           & Gaia DR3 unique identification number \\
TILEID           & STRING  & -           & DESI Tiles contributing to the spectrum \\
SN\_B             & FLOAT   & -           & Signal-to-noise ratio in the blue arm of the DESI spectrograph \\
SN\_R             & FLOAT   & -           & Signal-to-noise ratio in the red arm of the DESI spectrograph \\
SN\_Z             & FLOAT   & -           & Signal-to-noise ratio in the near-infrared arm of the DESI spectrograph \\
VRAD\_REDROCK     & DOUBLE  & ${\rm km\,s^{-1}}$ & Heliocentric radial velocity from the DESI Redrock pipeline in ${\rm km\,s^{-1}}$ \\
VRAD\_REDROCK\_ERR & DOUBLE  & ${\rm km\,s^{-1}}$ & Uncertainty on the heliocentric radial velocity from the DESI Redrock pipeline in ${\rm km\,s^{-1}}$ \\
REDROCK\_ZWARN    & LONG64  & -           & Warning flags for Redrock pipeline reductions (0=good) \\
RVS\_WARN         & INT     & -           & Warning flags for the RVS pipeline reductions (0=good) \\
VRAD\_VI          & DOUBLE  & ${\rm km\,s^{-1}}$ & Heliocentric radial velocity from the Visual Inspection in ${\rm km\,s^{-1}}$ \\
VI\_QUALITY       & INT     & -           & Quality flag from the Visual Inspection ($\ge$3 is good; $\le$2 is bad; $<$0 is not inspected) \\
VI\_SPECTYPE      & STRING  & -           & Spectral type from the Visual Inspection \\
M31DIST          & DOUBLE  & deg         & Projected radial distance from the center of M31\tablenotemark{a} in degrees \\
GAIA\_G           & FLOAT   & mag         & Gaia $G$ magnitude from the Gaia DR3 catalog \\
PANDAS\_G         & FLOAT   & mag         & PAndAS $g$-band magnitude from the PAndAS catalog \\
PANDAS\_I         & FLOAT   & mag         & PAndAS $i$-band magnitude from the PAndAS catalog \\
PRED\_Z           & FLOAT   & mag         & Predicted Legacy Surveys $z$-band magnitude \\
XI               & DOUBLE  & deg         & Standard $\xi$ coordinate in frame centered on M31\tablenotemark{a} in degrees \\
ETA              & DOUBLE  & deg         & Standard $\eta$ coordinate in frame centered on M31\tablenotemark{a} in degrees \\
THETA            & DOUBLE  & deg         & Polar angle centered on M31 and measured clockwise from the west (as in Figures 6,7,11, and 12) \\
\enddata
\tablenotetext{a}{The center of M31 is assumed to lie at RA=10.68470833$^\circ$ and DEC=41.26875$^\circ$.}
\end{deluxetable}


\end{CJK*} 

\end{document}

%% file: author_list.tex
\author[0000-0002-4928-4003]{Arjun~Dey}
\affiliation{NSF's NOIRLab, 950 N. Cherry Avenue, Tucson, AZ 85719, USA}
\author[0000-0002-5758-150X]{Joan~R.~Najita}
\affiliation{NSF's NOIRLab, 950 N. Cherry Avenue, Tucson, AZ 85719, USA}
\author[0000-0003-2644-135X]{Sergey~E.~Koposov}
\affiliation{Institute for Astronomy, University of Edinburgh, Royal Observatory, Blackford Hill, Edinburgh EH9 3HJ, UK}
\affiliation{Institute of Astronomy, University of Cambridge, Madingley road,  Cambridge, CB3 0HA, UK}
\author{J.~Josephy-Zack}
\affiliation{Harvard College, Cambridge, MA 02138, USA}
\author{Gabriel~Maxemin}
\affiliation{Harvard College, Cambridge, MA 02138, USA}
\author[0000-0002-5564-9873]{Eric~F.~Bell}
\affiliation{Department of Astronomy, University of Michigan, Ann Arbor, MI 48109, USA}
\author{C.~Poppett}
\affiliation{Lawrence Berkeley National Laboratory, 1 Cyclotron Road, Berkeley, CA 94720, USA}
\affiliation{Space Sciences Laboratory, University of California, Berkeley, 7 Gauss Way, Berkeley, CA  94720, USA}
\affiliation{University of California, Berkeley, 110 Sproul Hall \#5800 Berkeley, CA 94720, USA}
\author{E.~Patel}
\affiliation{Astronomy Dept., University of California at Berkeley, Berkeley, CA 94720, USA}
\author{L.~{Beraldo~e~Silva}}
\affiliation{Department of Astronomy, University of Michigan, Ann Arbor, MI 48109, USA}
\author{A.~Raichoor}
\affiliation{Lawrence Berkeley National Laboratory, Berkeley, CA 94720, USA}
\author{D.~Schlegel}
\affiliation{Lawrence Berkeley National Laboratory, 1 Cyclotron Road, Berkeley, CA 94720, USA}
\author{D.~Lang}
\affiliation{Perimeter Institute for Theoretical Physics, 31 Caroline Street N, Waterloo, ON N25 2YL, Canada}
\affiliation{Department of Physics and Astronomy, University of Waterloo, Waterloo, ON N2L 3G1, Canada}
\author[0000-0002-1125-7384]{A.~Meisner}
\affiliation{NSF's NOIRLab, 950 N. Cherry Avenue, Tucson, AZ 85719, USA}
\author{Adam~D.~Myers}
\affiliation{Department of Physics \& Astronomy, University  of Wyoming, 1000 E. University, Dept.~3905, Laramie, WY 82071, USA}
\author{J.~Aguilar}
\affiliation{Lawrence Berkeley National Laboratory, 1 Cyclotron Road, Berkeley, CA 94720, USA}
\author[0000-0001-6098-7247]{S.~Ahlen}
\affiliation{Physics Dept., Boston University, 590 Commonwealth Avenue, Boston, MA 02215, USA}
\author{C.~Allende~Prieto}
\affiliation{Instituto de Astrof\'{i}sica de Canarias, C/ Vía L\'{a}ctea, s/n, 38205 San Crist\'{o}bal de La Laguna, Santa Cruz de Tenerife, Spain}
\author{D.~Brooks}
\affiliation{Department of Physics \& Astronomy, University College London, Gower Street, London, WC1E 6BT, UK}
\author[0000-0001-8274-158X]{A.P.~Cooper}
\affiliation{Institute of Astronomy and Department of Physics, National Tsing Hua University, 101 Kuang-Fu Rd. Sec. 2, Hsinchu 30013, Taiwan}
\affiliation{Center for Informatics and Computation in Astronomy, NTHU, 101 Kuang-Fu Rd. Sec. 2, Hsinchu 30013, Taiwan}
\affiliation{Physics Division, National Center for Theoretical Sciences, Taipei 10617, Taiwan}
\author{K.~S.~Dawson}
\affiliation{Department of Physics and Astronomy, The University of Utah, 115 South 1400 East, Salt Lake City, UT 84112, USA}
\author{A.~de la Macorra}
\affiliation{Instituto de F\'{\i}sica, Universidad Nacional Aut\'{o}noma de M\'{e}xico,  Cd. de M\'{e}xico  C.P. 04510,  M\'{e}xico}
\author{P.~Doel}
\affiliation{Department of Physics \& Astronomy, University College London, Gower Street, London, WC1E 6BT, UK}
\author[0000-0002-3033-7312]{A.~Font-Ribera}
\affiliation{Institut de F\'{i}sica d’Altes Energies (IFAE), The Barcelona Institute of Science and Technology, Campus UAB, 08193 Bellaterra Barcelona, Spain}
\author[0000-0002-9370-8360]{Juan~Garc\'{i}a-Bellido}
\affiliation{Instituto de F\'{\i}sica Te\'{o}rica (IFT) UAM/CSIC, Universidad Aut\'{o}noma de Madrid, Cantoblanco, E-28049, Madrid, Spain}
\author[0000-0003-3142-233X]{{S.}~Gontcho A Gontcho}
\affiliation{Lawrence Berkeley National Laboratory, One Cyclotron Road, Berkeley, CA 94720, USA}
\author{J.~Guy}
\affiliation{Lawrence Berkeley National Laboratory, 1 Cyclotron Road, Berkeley, CA 94720, USA}
\author{K.~Honscheid}
\affiliation{Center for Cosmology and AstroParticle Physics, The Ohio State University, 191 West Woodruff Avenue, Columbus, OH 43210, USA}
\affiliation{Department of Physics, The Ohio State University, 191 West Woodruff Avenue, Columbus, OH 43210, USA}
\author{R.~Kehoe}
\affiliation{Department of Physics, Southern Methodist University, 3215 Daniel Avenue, Dallas, TX 75275, USA}
\author[0000-0003-3510-7134]{T.~Kisner}
\affiliation{Lawrence Berkeley National Laboratory, 1 Cyclotron Road, Berkeley, CA 94720, USA}
\author[0000-0001-6356-7424]{A.~Kremin}
\affiliation{Lawrence Berkeley National Laboratory, 1 Cyclotron Road, Berkeley, CA 94720, USA}
\author[0000-0003-1838-8528]{M.~Landriau}
\affiliation{Lawrence Berkeley National Laboratory, 1 Cyclotron Road, Berkeley, CA 94720, USA}
\author[0000-0001-7178-8868]{L.~Le~Guillou}
\affiliation{Sorbonne Universit\'{e}, CNRS/IN2P3, Laboratoire de Physique Nucl\'{e}aire et de Hautes Energies (LPNHE), FR-75005 Paris, France}
\author[0000-0003-1887-1018]{Michael E.~Levi}
\affiliation{Lawrence Berkeley National Laboratory, 1 Cyclotron Road, Berkeley, CA 94720, USA}
\author[0000-0002-9110-6163]{T.~S.~Li}
\affiliation{Department of Astronomy \& Astrophysics, University of Toronto, Toronto, ON M5S 3H4, Canada}
\author[0000-0002-4279-4182]{Paul~Martini}
\affiliation{Center for Cosmology and AstroParticle Physics, The Ohio State University, 191 West Woodruff Avenue, Columbus, OH 43210, USA}
\affiliation{Department of Astronomy, The Ohio State University, 4055 McPherson Laboratory, 140 W 18th Avenue, Columbus, OH 43210, USA}
\author{R.~Miquel}
\affiliation{Instituci\'{o} Catalana de Recerca i Estudis Avan\c{c}ats, Passeig de Llu\'{\i}s Companys, 23, 08010 Barcelona, Spain}
\affiliation{Institut de F\'{i}sica d’Altes Energies (IFAE), The Barcelona Institute of Science and Technology, Campus UAB, 08193 Bellaterra Barcelona, Spain}
\author[0000-0002-2733-4559]{J.~Moustakas}
\affiliation{Department of Physics and Astronomy, Siena College, 515 Loudon Road, Loudonville, NY 12211, USA}
\author[0000-0001-6590-8122]{Jundan Nie}
\affiliation{National Astronomical Observatories, Chinese Academy of Sciences, A20 Datun Rd., Chaoyang District, Beijing, 100012, P.R. China}
\author[0000-0003-3188-784X]{N.~Palanque-Delabrouille}
\affiliation{IRFU, CEA, Universit\'{e} Paris-Saclay, F-91191 Gif-sur-Yvette, France}
\affiliation{Lawrence Berkeley National Laboratory, 1 Cyclotron Road, Berkeley, CA 94720, USA}
\author{F.~Prada}
\affiliation{Instituto de Astrof\'isica de Andaluc\'ia (CSIC), Glorieta de la Astronom\'ia, E-18080 Granada, Spain}
\author[0000-0002-3569-7421]{E.~F.~Schlafly}
\affiliation{Space Telescope Science Institute, 3700 San Martin Dr, Baltimore MD 21218, USA}
\author{Ray~M.~Sharples}
\affiliation{Centre for Advanced Instrumentation, Department of Physics, Durham University, South Road, Durham DH1 3LE, UK}
\author[0000-0003-1704-0781]{Gregory~Tarl\'{e}}
\affiliation{Department of Physics, University of Michigan, Ann Arbor, MI 48109, USA}
\author[0000-0001-5082-9536]{Yuan-Sen~Ting~(丁源森)}
\affiliation{Research School of Astronomy \& Astrophysics, Australian National University, Cotter Road, Weston, ACT 2611, Australia}
\affiliation{School of Computing, Australian National University, Acton ACT 2601, Australia"}
\author{L.~Tyas}
\affiliation{Lawrence Berkeley National Laboratory, 1 Cyclotron Road, Berkeley, CA 94720, USA}
\author[0000-0002-6257-2341]{M.~Valluri}
\affiliation{Department of Astronomy, University of Michigan, Ann Arbor, MI 48109, USA}
\author[0000-0003-2229-011X]{Risa~H.~Wechsler}
\affiliation{Kavli Institute for Particle Astrophysics and Cosmology, Stanford University, 452 Lomita Mall, Stanford, CA 94305, USA}
\affiliation{Physics Department, Stanford University, Stanford, CA 93405, USA}
\affiliation{SLAC National Accelerator Laboratory, 2575 Sand Hill Road, Menlo Park, CA 94025, USA}
\author[0000-0002-6684-3997]{H.~Zou}
\affiliation{National Astronomical Observatories, Chinese Academy of Sciences, A20 Datun Rd., Chaoyang District, Beijing, 100012, P.R. China}

%% file: main.bbl
\begin{thebibliography}{}
\expandafter\ifx\csname natexlab\endcsname\relax\def\natexlab#1{#1}\fi
\providecommand{\url}[1]{\href{#1}{#1}}
\providecommand{\dodoi}[1]{doi:~\href{http://doi.org/#1}{\nolinkurl{#1}}}
\providecommand{\doeprint}[1]{\href{http://ascl.net/#1}{\nolinkurl{http://ascl.net/#1}}}
\providecommand{\doarXiv}[1]{\href{https://arxiv.org/abs/#1}{\nolinkurl{https://arxiv.org/abs/#1}}}

\bibitem[{{Alexander} {et~al.}(2022){Alexander}, {Davis}, {Chaussidon},
  {Fawcett}, {Gonzalez-Morales}, {Lan}, {Yeche}, {Ahlen}, {Aguilar},
  {Armengaud}, {Bailey}, {Brooks}, {Cai}, {Canning}, {Carr}, {Chabanier},
  {Cousinou}, {Dawson}, {de la Macorra}, {Dey}, {Dey}, {Dhungana}, {Edge},
  {Eftekharzadeh}, {Fanning}, {Farr}, {Font-Ribera}, {Garcia-Bellido},
  {Garrison}, {Gaztanaga}, {Gontcho}, {Gordon}, {Guadalupe Medellin Gonzalez},
  {Guy}, {Herrera-Alcantar}, {Jiang}, {Juneau}, {Karacayli}, {Kehoe}, {Kisner},
  {Kovacs}, {Landriau}, {Levi}, {Magneville}, {Martini}, {Meisner}, {Mezcua},
  {Miquel}, {Montero Camacho}, {Moustakas}, {Munoz-Gutierrez}, {Myers},
  {Nadathur}, {Napolitano}, {Nie}, {Palanque-Delabrouille}, {Pan}, {Percival},
  {Perez-Rafols}, {Poppett}, {Prada}, {Ramirez-Perez}, {Ravoux}, {Rosario},
  {Schubnell}, {Tarle}, {Walther}, {Weiner}, {Youles}, {Zhou}, {Zou}, \&
  {Zou}}]{alexander22a}
{Alexander}, D.~M., {Davis}, T.~M., {Chaussidon}, E., {et~al.} 2022, arXiv
  e-prints, arXiv:2208.08517.
\newblock \doarXiv{2208.08517}

\bibitem[{{Arnaboldi} {et~al.}(2022){Arnaboldi}, {Bhattacharya}, {Gerhard},
  {Kobayashi}, {Freeman}, {Caldwell}, {Hartke}, {McConnachie}, \&
  {Guhathakurta}}]{Arnaboldi2022}
{Arnaboldi}, M., {Bhattacharya}, S., {Gerhard}, O., {et~al.} 2022, \aap, 666,
  A109, \dodoi{10.1051/0004-6361/202244258}

\bibitem[{{Astropy Collaboration} {et~al.}(2013){Astropy Collaboration},
  {Robitaille}, {Tollerud}, {Greenfield}, {Droettboom}, {Bray}, {Aldcroft},
  {Davis}, {Ginsburg}, {Price-Whelan}, {Kerzendorf}, {Conley}, {Crighton},
  {Barbary}, {Muna}, {Ferguson}, {Grollier}, {Parikh}, {Nair}, {Unther},
  {Deil}, {Woillez}, {Conseil}, {Kramer}, {Turner}, {Singer}, {Fox}, {Weaver},
  {Zabalza}, {Edwards}, {Azalee Bostroem}, {Burke}, {Casey}, {Crawford},
  {Dencheva}, {Ely}, {Jenness}, {Labrie}, {Lim}, {Pierfederici}, {Pontzen},
  {Ptak}, {Refsdal}, {Servillat}, \& {Streicher}}]{astropy:2013}
{Astropy Collaboration}, {Robitaille}, T.~P., {Tollerud}, E.~J., {et~al.} 2013,
  \aap, 558, A33, \dodoi{10.1051/0004-6361/201322068}

\bibitem[{{Astropy Collaboration} {et~al.}(2018){Astropy Collaboration},
  {Price-Whelan}, {Sip{\H{o}}cz}, {G{\"u}nther}, {Lim}, {Crawford}, {Conseil},
  {Shupe}, {Craig}, {Dencheva}, {Ginsburg}, {VanderPlas}, {Bradley},
  {P{\'e}rez-Su{\'a}rez}, {de Val-Borro}, {Aldcroft}, {Cruz}, {Robitaille},
  {Tollerud}, {Ardelean}, {Babej}, {Bach}, {Bachetti}, {Bakanov}, {Bamford},
  {Barentsen}, {Barmby}, {Baumbach}, {Berry}, {Biscani}, {Boquien}, {Bostroem},
  {Bouma}, {Brammer}, {Bray}, {Breytenbach}, {Buddelmeijer}, {Burke},
  {Calderone}, {Cano Rodr{\'\i}guez}, {Cara}, {Cardoso}, {Cheedella}, {Copin},
  {Corrales}, {Crichton}, {D'Avella}, {Deil}, {Depagne}, {Dietrich}, {Donath},
  {Droettboom}, {Earl}, {Erben}, {Fabbro}, {Ferreira}, {Finethy}, {Fox},
  {Garrison}, {Gibbons}, {Goldstein}, {Gommers}, {Greco}, {Greenfield},
  {Groener}, {Grollier}, {Hagen}, {Hirst}, {Homeier}, {Horton}, {Hosseinzadeh},
  {Hu}, {Hunkeler}, {Ivezi{\'c}}, {Jain}, {Jenness}, {Kanarek}, {Kendrew},
  {Kern}, {Kerzendorf}, {Khvalko}, {King}, {Kirkby}, {Kulkarni}, {Kumar},
  {Lee}, {Lenz}, {Littlefair}, {Ma}, {Macleod}, {Mastropietro}, {McCully},
  {Montagnac}, {Morris}, {Mueller}, {Mumford}, {Muna}, {Murphy}, {Nelson},
  {Nguyen}, {Ninan}, {N{\"o}the}, {Ogaz}, {Oh}, {Parejko}, {Parley}, {Pascual},
  {Patil}, {Patil}, {Plunkett}, {Prochaska}, {Rastogi}, {Reddy Janga},
  {Sabater}, {Sakurikar}, {Seifert}, {Sherbert}, {Sherwood-Taylor}, {Shih},
  {Sick}, {Silbiger}, {Singanamalla}, {Singer}, {Sladen}, {Sooley},
  {Sornarajah}, {Streicher}, {Teuben}, {Thomas}, {Tremblay}, {Turner},
  {Terr{\'o}n}, {van Kerkwijk}, {de la Vega}, {Watkins}, {Weaver}, {Whitmore},
  {Woillez}, {Zabalza}, \& {Astropy Contributors}}]{astropy:2018}
{Astropy Collaboration}, {Price-Whelan}, A.~M., {Sip{\H{o}}cz}, B.~M., {et~al.}
  2018, \aj, 156, 123, \dodoi{10.3847/1538-3881/aabc4f}

\bibitem[{{Bailey}(2012)}]{Bailey2012}
{Bailey}, S. 2012, \pasp, 124, 1015, \dodoi{10.1086/668105}

\bibitem[{Bailey({2022})}]{Bailey_Redrock:inprep}
Bailey, S. {2022}, in preparation

\bibitem[{{Barmby} {et~al.}(2006){Barmby}, {Ashby}, {Bianchi}, {Engelbracht},
  {Gehrz}, {Gordon}, {Hinz}, {Huchra}, {Humphreys}, {Pahre},
  {P{\'e}rez-Gonz{\'a}lez}, {Polomski}, {Rieke}, {Thilker}, {Willner}, \&
  {Woodward}}]{barmby2006}
{Barmby}, P., {Ashby}, M.~L.~N., {Bianchi}, L., {et~al.} 2006, \apjl, 650, L45,
  \dodoi{10.1086/508626}

\bibitem[{{Beaumont} {et~al.}(2015){Beaumont}, {Goodman}, \&
  {Greenfield}}]{Beaumont2015}
{Beaumont}, C., {Goodman}, A., \& {Greenfield}, P. 2015, in Astronomical
  Society of the Pacific Conference Series, Vol. 495, Astronomical Data
  Analysis Software an Systems XXIV (ADASS XXIV), ed. A.~R. {Taylor} \&
  E.~{Rosolowsky}, 101

\bibitem[{{Bell} {et~al.}(2008){Bell}, {Zucker}, {Belokurov}, {Sharma},
  {Johnston}, {Bullock}, {Hogg}, {Jahnke}, {de Jong}, {Beers}, {Evans},
  {Grebel}, {Ivezi{\'c}}, {Koposov}, {Rix}, {Schneider}, {Steinmetz}, \&
  {Zolotov}}]{bell2008}
{Bell}, E.~F., {Zucker}, D.~B., {Belokurov}, V., {et~al.} 2008, \apj, 680, 295,
  \dodoi{10.1086/588032}

\bibitem[{{Belokurov} {et~al.}(2018){Belokurov}, {Erkal}, {Evans}, {Koposov},
  \& {Deason}}]{Belokurov2018}
{Belokurov}, V., {Erkal}, D., {Evans}, N.~W., {Koposov}, S.~E., \& {Deason},
  A.~J. 2018, \mnras, 478, 611, \dodoi{10.1093/mnras/sty982}

\bibitem[{{Belokurov} {et~al.}(2022){Belokurov}, {Vasiliev}, {Deason},
  {Koposov}, {Fattahi}, {Dillamore}, {Davies}, \& {Grand}}]{Belokurov2022a}
{Belokurov}, V., {Vasiliev}, E., {Deason}, A.~J., {et~al.} 2022, arXiv
  e-prints, arXiv:2208.11135.
\newblock \doarXiv{2208.11135}

\bibitem[{{Beraldo e Silva} {et~al.}(2019){Beraldo e Silva}, {de Siqueira
  Pedra}, \& {Valluri}}]{Beraldo_e_Silva2019}
{Beraldo e Silva}, L., {de Siqueira Pedra}, W., \& {Valluri}, M. 2019, \apj,
  872, 20, \dodoi{10.3847/1538-4357/aaf8a7}

\bibitem[{{Bhattacharya} {et~al.}(2019){Bhattacharya}, {Arnaboldi}, {Caldwell},
  {Gerhard}, {Bla{\~n}a}, {McConnachie}, {Hartke}, {Guhathakurta}, {Pulsoni},
  \& {Freeman}}]{Bhattacharya2019}
{Bhattacharya}, S., {Arnaboldi}, M., {Caldwell}, N., {et~al.} 2019, \aap, 631,
  A56, \dodoi{10.1051/0004-6361/201935898}

\bibitem[{{Boldrini} {et~al.}(2021){Boldrini}, {Mohayaee}, \&
  {Silk}}]{Boldrini2021}
{Boldrini}, P., {Mohayaee}, R., \& {Silk}, J. 2021, \apj, 919, 86,
  \dodoi{10.3847/1538-4357/ac12d3}

\bibitem[{{Bonaca} {et~al.}(2020){Bonaca}, {Conroy}, {Cargile}, {Naidu},
  {Johnson}, {Zaritsky}, {Ting}, {Caldwell}, {Han}, \& {van
  Dokkum}}]{Bonaca2020}
{Bonaca}, A., {Conroy}, C., {Cargile}, P.~A., {et~al.} 2020, \apjl, 897, L18,
  \dodoi{10.3847/2041-8213/ab9caa}

\bibitem[{Breiman(2001)}]{Breiman2001}
Breiman, L. 2001, Machine Learning, 45, 5, \dodoi{10.1023/A:1010933404324}

\bibitem[{{Bressan} {et~al.}(2012){Bressan}, {Marigo}, {Girardi}, {Salasnich},
  {Dal Cero}, {Rubele}, \& {Nanni}}]{PARSEC2012}
{Bressan}, A., {Marigo}, P., {Girardi}, L., {et~al.} 2012, \mnras, 427, 127,
  \dodoi{10.1111/j.1365-2966.2012.21948.x}

\bibitem[{{Brown} {et~al.}(2006){Brown}, {Smith}, {Ferguson}, {Rich},
  {Guhathakurta}, {Renzini}, {Sweigart}, \& {Kimble}}]{Brown2006}
{Brown}, T.~M., {Smith}, E., {Ferguson}, H.~C., {et~al.} 2006, \apj, 652, 323,
  \dodoi{10.1086/508015}

\bibitem[{{Bullock} \& {Johnston}(2005)}]{bullock_johnston_2005}
{Bullock}, J.~S., \& {Johnston}, K.~V. 2005, \apj, 635, 931,
  \dodoi{10.1086/497422}

\bibitem[{{Bullock} {et~al.}(2001){Bullock}, {Kravtsov}, \&
  {Weinberg}}]{bullock2001}
{Bullock}, J.~S., {Kravtsov}, A.~V., \& {Weinberg}, D.~H. 2001, \apj, 548, 33,
  \dodoi{10.1086/318681}

\bibitem[{{Caldwell} \& {Romanowsky}(2016)}]{Caldwell2016}
{Caldwell}, N., \& {Romanowsky}, A.~J. 2016, \apj, 824, 42,
  \dodoi{10.3847/0004-637X/824/1/42}

\bibitem[{{Chemin} {et~al.}(2009){Chemin}, {Carignan}, \&
  {Foster}}]{Chemin2009}
{Chemin}, L., {Carignan}, C., \& {Foster}, T. 2009, \apj, 705, 1395,
  \dodoi{10.1088/0004-637X/705/2/1395}

\bibitem[{{Chen} {et~al.}(2016){Chen}, {Liu}, {Xiang}, {Yuan}, {Huang}, {Shi},
  {Fan}, {Huo}, {Wang}, {Ren}, {Tian}, {Zhang}, {Liu}, {Cao}, {Zhang}, {Hou},
  \& {Wang}}]{Chen2016}
{Chen}, B., {Liu}, X., {Xiang}, M., {et~al.} 2016, \aj, 152, 45,
  \dodoi{10.3847/0004-6256/152/2/45}

\bibitem[{{Chen} {et~al.}(2015){Chen}, {Liu}, {Xiang}, {Yuan}, {Huang}, {Huo},
  {Sun}, {Wang}, {Ren}, {Zhang}, {Rebassa-Mansergas}, {Yang}, {Zhang}, {Hou},
  \& {Wang}}]{Chen2015}
{Chen}, B.-Q., {Liu}, X.-W., {Xiang}, M.-S., {et~al.} 2015, Research in
  Astronomy and Astrophysics, 15, 1392, \dodoi{10.1088/1674-4527/15/8/020}

\bibitem[{{Conn} {et~al.}(2016){Conn}, {McMonigal}, {Bate}, {Lewis}, {Ibata},
  {Martin}, {McConnachie}, {Ferguson}, {Irwin}, {Elahi}, {Venn}, \&
  {Mackey}}]{Conn2016}
{Conn}, A.~R., {McMonigal}, B., {Bate}, N.~F., {et~al.} 2016, \mnras, 458,
  3282, \dodoi{10.1093/mnras/stw513}

\bibitem[{{Cooper} {et~al.}(2010){Cooper}, {Cole}, {Frenk}, {White}, {Helly},
  {Benson}, {De Lucia}, {Helmi}, {Jenkins}, {Navarro}, {Springel}, \&
  {Wang}}]{Cooper2010}
{Cooper}, A.~P., {Cole}, S., {Frenk}, C.~S., {et~al.} 2010, \mnras, 406, 744,
  \dodoi{10.1111/j.1365-2966.2010.16740.x}

\bibitem[{{Cooper} {et~al.}(2022){Cooper}, {Koposov}, {Allende Prieto},
  {Manser}, {Kizhuprakkat}, {Myers}, {Dey}, {Gaensicke}, {Li}, {Rockosi},
  {Valluri}, {Najita}, {Deason}, {Raichoor}, {Wang}, {Ting}, {Kim}, {Carrillo},
  {Wang}, {Beraldo e Silva}, {Han}, {Ding}, {Sanchez-Conde}, {Aguilar},
  {Ahlen}, {Bailey}, {Belokurov}, {Brooks}, {Cunha}, {Dawson}, {Font-Ribera},
  {Forero-Romero}, {Gaztanaga}, {Gontcho}, {Guy}, {Honscheid}, {Kehoe},
  {Kisner}, {Kremin}, {Landriau}, {Levi}, {Martini}, {Meisner}, {Miquel},
  {Poppett}, {Prada}, {Rehemtulla}, {Schlafly}, {Schlegel}, {Weinberg}, {Zhou},
  \& {Zou}}]{Cooper2022}
{Cooper}, A.~P., {Koposov}, S.~E., {Allende Prieto}, C., {et~al.} 2022, arXiv
  e-prints, arXiv:2208.08514.
\newblock \doarXiv{2208.08514}

\bibitem[{{de Vaucouleurs} {et~al.}(1991){de Vaucouleurs}, {de Vaucouleurs},
  {Corwin}, {Buta}, {Paturel}, \& {Fouque}}]{rc3}
{de Vaucouleurs}, G., {de Vaucouleurs}, A., {Corwin}, Herold~G., J., {et~al.}
  1991, {Third Reference Catalogue of Bright Galaxies} (Springer-Verlag: New
  York)

\bibitem[{{Dehnen}(2000)}]{dehnen2000}
{Dehnen}, W. 2000, \apjl, 536, L39, \dodoi{10.1086/312724}

\bibitem[{{Dehnen}(2014)}]{dehnen2014}
---. 2014, {gyrfalcON: N-body code}, Astrophysics Source Code Library, record
  ascl:1402.031.
\newblock \doeprint{1402.031}

\bibitem[{{DESI Collaboration} {et~al.}(2016{\natexlab{a}}){DESI
  Collaboration}, {Aghamousa}, {Aguilar}, {Ahlen}, {Alam}, {Allen}, {Allende
  Prieto}, {Annis}, {Bailey}, {Balland}, {Ballester}, {Baltay}, {Beaufore},
  {Bebek}, {Beers}, {Bell}, {Bernal}, {Besuner}, {Beutler}, {Blake}, {Bleuler},
  {Blomqvist}, {Blum}, {Bolton}, {Briceno}, {Brooks}, {Brownstein},
  {Buckley-Geer}, {Burden}, {Burtin}, {Busca}, {Cahn}, {Cai}, {Cardiel-Sas},
  {Carlberg}, {Carton}, {Casas}, {Castander}, {Cervantes-Cota}, {Claybaugh},
  {Close}, {Coker}, {Cole}, {Comparat}, {Cooper}, {Cousinou}, {Crocce}, {Cuby},
  {Cunningham}, {Davis}, {Dawson}, {de la Macorra}, {De Vicente}, {Delubac},
  {Derwent}, {Dey}, {Dhungana}, {Ding}, {Doel}, {Duan}, {Ealet}, {Edelstein},
  {Eftekharzadeh}, {Eisenstein}, {Elliott}, {Escoffier}, {Evatt}, {Fagrelius},
  {Fan}, {Fanning}, {Farahi}, {Farihi}, {Favole}, {Feng}, {Fernandez},
  {Findlay}, {Finkbeiner}, {Fitzpatrick}, {Flaugher}, {Flender}, {Font-Ribera},
  {Forero-Romero}, {Fosalba}, {Frenk}, {Fumagalli}, {Gaensicke}, {Gallo},
  {Garcia-Bellido}, {Gaztanaga}, {Pietro Gentile Fusillo}, {Gerard},
  {Gershkovich}, {Giannantonio}, {Gillet}, {Gonzalez-de-Rivera},
  {Gonzalez-Perez}, {Gott}, {Graur}, {Gutierrez}, {Guy}, {Habib}, {Heetderks},
  {Heetderks}, {Heitmann}, {Hellwing}, {Herrera}, {Ho}, {Holland}, {Honscheid},
  {Huff}, {Hutchinson}, {Huterer}, {Hwang}, {Illa Laguna}, {Ishikawa},
  {Jacobs}, {Jeffrey}, {Jelinsky}, {Jennings}, {Jiang}, {Jimenez}, {Johnson},
  {Joyce}, {Jullo}, {Juneau}, {Kama}, {Karcher}, {Karkar}, {Kehoe}, {Kennamer},
  {Kent}, {Kilbinger}, {Kim}, {Kirkby}, {Kisner}, {Kitanidis}, {Kneib},
  {Koposov}, {Kovacs}, {Koyama}, {Kremin}, {Kron}, {Kronig}, {Kueter-Young},
  {Lacey}, {Lafever}, {Lahav}, {Lambert}, {Lampton}, {Landriau}, {Lang},
  {Lauer}, {Le Goff}, {Le Guillou}, {Le Van Suu}, {Lee}, {Lee}, {Leitner},
  {Lesser}, {Levi}, {L'Huillier}, {Li}, {Liang}, {Lin}, {Linder}, {Loebman},
  {Luki{\'c}}, {Ma}, {MacCrann}, {Magneville}, {Makarem}, {Manera}, {Manser},
  {Marshall}, {Martini}, {Massey}, {Matheson}, {McCauley}, {McDonald},
  {McGreer}, {Meisner}, {Metcalfe}, {Miller}, {Miquel}, {Moustakas}, {Myers},
  {Naik}, {Newman}, {Nichol}, {Nicola}, {Nicolati da Costa}, {Nie}, {Niz},
  {Norberg}, {Nord}, {Norman}, {Nugent}, {O'Brien}, {Oh}, {Olsen}, {Padilla},
  {Padmanabhan}, {Padmanabhan}, {Palanque-Delabrouille}, {Palmese},
  {Pappalardo}, {P{\^a}ris}, {Park}, {Patej}, {Peacock}, {Peiris}, {Peng},
  {Percival}, {Perruchot}, {Pieri}, {Pogge}, {Pollack}, {Poppett}, {Prada},
  {Prakash}, {Probst}, {Rabinowitz}, {Raichoor}, {Ree}, {Refregier}, {Regal},
  {Reid}, {Reil}, {Rezaie}, {Rockosi}, {Roe}, {Ronayette}, {Roodman}, {Ross},
  {Ross}, {Rossi}, {Rozo}, {Ruhlmann-Kleider}, {Rykoff}, {Sabiu}, {Samushia},
  {Sanchez}, {Sanchez}, {Schlegel}, {Schneider}, {Schubnell}, {Secroun},
  {Seljak}, {Seo}, {Serrano}, {Shafieloo}, {Shan}, {Sharples}, {Sholl},
  {Shourt}, {Silber}, {Silva}, {Sirk}, {Slosar}, {Smith}, {Smoot}, {Som},
  {Song}, {Sprayberry}, {Staten}, {Stefanik}, {Tarle}, {Sien Tie}, {Tinker},
  {Tojeiro}, {Valdes}, {Valenzuela}, {Valluri}, {Vargas-Magana}, {Verde},
  {Walker}, {Wang}, {Wang}, {Weaver}, {Weaverdyck}, {Wechsler}, {Weinberg},
  {White}, {Yang}, {Yeche}, {Zhang}, {Zhao}, {Zheng}, {Zhou}, {Zhou}, {Zhu},
  {Zou}, \& {Zu}}]{DESI_Tech_FDR}
{DESI Collaboration}, {Aghamousa}, A., {Aguilar}, J., {et~al.}
  2016{\natexlab{a}}, arXiv e-prints, arXiv:1611.00037.
\newblock \doarXiv{1611.00037}

\bibitem[{{DESI Collaboration} {et~al.}(2016{\natexlab{b}}){DESI
  Collaboration}, {Aghamousa}, {Aguilar}, {Ahlen}, {Alam}, {Allen}, {Allende
  Prieto}, {Annis}, {Bailey}, {Balland}, {Ballester}, {Baltay}, {Beaufore},
  {Bebek}, {Beers}, {Bell}, {Bernal}, {Besuner}, {Beutler}, {Blake}, {Bleuler},
  {Blomqvist}, {Blum}, {Bolton}, {Briceno}, {Brooks}, {Brownstein},
  {Buckley-Geer}, {Burden}, {Burtin}, {Busca}, {Cahn}, {Cai}, {Cardiel-Sas},
  {Carlberg}, {Carton}, {Casas}, {Castander}, {Cervantes-Cota}, {Claybaugh},
  {Close}, {Coker}, {Cole}, {Comparat}, {Cooper}, {Cousinou}, {Crocce}, {Cuby},
  {Cunningham}, {Davis}, {Dawson}, {de la Macorra}, {De Vicente}, {Delubac},
  {Derwent}, {Dey}, {Dhungana}, {Ding}, {Doel}, {Duan}, {Ealet}, {Edelstein},
  {Eftekharzadeh}, {Eisenstein}, {Elliott}, {Escoffier}, {Evatt}, {Fagrelius},
  {Fan}, {Fanning}, {Farahi}, {Farihi}, {Favole}, {Feng}, {Fernandez},
  {Findlay}, {Finkbeiner}, {Fitzpatrick}, {Flaugher}, {Flender}, {Font-Ribera},
  {Forero-Romero}, {Fosalba}, {Frenk}, {Fumagalli}, {Gaensicke}, {Gallo},
  {Garcia-Bellido}, {Gaztanaga}, {Pietro Gentile Fusillo}, {Gerard},
  {Gershkovich}, {Giannantonio}, {Gillet}, {Gonzalez-de-Rivera},
  {Gonzalez-Perez}, {Gott}, {Graur}, {Gutierrez}, {Guy}, {Habib}, {Heetderks},
  {Heetderks}, {Heitmann}, {Hellwing}, {Herrera}, {Ho}, {Holland}, {Honscheid},
  {Huff}, {Hutchinson}, {Huterer}, {Hwang}, {Illa Laguna}, {Ishikawa},
  {Jacobs}, {Jeffrey}, {Jelinsky}, {Jennings}, {Jiang}, {Jimenez}, {Johnson},
  {Joyce}, {Jullo}, {Juneau}, {Kama}, {Karcher}, {Karkar}, {Kehoe}, {Kennamer},
  {Kent}, {Kilbinger}, {Kim}, {Kirkby}, {Kisner}, {Kitanidis}, {Kneib},
  {Koposov}, {Kovacs}, {Koyama}, {Kremin}, {Kron}, {Kronig}, {Kueter-Young},
  {Lacey}, {Lafever}, {Lahav}, {Lambert}, {Lampton}, {Landriau}, {Lang},
  {Lauer}, {Le Goff}, {Le Guillou}, {Le Van Suu}, {Lee}, {Lee}, {Leitner},
  {Lesser}, {Levi}, {L'Huillier}, {Li}, {Liang}, {Lin}, {Linder}, {Loebman},
  {Luki{\'c}}, {Ma}, {MacCrann}, {Magneville}, {Makarem}, {Manera}, {Manser},
  {Marshall}, {Martini}, {Massey}, {Matheson}, {McCauley}, {McDonald},
  {McGreer}, {Meisner}, {Metcalfe}, {Miller}, {Miquel}, {Moustakas}, {Myers},
  {Naik}, {Newman}, {Nichol}, {Nicola}, {Nicolati da Costa}, {Nie}, {Niz},
  {Norberg}, {Nord}, {Norman}, {Nugent}, {O'Brien}, {Oh}, {Olsen}, {Padilla},
  {Padmanabhan}, {Padmanabhan}, {Palanque-Delabrouille}, {Palmese},
  {Pappalardo}, {P{\^a}ris}, {Park}, {Patej}, {Peacock}, {Peiris}, {Peng},
  {Percival}, {Perruchot}, {Pieri}, {Pogge}, {Pollack}, {Poppett}, {Prada},
  {Prakash}, {Probst}, {Rabinowitz}, {Raichoor}, {Ree}, {Refregier}, {Regal},
  {Reid}, {Reil}, {Rezaie}, {Rockosi}, {Roe}, {Ronayette}, {Roodman}, {Ross},
  {Ross}, {Rossi}, {Rozo}, {Ruhlmann-Kleider}, {Rykoff}, {Sabiu}, {Samushia},
  {Sanchez}, {Sanchez}, {Schlegel}, {Schneider}, {Schubnell}, {Secroun},
  {Seljak}, {Seo}, {Serrano}, {Shafieloo}, {Shan}, {Sharples}, {Sholl},
  {Shourt}, {Silber}, {Silva}, {Sirk}, {Slosar}, {Smith}, {Smoot}, {Som},
  {Song}, {Sprayberry}, {Staten}, {Stefanik}, {Tarle}, {Sien Tie}, {Tinker},
  {Tojeiro}, {Valdes}, {Valenzuela}, {Valluri}, {Vargas-Magana}, {Verde},
  {Walker}, {Wang}, {Wang}, {Weaver}, {Weaverdyck}, {Wechsler}, {Weinberg},
  {White}, {Yang}, {Yeche}, {Zhang}, {Zhao}, {Zheng}, {Zhou}, {Zhou}, {Zhu},
  {Zou}, \& {Zu}}]{DESI_Science_FDR}
---. 2016{\natexlab{b}}, arXiv e-prints, arXiv:1611.00036.
\newblock \doarXiv{1611.00036}

\bibitem[{{DESI Collaboration} {et~al.}(2022){DESI Collaboration}, {Abareshi},
  {Aguilar}, {Ahlen}, {Alam}, {Alexander}, {Alfarsy}, {Allen}, {Allende
  Prieto}, {Alves}, {Ameel}, {Armengaud}, {Asorey}, {Aviles}, {Bailey},
  {Balaguera-Antol{\'\i}nez}, {Ballester}, {Baltay}, {Bault}, {Beltran},
  {Benavides}, {BenZvi}, {Berti}, {Besuner}, {Beutler}, {Bianchi}, {Blake},
  {Blanc}, {Blum}, {Bolton}, {Bose}, {Bramall}, {Brieden}, {Brodzeller},
  {Brooks}, {Brownewell}, {Buckley-Geer}, {Cahn}, {Cai}, {Canning}, {Carnero
  Rosell}, {Carton}, {Casas}, {Castander}, {Cervantes-Cota}, {Chabanier},
  {Chaussidon}, {Chuang}, {Circosta}, {Cole}, {Cooper}, {da Costa}, {Cousinou},
  {Cuceu}, {Davis}, {Dawson}, {de la Cruz-Noriega}, {de la Macorra}, {de
  Mattia}, {Della Costa}, {Demmer}, {Derwent}, {Dey}, {Dey}, {Dhungana},
  {Ding}, {Dobson}, {Doel}, {Donald-McCann}, {Donaldson}, {Douglass}, {Duan},
  {Dunlop}, {Edelstein}, {Eftekharzadeh}, {Eisenstein}, {Enriquez-Vargas},
  {Escoffier}, {Evatt}, {Fagrelius}, {Fan}, {Fanning}, {Fawcett}, {Ferraro},
  {Ereza}, {Flaugher}, {Font-Ribera}, {Forero-Romero}, {Frenk}, {Fromenteau},
  {G{\"a}nsicke}, {Garcia-Quintero}, {Garrison}, {Gazta{\~n}aga}, {Gerardi},
  {Gil-Mar{\'\i}n}, {Gontcho}, {Gonzalez-Morales}, {Gonzalez-de-Rivera},
  {Gonzalez-Perez}, {Gordon}, {Graur}, {Green}, {Grove}, {Gruen}, {Gutierrez},
  {Guy}, {Hahn}, {Harris}, {Herrera}, {Herrera-Alcantar}, {Honscheid},
  {Howlett}, {Huterer}, {Ir{\v{s}}i{\v{c}}}, {Ishak}, {Jelinsky}, {Jiang},
  {Jimenez}, {Jing}, {Joyce}, {Jullo}, {Juneau}, {Kara{\c{c}}ayl{\i}},
  {Karamanis}, {Karcher}, {Karim}, {Kehoe}, {Kent}, {Kirkby}, {Kisner},
  {Kitaura}, {Koposov}, {Kov{\'a}cs}, {Kremin}, {Krolewski}, {L'Huillier},
  {Lahav}, {Lambert}, {Lamman}, {Lan}, {Landriau}, {Lane}, {Lang}, {Lange},
  {Lasker}, {Le Guillou}, {Leauthaud}, {Le Van Suu}, {Levi}, {Li},
  {Magneville}, {Manera}, {Manser}, {Marshall}, {McCollam}, {McDonald},
  {Meisner}, {Mezcua}, {Miller}, {Miquel}, {Montero-Camacho}, {Moon},
  {Martini}, {Meneses-Rizo}, {Moustakas}, {Mueller}, {Mu{\~n}oz-Guti{\'e}rrez},
  {Myers}, {Nadathur}, {Najita}, {Napolitano}, {Neilsen}, {Newman}, {Nie},
  {Ning}, {Niz}, {Norberg}, {Noriega}, {O'Brien}, {Obuljen},
  {Palanque-Delabrouille}, {Palmese}, {Zhiwei}, {Pappalardo}, {Peng},
  {Percival}, {Perruchot}, {Pogge}, {Poppett}, {Porredon}, {Prada},
  {Prochaska}, {Pucha}, {P{\'e}rez-Fern{\'a}ndez}, {P{\'e}rez-R{\'a}fols},
  {Rabinowitz}, {Raichoor}, {Ramirez-Solano}, {Ram{\'\i}rez-P{\'e}rez},
  {Ravoux}, {Reil}, {Rezaie}, {Rocher}, {Rockosi}, {Roe}, {Roodman}, {Ross},
  {Rossi}, {Ruggeri}, {Ruhlmann-Kleider}, {Sabiu}, {Safonova}, {Said},
  {Saintonge}, {Salas Catonga}, {Samushia}, {Sanchez}, {Saulder}, {Schaan},
  {Schlafly}, {Schlegel}, {Schmoll}, {Scholte}, {Schubnell}, {Secroun}, {Seo},
  {Serrano}, {Sharples}, {Sholl}, {Silber}, {Silva}, {Sirk}, {Siudek}, {Smith},
  {Sprayberry}, {Staten}, {Stupak}, {Tan}, {Tarl{\'e}}, {Sien Tie}, {Tojeiro},
  {Ure{\~n}a-L{\'o}pez}, {Valdes}, {Valenzuela}, {Valluri},
  {Vargas-Maga{\~n}a}, {Verde}, {Walther}, {Wang}, {Wang}, {Weaver},
  {Weaverdyck}, {Wechsler}, {Wilson}, {Yang}, {Yu}, {Yuan}, {Y{\`e}che},
  {Zhang}, {Zhang}, {Zhao}, {Zhou}, {Zhou}, {Zou}, {Zou}, {Zou}, \&
  {Zu}}]{DESI_Instr_Overview2022}
{DESI Collaboration}, {Abareshi}, B., {Aguilar}, J., {et~al.} 2022, arXiv
  e-prints, arXiv:2205.10939.
\newblock \doarXiv{2205.10939}

\bibitem[{{Dey} {et~al.}(2019){Dey}, {Schlegel}, {Lang}, {Blum}, {Burleigh},
  {Fan}, {Findlay}, {Finkbeiner}, {Herrera}, {Juneau}, {Landriau}, {Levi},
  {McGreer}, {Meisner}, {Myers}, {Moustakas}, {Nugent}, {Patej}, {Schlafly},
  {Walker}, {Valdes}, {Weaver}, {Y{\`e}che}, {Zou}, {Zhou}, {Abareshi},
  {Abbott}, {Abolfathi}, {Aguilera}, {Alam}, {Allen}, {Alvarez}, {Annis},
  {Ansarinejad}, {Aubert}, {Beechert}, {Bell}, {BenZvi}, {Beutler}, {Bielby},
  {Bolton}, {Brice{\~n}o}, {Buckley-Geer}, {Butler}, {Calamida}, {Carlberg},
  {Carter}, {Casas}, {Castander}, {Choi}, {Comparat}, {Cukanovaite}, {Delubac},
  {DeVries}, {Dey}, {Dhungana}, {Dickinson}, {Ding}, {Donaldson}, {Duan},
  {Duckworth}, {Eftekharzadeh}, {Eisenstein}, {Etourneau}, {Fagrelius},
  {Farihi}, {Fitzpatrick}, {Font-Ribera}, {Fulmer}, {G{\"a}nsicke},
  {Gaztanaga}, {George}, {Gerdes}, {Gontcho}, {Gorgoni}, {Green}, {Guy},
  {Harmer}, {Hernandez}, {Honscheid}, {Huang}, {James}, {Jannuzi}, {Jiang},
  {Joyce}, {Karcher}, {Karkar}, {Kehoe}, {Kneib}, {Kueter-Young}, {Lan},
  {Lauer}, {Le Guillou}, {Le Van Suu}, {Lee}, {Lesser}, {Perreault Levasseur},
  {Li}, {Mann}, {Marshall}, {Mart{\'\i}nez-V{\'a}zquez}, {Martini}, {du Mas des
  Bourboux}, {McManus}, {Meier}, {M{\'e}nard}, {Metcalfe},
  {Mu{\~n}oz-Guti{\'e}rrez}, {Najita}, {Napier}, {Narayan}, {Newman}, {Nie},
  {Nord}, {Norman}, {Olsen}, {Paat}, {Palanque-Delabrouille}, {Peng},
  {Poppett}, {Poremba}, {Prakash}, {Rabinowitz}, {Raichoor}, {Rezaie},
  {Robertson}, {Roe}, {Ross}, {Ross}, {Rudnick}, {Safonova}, {Saha},
  {S{\'a}nchez}, {Savary}, {Schweiker}, {Scott}, {Seo}, {Shan}, {Silva},
  {Slepian}, {Soto}, {Sprayberry}, {Staten}, {Stillman}, {Stupak}, {Summers},
  {Sien Tie}, {Tirado}, {Vargas-Maga{\~n}a}, {Vivas}, {Wechsler}, {Williams},
  {Yang}, {Yang}, {Yapici}, {Zaritsky}, {Zenteno}, {Zhang}, {Zhang}, {Zhou}, \&
  {Zhou}}]{Dey2019}
{Dey}, A., {Schlegel}, D.~J., {Lang}, D., {et~al.} 2019, \aj, 157, 168,
  \dodoi{10.3847/1538-3881/ab089d}

\bibitem[{{Dong-P{\'a}ez} {et~al.}(2022){Dong-P{\'a}ez}, {Vasiliev}, \&
  {Evans}}]{DongPaez2022}
{Dong-P{\'a}ez}, C.~A., {Vasiliev}, E., \& {Evans}, N.~W. 2022, \mnras, 510,
  230, \dodoi{10.1093/mnras/stab3361}

\bibitem[{{Dorman} {et~al.}(2012){Dorman}, {Guhathakurta}, {Fardal}, {Lang},
  {Geha}, {Howley}, {Kalirai}, {Bullock}, {Cuillandre}, {Dalcanton}, {Gilbert},
  {Seth}, {Tollerud}, {Williams}, \& {Yniguez}}]{Dorman2012}
{Dorman}, C.~E., {Guhathakurta}, P., {Fardal}, M.~A., {et~al.} 2012, \apj, 752,
  147, \dodoi{10.1088/0004-637X/752/2/147}

\bibitem[{{Dorman} {et~al.}(2015){Dorman}, {Guhathakurta}, {Seth}, {Weisz},
  {Bell}, {Dalcanton}, {Gilbert}, {Hamren}, {Lewis}, {Skillman}, {Toloba}, \&
  {Williams}}]{Dorman2015}
{Dorman}, C.~E., {Guhathakurta}, P., {Seth}, A.~C., {et~al.} 2015, \apj, 803,
  24, \dodoi{10.1088/0004-637X/803/1/24}

\bibitem[{{D'Souza} \& {Bell}(2018)}]{dsouza2018}
{D'Souza}, R., \& {Bell}, E.~F. 2018, Nature Astronomy, 2, 737,
  \dodoi{10.1038/s41550-018-0533-x}

\bibitem[{{D'Souza} \& {Bell}(2021)}]{dsouza2021}
---. 2021, \mnras, 504, 5270, \dodoi{10.1093/mnras/stab1283}

\bibitem[{{Escala} {et~al.}(2022){Escala}, {Gilbert}, {Fardal}, {Guhathakurta},
  {Sanderson}, {Kalirai}, \& {Mobasher}}]{Escala2022}
{Escala}, I., {Gilbert}, K.~M., {Fardal}, M., {et~al.} 2022, \aj, 164, 20,
  \dodoi{10.3847/1538-3881/ac7146}

\bibitem[{{Escala} {et~al.}(2020{\natexlab{a}}){Escala}, {Gilbert}, {Kirby},
  {Wojno}, {Cunningham}, \& {Guhathakurta}}]{Escala2020a}
{Escala}, I., {Gilbert}, K.~M., {Kirby}, E.~N., {et~al.} 2020{\natexlab{a}},
  \apj, 889, 177, \dodoi{10.3847/1538-4357/ab6659}

\bibitem[{{Escala} {et~al.}(2021){Escala}, {Gilbert}, {Wojno}, {Kirby}, \&
  {Guhathakurta}}]{Escala2021}
{Escala}, I., {Gilbert}, K.~M., {Wojno}, J., {Kirby}, E.~N., \& {Guhathakurta},
  P. 2021, \aj, 162, 45, \dodoi{10.3847/1538-3881/abfec4}

\bibitem[{{Escala} {et~al.}(2019){Escala}, {Kirby}, {Gilbert}, {Cunningham}, \&
  {Wojno}}]{Escala2019}
{Escala}, I., {Kirby}, E.~N., {Gilbert}, K.~M., {Cunningham}, E.~C., \&
  {Wojno}, J. 2019, \apj, 878, 42, \dodoi{10.3847/1538-4357/ab1eac}

\bibitem[{{Escala} {et~al.}(2020{\natexlab{b}}){Escala}, {Kirby}, {Gilbert},
  {Wojno}, {Cunningham}, \& {Guhathakurta}}]{Escala2020b}
{Escala}, I., {Kirby}, E.~N., {Gilbert}, K.~M., {et~al.} 2020{\natexlab{b}},
  \apj, 902, 51, \dodoi{10.3847/1538-4357/abb474}

\bibitem[{{Faber} {et~al.}(2003){Faber}, {Phillips}, {Kibrick}, {Alcott},
  {Allen}, {Burrous}, {Cantrall}, {Clarke}, {Coil}, {Cowley}, {Davis}, {Deich},
  {Dietsch}, {Gilmore}, {Harper}, {Hilyard}, {Lewis}, {McVeigh}, {Newman},
  {Osborne}, {Schiavon}, {Stover}, {Tucker}, {Wallace}, {Wei}, {Wirth}, \&
  {Wright}}]{DEIMOS}
{Faber}, S.~M., {Phillips}, A.~C., {Kibrick}, R.~I., {et~al.} 2003, in Society
  of Photo-Optical Instrumentation Engineers (SPIE) Conference Series, Vol.
  4841, Instrument Design and Performance for Optical/Infrared Ground-based
  Telescopes, ed. M.~{Iye} \& A.~F.~M. {Moorwood}, 1657--1669,
  \dodoi{10.1117/12.460346}

\bibitem[{{Fardal} {et~al.}(2006){Fardal}, {Babul}, {Geehan}, \&
  {Guhathakurta}}]{Fardal2006}
{Fardal}, M.~A., {Babul}, A., {Geehan}, J.~J., \& {Guhathakurta}, P. 2006,
  \mnras, 366, 1012, \dodoi{10.1111/j.1365-2966.2005.09864.x}

\bibitem[{{Fardal} {et~al.}(2008){Fardal}, {Babul}, {Guhathakurta}, {Gilbert},
  \& {Dodge}}]{Fardal2008}
{Fardal}, M.~A., {Babul}, A., {Guhathakurta}, P., {Gilbert}, K.~M., \& {Dodge},
  C. 2008, \apjl, 682, L33, \dodoi{10.1086/590386}

\bibitem[{{Fardal} {et~al.}(2007){Fardal}, {Guhathakurta}, {Babul}, \&
  {McConnachie}}]{Fardal2007}
{Fardal}, M.~A., {Guhathakurta}, P., {Babul}, A., \& {McConnachie}, A.~W. 2007,
  \mnras, 380, 15, \dodoi{10.1111/j.1365-2966.2007.11929.x}

\bibitem[{{Fardal} {et~al.}(2012){Fardal}, {Guhathakurta}, {Gilbert},
  {Tollerud}, {Kalirai}, {Tanaka}, {Beaton}, {Chiba}, {Komiyama}, \&
  {Iye}}]{Fardal2012}
{Fardal}, M.~A., {Guhathakurta}, P., {Gilbert}, K.~M., {et~al.} 2012, \mnras,
  423, 3134, \dodoi{10.1111/j.1365-2966.2012.21094.x}

\bibitem[{{Fardal} {et~al.}(2013){Fardal}, {Weinberg}, {Babul}, {Irwin},
  {Guhathakurta}, {Gilbert}, {Ferguson}, {Ibata}, {Lewis}, {Tanvir}, \&
  {Huxor}}]{Fardal2013}
{Fardal}, M.~A., {Weinberg}, M.~D., {Babul}, A., {et~al.} 2013, \mnras, 434,
  2779, \dodoi{10.1093/mnras/stt1121}

\bibitem[{{Ferguson} \& {Mackey}(2016)}]{Ferguson2016}
{Ferguson}, A. M.~N., \& {Mackey}, A.~D. 2016, in Astrophysics and Space
  Science Library, Vol. 420, Tidal Streams in the Local Group and Beyond, ed.
  H.~J. {Newberg} \& J.~L. {Carlin}, 191, \dodoi{10.1007/978-3-319-19336-6\_8}

\bibitem[{{Font} {et~al.}(2006){Font}, {Johnston}, {Guhathakurta}, {Majewski},
  \& {Rich}}]{font2006}
{Font}, A.~S., {Johnston}, K.~V., {Guhathakurta}, P., {Majewski}, S.~R., \&
  {Rich}, R.~M. 2006, \aj, 131, 1436, \dodoi{10.1086/499564}

\bibitem[{{Gaia Collaboration} {et~al.}(2016{\natexlab{a}}){Gaia
  Collaboration}, {Prusti}, {de Bruijne}, {Brown}, {Vallenari}, {Babusiaux},
  {Bailer-Jones}, {Bastian}, {Biermann}, {Evans}, \& et~al.}]{gaia}
{Gaia Collaboration}, {Prusti}, T., {de Bruijne}, J.~H.~J., {et~al.}
  2016{\natexlab{a}}, \aap, 595, A1, \dodoi{10.1051/0004-6361/201629272}

\bibitem[{{Gaia Collaboration} {et~al.}(2016{\natexlab{b}}){Gaia
  Collaboration}, {Prusti}, {de Bruijne}, {Brown}, {Vallenari}, {Babusiaux},
  {Bailer-Jones}, {Bastian}, {Biermann}, {Evans}, {Eyer}, {Jansen}, {Jordi},
  {Klioner}, {Lammers}, {Lindegren}, {Luri}, {Mignard}, {Milligan}, {Panem},
  {Poinsignon}, {Pourbaix}, {Randich}, {Sarri}, {Sartoretti}, {Siddiqui},
  {Soubiran}, {Valette}, {van Leeuwen}, {Walton}, {Aerts}, {Arenou}, {Cropper},
  {Drimmel}, {H{\o}g}, {Katz}, {Lattanzi}, {O'Mullane}, {Grebel}, {Holland},
  {Huc}, {Passot}, {Bramante}, {Cacciari}, {Casta{\~n}eda}, {Chaoul}, {Cheek},
  {De Angeli}, {Fabricius}, {Guerra}, {Hern{\'a}ndez}, {Jean-Antoine-Piccolo},
  {Masana}, {Messineo}, {Mowlavi}, {Nienartowicz}, {Ord{\'o}{\~n}ez-Blanco},
  {Panuzzo}, {Portell}, {Richards}, {Riello}, {Seabroke}, {Tanga},
  {Th{\'e}venin}, {Torra}, {Els}, {Gracia-Abril}, {Comoretto},
  {Garcia-Reinaldos}, {Lock}, {Mercier}, {Altmann}, {Andrae}, {Astraatmadja},
  {Bellas-Velidis}, {Benson}, {Berthier}, {Blomme}, {Busso}, {Carry},
  {Cellino}, {Clementini}, {Cowell}, {Creevey}, {Cuypers}, {Davidson}, {De
  Ridder}, {de Torres}, {Delchambre}, {Dell'Oro}, {Ducourant}, {Fr{\'e}mat},
  {Garc{\'\i}a-Torres}, {Gosset}, {Halbwachs}, {Hambly}, {Harrison}, {Hauser},
  {Hestroffer}, {Hodgkin}, {Huckle}, {Hutton}, {Jasniewicz}, {Jordan},
  {Kontizas}, {Korn}, {Lanzafame}, {Manteiga}, {Moitinho}, {Muinonen},
  {Osinde}, {Pancino}, {Pauwels}, {Petit}, {Recio-Blanco}, {Robin}, {Sarro},
  {Siopis}, {Smith}, {Smith}, {Sozzetti}, {Thuillot}, {van Reeven}, {Viala},
  {Abbas}, {Abreu Aramburu}, {Accart}, {Aguado}, {Allan}, {Allasia},
  {Altavilla}, {{\'A}lvarez}, {Alves}, {Anderson}, {Andrei}, {Anglada Varela},
  {Antiche}, {Antoja}, {Ant{\'o}n}, {Arcay}, {Atzei}, {Ayache}, {Bach},
  {Baker}, {Balaguer-N{\'u}{\~n}ez}, {Barache}, {Barata}, {Barbier}, {Barblan},
  {Baroni}, {Barrado y Navascu{\'e}s}, {Barros}, {Barstow}, {Becciani},
  {Bellazzini}, {Bellei}, {Bello Garc{\'\i}a}, {Belokurov}, {Bendjoya},
  {Berihuete}, {Bianchi}, {Bienaym{\'e}}, {Billebaud}, {Blagorodnova},
  {Blanco-Cuaresma}, {Boch}, {Bombrun}, {Borrachero}, {Bouquillon}, {Bourda},
  {Bouy}, {Bragaglia}, {Breddels}, {Brouillet}, {Br{\"u}semeister},
  {Bucciarelli}, {Budnik}, {Burgess}, {Burgon}, {Burlacu}, {Busonero}, {Buzzi},
  {Caffau}, {Cambras}, {Campbell}, {Cancelliere}, {Cantat-Gaudin}, {Carlucci},
  {Carrasco}, {Castellani}, {Charlot}, {Charnas}, {Charvet}, {Chassat},
  {Chiavassa}, {Clotet}, {Cocozza}, {Collins}, {Collins}, {Costigan}, {Crifo},
  {Cross}, {Crosta}, {Crowley}, {Dafonte}, {Damerdji}, {Dapergolas}, {David},
  {David}, {De Cat}, {de Felice}, {de Laverny}, {De Luise}, {De March}, {de
  Martino}, {de Souza}, {Debosscher}, {del Pozo}, {Delbo}, {Delgado},
  {Delgado}, {di Marco}, {Di Matteo}, {Diakite}, {Distefano}, {Dolding}, {Dos
  Anjos}, {Drazinos}, {Dur{\'a}n}, {Dzigan}, {Ecale}, {Edvardsson}, {Enke},
  {Erdmann}, {Escolar}, {Espina}, {Evans}, {Eynard Bontemps}, {Fabre},
  {Fabrizio}, {Faigler}, {Falc{\~a}o}, {Farr{\`a}s Casas}, {Faye}, {Federici},
  {Fedorets}, {Fern{\'a}ndez-Hern{\'a}ndez}, {Fernique}, {Fienga}, {Figueras},
  {Filippi}, {Findeisen}, {Fonti}, {Fouesneau}, {Fraile}, {Fraser}, {Fuchs},
  {Furnell}, {Gai}, {Galleti}, {Galluccio}, {Garabato}, {Garc{\'\i}a-Sedano},
  {Gar{\'e}}, {Garofalo}, {Garralda}, {Gavras}, {Gerssen}, {Geyer}, {Gilmore},
  {Girona}, {Giuffrida}, {Gomes}, {Gonz{\'a}lez-Marcos},
  {Gonz{\'a}lez-N{\'u}{\~n}ez}, {Gonz{\'a}lez-Vidal}, {Granvik}, {Guerrier},
  {Guillout}, {Guiraud}, {G{\'u}rpide}, {Guti{\'e}rrez-S{\'a}nchez}, {Guy},
  {Haigron}, {Hatzidimitriou}, {Haywood}, {Heiter}, {Helmi}, {Hobbs},
  {Hofmann}, {Holl}, {Holland}, {Hunt}, {Hypki}, {Icardi}, {Irwin}, {Jevardat
  de Fombelle}, {Jofr{\'e}}, {Jonker}, {Jorissen}, {Julbe}, {Karampelas},
  {Kochoska}, {Kohley}, {Kolenberg}, {Kontizas}, {Koposov}, {Kordopatis},
  {Koubsky}, {Kowalczyk}, {Krone-Martins}, {Kudryashova}, {Kull}, {Bachchan},
  {Lacoste-Seris}, {Lanza}, {Lavigne}, {Le Poncin-Lafitte}, {Lebreton},
  {Lebzelter}, {Leccia}, {Leclerc}, {Lecoeur-Taibi}, {Lemaitre}, {Lenhardt},
  {Leroux}, {Liao}, {Licata}, {Lindstr{\o}m}, {Lister}, {Livanou}, {Lobel},
  {L{\"o}ffler}, {L{\'o}pez}, {Lopez-Lozano}, {Lorenz}, {Loureiro},
  {MacDonald}, {Magalh{\~a}es Fernandes}, {Managau}, {Mann}, {Mantelet},
  {Marchal}, {Marchant}, {Marconi}, {Marie}, {Marinoni}, {Marrese},
  {Marschalk{\'o}}, {Marshall}, {Mart{\'\i}n-Fleitas}, {Martino}, {Mary},
  {Matijevi{\v{c}}}, {Mazeh}, {McMillan}, {Messina}, {Mestre}, {Michalik},
  {Millar}, {Miranda}, {Molina}, {Molinaro}, {Molinaro}, {Moln{\'a}r},
  {Moniez}, {Montegriffo}, {Monteiro}, {Mor}, {Mora}, {Morbidelli}, {Morel},
  {Morgenthaler}, {Morley}, {Morris}, {Mulone}, {Muraveva}, {Musella},
  {Narbonne}, {Nelemans}, {Nicastro}, {Noval}, {Ord{\'e}novic},
  {Ordieres-Mer{\'e}}, {Osborne}, {Pagani}, {Pagano}, {Pailler}, {Palacin},
  {Palaversa}, {Parsons}, {Paulsen}, {Pecoraro}, {Pedrosa}, {Pentik{\"a}inen},
  {Pereira}, {Pichon}, {Piersimoni}, {Pineau}, {Plachy}, {Plum}, {Poujoulet},
  {Pr{\v{s}}a}, {Pulone}, {Ragaini}, {Rago}, {Rambaux}, {Ramos-Lerate},
  {Ranalli}, {Rauw}, {Read}, {Regibo}, {Renk}, {Reyl{\'e}}, {Ribeiro},
  {Rimoldini}, {Ripepi}, {Riva}, {Rixon}, {Roelens}, {Romero-G{\'o}mez},
  {Rowell}, {Royer}, {Rudolph}, {Ruiz-Dern}, {Sadowski}, {Sagrist{\`a}
  Sell{\'e}s}, {Sahlmann}, {Salgado}, {Salguero}, {Sarasso}, {Savietto},
  {Schnorhk}, {Schultheis}, {Sciacca}, {Segol}, {Segovia}, {Segransan},
  {Serpell}, {Shih}, {Smareglia}, {Smart}, {Smith}, {Solano}, {Solitro},
  {Sordo}, {Soria Nieto}, {Souchay}, {Spagna}, {Spoto}, {Stampa}, {Steele},
  {Steidelm{\"u}ller}, {Stephenson}, {Stoev}, {Suess}, {S{\"u}veges}, {Surdej},
  {Szabados}, {Szegedi-Elek}, {Tapiador}, {Taris}, {Tauran}, {Taylor},
  {Teixeira}, {Terrett}, {Tingley}, {Trager}, {Turon}, {Ulla}, {Utrilla},
  {Valentini}, {van Elteren}, {Van Hemelryck}, {van Leeuwen}, {Varadi},
  {Vecchiato}, {Veljanoski}, {Via}, {Vicente}, {Vogt}, {Voss}, {Votruba},
  {Voutsinas}, {Walmsley}, {Weiler}, {Weingrill}, {Werner}, {Wevers},
  {Whitehead}, {Wyrzykowski}, {Yoldas}, {{\v{Z}}erjal}, {Zucker}, {Zurbach},
  {Zwitter}, {Alecu}, {Allen}, {Allende Prieto}, {Amorim},
  {Anglada-Escud{\'e}}, {Arsenijevic}, {Azaz}, {Balm}, {Beck}, {Bernstein},
  {Bigot}, {Bijaoui}, {Blasco}, {Bonfigli}, {Bono}, {Boudreault}, {Bressan},
  {Brown}, {Brunet}, {Bunclark}, {Buonanno}, {Butkevich}, {Carret}, {Carrion},
  {Chemin}, {Ch{\'e}reau}, {Corcione}, {Darmigny}, {de Boer}, {de Teodoro}, {de
  Zeeuw}, {Delle Luche}, {Domingues}, {Dubath}, {Fodor}, {Fr{\'e}zouls},
  {Fries}, {Fustes}, {Fyfe}, {Gallardo}, {Gallegos}, {Gardiol}, {Gebran},
  {Gomboc}, {G{\'o}mez}, {Grux}, {Gueguen}, {Heyrovsky}, {Hoar}, {Iannicola},
  {Isasi Parache}, {Janotto}, {Joliet}, {Jonckheere}, {Keil}, {Kim},
  {Klagyivik}, {Klar}, {Knude}, {Kochukhov}, {Kolka}, {Kos}, {Kutka}, {Lainey},
  {LeBouquin}, {Liu}, {Loreggia}, {Makarov}, {Marseille}, {Martayan},
  {Martinez-Rubi}, {Massart}, {Meynadier}, {Mignot}, {Munari}, {Nguyen},
  {Nordlander}, {Ocvirk}, {O'Flaherty}, {Olias Sanz}, {Ortiz}, {Osorio},
  {Oszkiewicz}, {Ouzounis}, {Palmer}, {Park}, {Pasquato}, {Peltzer}, {Peralta},
  {P{\'e}turaud}, {Pieniluoma}, {Pigozzi}, {Poels}, {Prat}, {Prod'homme},
  {Raison}, {Rebordao}, {Risquez}, {Rocca-Volmerange}, {Rosen}, {Ruiz-Fuertes},
  {Russo}, {Sembay}, {Serraller Vizcaino}, {Short}, {Siebert}, {Silva},
  {Sinachopoulos}, {Slezak}, {Soffel}, {Sosnowska}, {Strai{\v{z}}ys}, {ter
  Linden}, {Terrell}, {Theil}, {Tiede}, {Troisi}, {Tsalmantza}, {Tur},
  {Vaccari}, {Vachier}, {Valles}, {Van Hamme}, {Veltz}, {Virtanen}, {Wallut},
  {Wichmann}, {Wilkinson}, {Ziaeepour}, \& {Zschocke}}]{GaiaMission2016}
---. 2016{\natexlab{b}}, \aap, 595, A1, \dodoi{10.1051/0004-6361/201629272}

\bibitem[{{Gaia Collaboration} {et~al.}(2018){Gaia Collaboration}, {Brown},
  {Vallenari}, {Prusti}, {de Bruijne}, {Babusiaux}, {Bailer-Jones}, {Biermann},
  {Evans}, {Eyer}, {Jansen}, {Jordi}, {Klioner}, {Lammers}, {Lindegren},
  {Luri}, {Mignard}, {Panem}, {Pourbaix}, {Randich}, {Sartoretti}, {Siddiqui},
  {Soubiran}, {van Leeuwen}, {Walton}, {Arenou}, {Bastian}, {Cropper},
  {Drimmel}, {Katz}, {Lattanzi}, {Bakker}, {Cacciari}, {Casta{\~n}eda},
  {Chaoul}, {Cheek}, {De Angeli}, {Fabricius}, {Guerra}, {Holl}, {Masana},
  {Messineo}, {Mowlavi}, {Nienartowicz}, {Panuzzo}, {Portell}, {Riello},
  {Seabroke}, {Tanga}, {Th{\'e}venin}, {Gracia-Abril}, {Comoretto},
  {Garcia-Reinaldos}, {Teyssier}, {Altmann}, {Andrae}, {Audard},
  {Bellas-Velidis}, {Benson}, {Berthier}, {Blomme}, {Burgess}, {Busso},
  {Carry}, {Cellino}, {Clementini}, {Clotet}, {Creevey}, {Davidson}, {De
  Ridder}, {Delchambre}, {Dell'Oro}, {Ducourant},
  {Fern{\'a}ndez-Hern{\'a}ndez}, {Fouesneau}, {Fr{\'e}mat}, {Galluccio},
  {Garc{\'\i}a-Torres}, {Gonz{\'a}lez-N{\'u}{\~n}ez}, {Gonz{\'a}lez-Vidal},
  {Gosset}, {Guy}, {Halbwachs}, {Hambly}, {Harrison}, {Hern{\'a}ndez},
  {Hestroffer}, {Hodgkin}, {Hutton}, {Jasniewicz}, {Jean-Antoine-Piccolo},
  {Jordan}, {Korn}, {Krone-Martins}, {Lanzafame}, {Lebzelter}, {L{\"o}ffler},
  {Manteiga}, {Marrese}, {Mart{\'\i}n-Fleitas}, {Moitinho}, {Mora}, {Muinonen},
  {Osinde}, {Pancino}, {Pauwels}, {Petit}, {Recio-Blanco}, {Richards},
  {Rimoldini}, {Robin}, {Sarro}, {Siopis}, {Smith}, {Sozzetti}, {S{\"u}veges},
  {Torra}, {van Reeven}, {Abbas}, {Abreu Aramburu}, {Accart}, {Aerts},
  {Altavilla}, {{\'A}lvarez}, {Alvarez}, {Alves}, {Anderson}, {Andrei},
  {Anglada Varela}, {Antiche}, {Antoja}, {Arcay}, {Astraatmadja}, {Bach},
  {Baker}, {Balaguer-N{\'u}{\~n}ez}, {Balm}, {Barache}, {Barata}, {Barbato},
  {Barblan}, {Barklem}, {Barrado}, {Barros}, {Barstow}, {Bartholom{\'e}
  Mu{\~n}oz}, {Bassilana}, {Becciani}, {Bellazzini}, {Berihuete}, {Bertone},
  {Bianchi}, {Bienaym{\'e}}, {Blanco-Cuaresma}, {Boch}, {Boeche}, {Bombrun},
  {Borrachero}, {Bossini}, {Bouquillon}, {Bourda}, {Bragaglia}, {Bramante},
  {Breddels}, {Bressan}, {Brouillet}, {Br{\"u}semeister}, {Brugaletta},
  {Bucciarelli}, {Burlacu}, {Busonero}, {Butkevich}, {Buzzi}, {Caffau},
  {Cancelliere}, {Cannizzaro}, {Cantat-Gaudin}, {Carballo}, {Carlucci},
  {Carrasco}, {Casamiquela}, {Castellani}, {Castro-Ginard}, {Charlot},
  {Chemin}, {Chiavassa}, {Cocozza}, {Costigan}, {Cowell}, {Crifo}, {Crosta},
  {Crowley}, {Cuypers}, {Dafonte}, {Damerdji}, {Dapergolas}, {David}, {David},
  {de Laverny}, {De Luise}, {De March}, {de Martino}, {de Souza}, {de Torres},
  {Debosscher}, {del Pozo}, {Delbo}, {Delgado}, {Delgado}, {Di Matteo},
  {Diakite}, {Diener}, {Distefano}, {Dolding}, {Drazinos}, {Dur{\'a}n},
  {Edvardsson}, {Enke}, {Eriksson}, {Esquej}, {Eynard Bontemps}, {Fabre},
  {Fabrizio}, {Faigler}, {Falc{\~a}o}, {Farr{\`a}s Casas}, {Federici},
  {Fedorets}, {Fernique}, {Figueras}, {Filippi}, {Findeisen}, {Fonti},
  {Fraile}, {Fraser}, {Fr{\'e}zouls}, {Gai}, {Galleti}, {Garabato},
  {Garc{\'\i}a-Sedano}, {Garofalo}, {Garralda}, {Gavel}, {Gavras}, {Gerssen},
  {Geyer}, {Giacobbe}, {Gilmore}, {Girona}, {Giuffrida}, {Glass}, {Gomes},
  {Granvik}, {Gueguen}, {Guerrier}, {Guiraud}, {Guti{\'e}rrez-S{\'a}nchez},
  {Haigron}, {Hatzidimitriou}, {Hauser}, {Haywood}, {Heiter}, {Helmi}, {Heu},
  {Hilger}, {Hobbs}, {Hofmann}, {Holland}, {Huckle}, {Hypki}, {Icardi},
  {Jan{\ss}en}, {Jevardat de Fombelle}, {Jonker}, {Juh{\'a}sz}, {Julbe},
  {Karampelas}, {Kewley}, {Klar}, {Kochoska}, {Kohley}, {Kolenberg},
  {Kontizas}, {Kontizas}, {Koposov}, {Kordopatis}, {Kostrzewa-Rutkowska},
  {Koubsky}, {Lambert}, {Lanza}, {Lasne}, {Lavigne}, {Le Fustec}, {Le
  Poncin-Lafitte}, {Lebreton}, {Leccia}, {Leclerc}, {Lecoeur-Taibi},
  {Lenhardt}, {Leroux}, {Liao}, {Licata}, {Lindstr{\o}m}, {Lister}, {Livanou},
  {Lobel}, {L{\'o}pez}, {Managau}, {Mann}, {Mantelet}, {Marchal}, {Marchant},
  {Marconi}, {Marinoni}, {Marschalk{\'o}}, {Marshall}, {Martino}, {Marton},
  {Mary}, {Massari}, {Matijevi{\v{c}}}, {Mazeh}, {McMillan}, {Messina},
  {Michalik}, {Millar}, {Molina}, {Molinaro}, {Moln{\'a}r}, {Montegriffo},
  {Mor}, {Morbidelli}, {Morel}, {Morris}, {Mulone}, {Muraveva}, {Musella},
  {Nelemans}, {Nicastro}, {Noval}, {O'Mullane}, {Ord{\'e}novic},
  {Ord{\'o}{\~n}ez-Blanco}, {Osborne}, {Pagani}, {Pagano}, {Pailler},
  {Palacin}, {Palaversa}, {Panahi}, {Pawlak}, {Piersimoni}, {Pineau}, {Plachy},
  {Plum}, {Poggio}, {Poujoulet}, {Pr{\v{s}}a}, {Pulone}, {Racero}, {Ragaini},
  {Rambaux}, {Ramos-Lerate}, {Regibo}, {Reyl{\'e}}, {Riclet}, {Ripepi}, {Riva},
  {Rivard}, {Rixon}, {Roegiers}, {Roelens}, {Romero-G{\'o}mez}, {Rowell},
  {Royer}, {Ruiz-Dern}, {Sadowski}, {Sagrist{\`a} Sell{\'e}s}, {Sahlmann},
  {Salgado}, {Salguero}, {Sanna}, {Santana-Ros}, {Sarasso}, {Savietto},
  {Schultheis}, {Sciacca}, {Segol}, {Segovia}, {S{\'e}gransan}, {Shih},
  {Siltala}, {Silva}, {Smart}, {Smith}, {Solano}, {Solitro}, {Sordo}, {Soria
  Nieto}, {Souchay}, {Spagna}, {Spoto}, {Stampa}, {Steele},
  {Steidelm{\"u}ller}, {Stephenson}, {Stoev}, {Suess}, {Surdej}, {Szabados},
  {Szegedi-Elek}, {Tapiador}, {Taris}, {Tauran}, {Taylor}, {Teixeira},
  {Terrett}, {Teyssandier}, {Thuillot}, {Titarenko}, {Torra Clotet}, {Turon},
  {Ulla}, {Utrilla}, {Uzzi}, {Vaillant}, {Valentini}, {Valette}, {van Elteren},
  {Van Hemelryck}, {van Leeuwen}, {Vaschetto}, {Vecchiato}, {Veljanoski},
  {Viala}, {Vicente}, {Vogt}, {von Essen}, {Voss}, {Votruba}, {Voutsinas},
  {Walmsley}, {Weiler}, {Wertz}, {Wevers}, {Wyrzykowski}, {Yoldas},
  {{\v{Z}}erjal}, {Ziaeepour}, {Zorec}, {Zschocke}, {Zucker}, {Zurbach}, \&
  {Zwitter}}]{GaiaDR2summary}
{Gaia Collaboration}, {Brown}, A.~G.~A., {Vallenari}, A., {et~al.} 2018, \aap,
  616, A1, \dodoi{10.1051/0004-6361/201833051}

\bibitem[{{Gallart} {et~al.}(2019){Gallart}, {Bernard}, {Brook}, {Ruiz-Lara},
  {Cassisi}, {Hill}, \& {Monelli}}]{Gallart2019}
{Gallart}, C., {Bernard}, E.~J., {Brook}, C.~B., {et~al.} 2019, Nature
  Astronomy, 3, 932, \dodoi{10.1038/s41550-019-0829-5}

\bibitem[{{Galleti} {et~al.}(2007){Galleti}, {Bellazzini}, {Federici},
  {Buzzoni}, \& {Fusi Pecci}}]{galleti2007}
{Galleti}, S., {Bellazzini}, M., {Federici}, L., {Buzzoni}, A., \& {Fusi
  Pecci}, F. 2007, \aap, 471, 127, \dodoi{10.1051/0004-6361:20077788}

\bibitem[{{Galleti} {et~al.}(2014){Galleti}, {Federici}, {Bellazzini}, {Fusi
  Pecci}, {Macrina}, \& {Buzzoni}}]{galleti2014cat}
{Galleti}, S., {Federici}, L., {Bellazzini}, M., {et~al.} 2014, VizieR Online
  Data Catalog, V/143

\bibitem[{{Genel} {et~al.}(2015){Genel}, {Fall}, {Hernquist}, {Vogelsberger},
  {Snyder}, {Rodriguez-Gomez}, {Sijacki}, \& {Springel}}]{genel2015}
{Genel}, S., {Fall}, S.~M., {Hernquist}, L., {et~al.} 2015, \apjl, 804, L40,
  \dodoi{10.1088/2041-8205/804/2/L40}

\bibitem[{{Gilbert} {et~al.}(2009{\natexlab{a}}){Gilbert}, {Font}, {Johnston},
  \& {Guhathakurta}}]{Gilbert2009a}
{Gilbert}, K.~M., {Font}, A.~S., {Johnston}, K.~V., \& {Guhathakurta}, P.
  2009{\natexlab{a}}, \apj, 701, 776, \dodoi{10.1088/0004-637X/701/1/776}

\bibitem[{{Gilbert} {et~al.}(2019){Gilbert}, {Kirby}, {Escala}, {Wojno},
  {Kalirai}, \& {Guhathakurta}}]{Gilbert2019}
{Gilbert}, K.~M., {Kirby}, E.~N., {Escala}, I., {et~al.} 2019, \apj, 883, 128,
  \dodoi{10.3847/1538-4357/ab3807}

\bibitem[{{Gilbert} {et~al.}(2020){Gilbert}, {Wojno}, {Kirby}, {Escala},
  {Beaton}, {Guhathakurta}, \& {Majewski}}]{Gilbert2020}
{Gilbert}, K.~M., {Wojno}, J., {Kirby}, E.~N., {et~al.} 2020, \aj, 160, 41,
  \dodoi{10.3847/1538-3881/ab9602}

\bibitem[{{Gilbert} {et~al.}(2007){Gilbert}, {Fardal}, {Kalirai},
  {Guhathakurta}, {Geha}, {Isler}, {Majewski}, {Ostheimer}, {Patterson},
  {Reitzel}, {Kirby}, \& {Cooper}}]{Gilbert2007}
{Gilbert}, K.~M., {Fardal}, M., {Kalirai}, J.~S., {et~al.} 2007, \apj, 668,
  245, \dodoi{10.1086/521094}

\bibitem[{{Gilbert} {et~al.}(2009{\natexlab{b}}){Gilbert}, {Guhathakurta},
  {Kollipara}, {Beaton}, {Geha}, {Kalirai}, {Kirby}, {Majewski}, \&
  {Patterson}}]{Gilbert2009b}
{Gilbert}, K.~M., {Guhathakurta}, P., {Kollipara}, P., {et~al.}
  2009{\natexlab{b}}, \apj, 705, 1275, \dodoi{10.1088/0004-637X/705/2/1275}

\bibitem[{{Gilbert} {et~al.}(2014){Gilbert}, {Kalirai}, {Guhathakurta},
  {Beaton}, {Geha}, {Kirby}, {Majewski}, {Patterson}, {Tollerud}, {Bullock},
  {Tanaka}, \& {Chiba}}]{Gilbert2014}
{Gilbert}, K.~M., {Kalirai}, J.~S., {Guhathakurta}, P., {et~al.} 2014, \apj,
  796, 76, \dodoi{10.1088/0004-637X/796/2/76}

\bibitem[{{Guhathakurta} {et~al.}(2006){Guhathakurta}, {Rich}, {Reitzel},
  {Cooper}, {Gilbert}, {Majewski}, {Ostheimer}, {Geha}, {Johnston}, \&
  {Patterson}}]{Guhathakurta2006}
{Guhathakurta}, P., {Rich}, R.~M., {Reitzel}, D.~B., {et~al.} 2006, \aj, 131,
  2497, \dodoi{10.1086/499562}

\bibitem[{{Guy} {et~al.}(2022){Guy}, {Bailey}, {Kremin}, {Alam}, {Allende
  Prieto}, {BenZvi}, {Bolton}, {Brooks}, {Chaussidon}, {Cooper}, {Dawson}, {de
  la Macorra}, {Dey}, {Dey}, {Dhungana}, {Eisenstein}, {Font-Ribera},
  {Forero-Romero}, {Gazta{\~n}aga}, {Gontcho}, {Green}, {Honscheid}, {Ishak},
  {Kehoe}, {Kirkby}, {Kisner}, {Koposov}, {Lan}, {Landriau}, {Le Guillou},
  {Levi}, {Magneville}, {Manser}, {Martini}, {Meisner}, {Miquel}, {Moustakas},
  {Myers}, {Newman}, {Nie}, {Palanque-Delabrouille}, {Percival}, {Poppett},
  {Prada}, {Raichoor}, {Ravoux}, {Ross}, {Schlafly}, {Schlegel}, {Schubnell},
  {Sharples}, {Tarl{\'e}}, {Weaver}, {Y{\`e}che}, {Zhou}, {Zhou}, \&
  {Zou}}]{Guy2022}
{Guy}, J., {Bailey}, S., {Kremin}, A., {et~al.} 2022, arXiv e-prints,
  arXiv:2209.14482.
\newblock \doarXiv{2209.14482}

\bibitem[{{Hammer} {et~al.}(2018){Hammer}, {Yang}, {Wang}, {Ibata}, {Flores},
  \& {Puech}}]{Hammer2018}
{Hammer}, F., {Yang}, Y.~B., {Wang}, J.~L., {et~al.} 2018, \mnras, 475, 2754,
  \dodoi{10.1093/mnras/stx3343}

\bibitem[{{Helmi} {et~al.}(2018){Helmi}, {Babusiaux}, {Koppelman}, {Massari},
  {Veljanoski}, \& {Brown}}]{Helmi2018}
{Helmi}, A., {Babusiaux}, C., {Koppelman}, H.~H., {et~al.} 2018, \nat, 563, 85,
  \dodoi{10.1038/s41586-018-0625-x}

\bibitem[{{Hernquist}(1990)}]{Hernquist1990}
{Hernquist}, L. 1990, \apj, 356, 359, \dodoi{10.1086/168845}

\bibitem[{{Hinton}(2016)}]{Hinton2016}
{Hinton}, S.~R. 2016, The Journal of Open Source Software, 1, 00045,
  \dodoi{10.21105/joss.00045}

\bibitem[{{Huo} {et~al.}(2010){Huo}, {Liu}, {Yuan}, {Zhang}, {Zhao}, {Chen},
  {Bai}, {Zhang}, {Zhang}, {Garc{\'\i}a-Benito}, {Xiang}, {Yan}, {Ren}, {Sun},
  {Zhang}, {Li}, {Lu}, {Wang}, {Ni}, \& {Wang}}]{Huo2010}
{Huo}, Z.-Y., {Liu}, X.-W., {Yuan}, H.-B., {et~al.} 2010, Research in Astronomy
  and Astrophysics, 10, 612, \dodoi{10.1088/1674-4527/10/7/002}

\bibitem[{{Huo} {et~al.}(2013){Huo}, {Liu}, {Xiang}, {Yuan}, {Huang}, {Zhang},
  {Yan}, {Bai}, {Chen}, {Chen}, {Chu}, {Chu}, {Cui}, {Du}, {Hou}, {Hu}, {Hu},
  {Jia}, {Jiang}, {Lei}, {Li}, {Li}, {Li}, {Li}, {Li}, {Li}, {Li}, {Liu},
  {Liu}, {Lu}, {Luo}, {Luo}, {Men}, {Ni}, {Qi}, {Qi}, {Shi}, {Shi}, {Sun},
  {Tang}, {Tian}, {Tu}, {Wang}, {Wang}, {Wang}, {Wang}, {Wang}, {Wang}, {Wang},
  {Wang}, {Wei}, {Wu}, {Xue}, {Yao}, {Yu}, {Yuan}, {Zhai}, {Zhang}, {Zhang},
  {Zhang}, {Zhang}, {Zhang}, {Zhang}, {Zhang}, {Zhao}, {Zhao}, {Zhao}, {Zhou},
  {Zhou}, {Zhu}, \& {Zou}}]{Huo2013}
{Huo}, Z.-Y., {Liu}, X.-W., {Xiang}, M.-S., {et~al.} 2013, \aj, 145, 159,
  \dodoi{10.1088/0004-6256/145/6/159}

\bibitem[{{Huo} {et~al.}(2015){Huo}, {Liu}, {Xiang}, {Shi}, {Yuan}, {Huang},
  {Zhang}, {Hou}, {Wang}, \& {Yang}}]{Huo2015}
---. 2015, Research in Astronomy and Astrophysics, 15, 1438,
  \dodoi{10.1088/1674-4527/15/8/023}

\bibitem[{{Husser} {et~al.}(2013){Husser}, {Wende-von Berg}, {Dreizler},
  {Homeier}, {Reiners}, {Barman}, \& {Hauschildt}}]{Phoenix_2013}
{Husser}, T.~O., {Wende-von Berg}, S., {Dreizler}, S., {et~al.} 2013, \aap,
  553, A6, \dodoi{10.1051/0004-6361/201219058}

\bibitem[{{Ibata} {et~al.}(2004){Ibata}, {Chapman}, {Ferguson}, {Irwin},
  {Lewis}, \& {McConnachie}}]{Ibata2004}
{Ibata}, R., {Chapman}, S., {Ferguson}, A.~M.~N., {et~al.} 2004, \mnras, 351,
  117, \dodoi{10.1111/j.1365-2966.2004.07759.x}

\bibitem[{{Ibata} {et~al.}(2001){Ibata}, {Irwin}, {Lewis}, {Ferguson}, \&
  {Tanvir}}]{Ibata2001}
{Ibata}, R., {Irwin}, M., {Lewis}, G., {Ferguson}, A. M.~N., \& {Tanvir}, N.
  2001, \nat, 412, 49.
\newblock \doarXiv{astro-ph/0107090}

\bibitem[{{Ibata} {et~al.}(2007){Ibata}, {Martin}, {Irwin}, {Chapman},
  {Ferguson}, {Lewis}, \& {McConnachie}}]{Ibata2007}
{Ibata}, R., {Martin}, N.~F., {Irwin}, M., {et~al.} 2007, \apj, 671, 1591,
  \dodoi{10.1086/522574}

\bibitem[{{Ibata} {et~al.}(1994){Ibata}, {Gilmore}, \& {Irwin}}]{Ibata1994}
{Ibata}, R.~A., {Gilmore}, G., \& {Irwin}, M.~J. 1994, \nat, 370, 194,
  \dodoi{10.1038/370194a0}

\bibitem[{{Ibata} {et~al.}(2014){Ibata}, {Lewis}, {McConnachie}, {Martin},
  {Irwin}, {Ferguson}, {Babul}, {Bernard}, {Chapman}, {Collins}, {Fardal},
  {Mackey}, {Navarro}, {Pe{\~n}arrubia}, {Rich}, {Tanvir}, \&
  {Widrow}}]{Ibata2014}
{Ibata}, R.~A., {Lewis}, G.~F., {McConnachie}, A.~W., {et~al.} 2014, \apj, 780,
  128, \dodoi{10.1088/0004-637X/780/2/128}

\bibitem[{{Kafle} {et~al.}(2018){Kafle}, {Sharma}, {Lewis}, {Robotham}, \&
  {Driver}}]{Kafle2018}
{Kafle}, P.~R., {Sharma}, S., {Lewis}, G.~F., {Robotham}, A. S.~G., \&
  {Driver}, S.~P. 2018, \mnras, 475, 4043, \dodoi{10.1093/mnras/sty082}

\bibitem[{{Kalirai} {et~al.}(2006{\natexlab{a}}){Kalirai}, {Guhathakurta},
  {Gilbert}, {Reitzel}, {Majewski}, {Rich}, \& {Cooper}}]{Kalirai2006a}
{Kalirai}, J.~S., {Guhathakurta}, P., {Gilbert}, K.~M., {et~al.}
  2006{\natexlab{a}}, \apj, 641, 268, \dodoi{10.1086/498700}

\bibitem[{{Kalirai} {et~al.}(2006{\natexlab{b}}){Kalirai}, {Gilbert},
  {Guhathakurta}, {Majewski}, {Ostheimer}, {Rich}, {Cooper}, {Reitzel}, \&
  {Patterson}}]{Kalirai2006b}
{Kalirai}, J.~S., {Gilbert}, K.~M., {Guhathakurta}, P., {et~al.}
  2006{\natexlab{b}}, \apj, 648, 389, \dodoi{10.1086/505697}

\bibitem[{{Kirby} {et~al.}(2020){Kirby}, {Gilbert}, {Escala}, {Wojno},
  {Guhathakurta}, {Majewski}, \& {Beaton}}]{Kirby2020}
{Kirby}, E.~N., {Gilbert}, K.~M., {Escala}, I., {et~al.} 2020, \aj, 159, 46,
  \dodoi{10.3847/1538-3881/ab5f0f}

\bibitem[{{Kirihara} {et~al.}(2017){Kirihara}, {Miki}, {Mori}, {Kawaguchi}, \&
  {Rich}}]{Kirihara2017a}
{Kirihara}, T., {Miki}, Y., {Mori}, M., {Kawaguchi}, T., \& {Rich}, R.~M. 2017,
  \mnras, 464, 3509, \dodoi{10.1093/mnras/stw2563}

\bibitem[{{Koch} {et~al.}(2015){Koch}, {Danforth}, {Rich}, {Ibata}, \&
  {Keeney}}]{Koch2015}
{Koch}, A., {Danforth}, C.~W., {Rich}, R.~M., {Ibata}, R., \& {Keeney}, B.~A.
  2015, \apj, 807, 153, \dodoi{10.1088/0004-637X/807/2/153}

\bibitem[{Koposov(2022)}]{sqlutilpy}
Koposov, S. 2022, segasai/sqlutilpy: sqlutilpy v0.19.0, 0.19.0,  Zenodo,
  \dodoi{10.5281/zenodo.6867957}

\bibitem[{{Koposov} \& {Bartunov}(2006)}]{Q3C}
{Koposov}, S., \& {Bartunov}, O. 2006, in Astronomical Society of the Pacific
  Conference Series, Vol. 351, Astronomical Data Analysis Software and Systems
  XV, ed. C.~{Gabriel}, C.~{Arviset}, D.~{Ponz}, \& S.~{Enrique}, 735

\bibitem[{Koposov {et~al.}(2022)Koposov, Speagle, Barbary, Ashton, Bennett,
  Buchner, Scheffler, Cook, Talbot, Guillochon, Cubillos, Ramos, Johnson, Lang,
  Ilya, Dartiailh, Nitz, McCluskey, Archibald, Deil, Foreman-Mackey, Goldstein,
  Tollerud, Leja, Kirk, Pitkin, Sheehan, Cargile, ruskin23, \& Angus}]{dynesty}
Koposov, S., Speagle, J., Barbary, K., {et~al.} 2022, joshspeagle/dynesty:
  v2.0.1, v2.0.1,  Zenodo, \dodoi{10.5281/zenodo.7215695}

\bibitem[{{Koposov}(2019)}]{rvspecfit}
{Koposov}, S.~E. 2019, {RVSpecFit: Radial velocity and stellar atmospheric
  parameter fitting}, Astrophysics Source Code Library, record ascl:1907.013.
\newblock \doeprint{1907.013}

\bibitem[{{Koposov} {et~al.}(2010){Koposov}, {Rix}, \& {Hogg}}]{Koposov2010}
{Koposov}, S.~E., {Rix}, H.-W., \& {Hogg}, D.~W. 2010, \apj, 712, 260,
  \dodoi{10.1088/0004-637X/712/1/260}

\bibitem[{{Koposov} {et~al.}(2011){Koposov}, {Gilmore}, {Walker}, {Belokurov},
  {Evans}, {Fellhauer}, {Gieren}, {Geisler}, {Monaco}, {Norris}, {Okamoto},
  {Pe{\~n}arrubia}, {Wilkinson}, {Wyse}, \& {Zucker}}]{Koposov2011}
{Koposov}, S.~E., {Gilmore}, G., {Walker}, M.~G., {et~al.} 2011, \apj, 736,
  146, \dodoi{10.1088/0004-637X/736/2/146}

\bibitem[{{Kruijssen} {et~al.}(2020){Kruijssen}, {Pfeffer}, {Chevance},
  {Bonaca}, {Trujillo-Gomez}, {Bastian}, {Reina-Campos}, {Crain}, \&
  {Hughes}}]{Kraken2020}
{Kruijssen}, J.~M.~D., {Pfeffer}, J.~L., {Chevance}, M., {et~al.} 2020, \mnras,
  498, 2472, \dodoi{10.1093/mnras/staa2452}

\bibitem[{{Lan} {et~al.}(2022){Lan}, {Tojeiro}, {Armengaud}, {Prochaska},
  {Davis}, {Alexander}, {Raichoor}, {Zhou}, {Yeche}, {Balland}, {BenZvi},
  {Berti}, {Canning}, {Carr}, {Chittenden}, {Cole}, {Cousinou}, {Dawson},
  {Dey}, {Douglass}, {Edge}, {Escoffier}, {Glanville}, {Gontcho}, {Guy},
  {Hahn}, {Howlett}, {Hwang}, {Jiang}, {Kovacs}, {Mezcua}, {Moore}, {Nadathur},
  {Oh}, {Parkinson}, {Rocher}, {Ross}, {Ruhlmann-Kleider}, {Sabiu}, {Said},
  {Saulder}, {Sierra-Porta}, {Weiner}, {Yu}, {Zarrouk}, {Zhang}, {Zou},
  {Ahlen}, {Bailey}, {Brooks}, {Cooper}, {de la Macorra}, {Dey}, {Dhungana},
  {Doel}, {Eftekharzadeh}, {Fanning}, {Font-Ribera}, {Garrison}, {Gaztanaga},
  {Kehoe}, {Kisner}, {Kremin}, {Landriau}, {Le Guillou}, {Levi}, {Magneville},
  {Meisner}, {Miquel}, {Moustakas}, {Myers}, {Newman}, {Nie},
  {Palanque-Delabrouille}, {Percival}, {Poppett}, {Prada}, {Schubnell},
  {Tarle}, {Weaver}, {Zhang}, \& {Zhou}}]{lan22a}
{Lan}, T.-W., {Tojeiro}, R., {Armengaud}, E., {et~al.} 2022, arXiv e-prints,
  arXiv:2208.08516.
\newblock \doarXiv{2208.08516}

\bibitem[{{Lang}(2014)}]{unWISEcoadd_2014}
{Lang}, D. 2014, \aj, 147, 108, \dodoi{10.1088/0004-6256/147/5/108}

\bibitem[{{Lewis} {et~al.}(2015){Lewis}, {Dolphin}, {Dalcanton}, {Weisz},
  {Williams}, {Bell}, {Seth}, {Simones}, {Skillman}, {Choi}, {Fouesneau},
  {Guhathakurta}, {Johnson}, {Kalirai}, {Leroy}, {Monachesi}, {Rix}, \&
  {Schruba}}]{lewis2015}
{Lewis}, A.~R., {Dolphin}, A.~E., {Dalcanton}, J.~J., {et~al.} 2015, \apj, 805,
  183, \dodoi{10.1088/0004-637X/805/2/183}

\bibitem[{{Mackey} {et~al.}(2019{\natexlab{a}}){Mackey}, {Ferguson}, {Huxor},
  {Veljanoski}, {Lewis}, {McConnachie}, {Martin}, {Ibata}, {Irwin},
  {C{\^o}t{\'e}}, {Collins}, {Tanvir}, \& {Bate}}]{mackey2019}
{Mackey}, A.~D., {Ferguson}, A.~M.~N., {Huxor}, A.~P., {et~al.}
  2019{\natexlab{a}}, \mnras, 484, 1756, \dodoi{10.1093/mnras/stz072}

\bibitem[{{Mackey} {et~al.}(2019{\natexlab{b}}){Mackey}, {Lewis}, {Brewer},
  {Ferguson}, {Veljanoski}, {Huxor}, {Collins}, {C{\^o}t{\'e}}, {Ibata},
  {Irwin}, {Martin}, {McConnachie}, {Pe{\~n}arrubia}, {Tanvir}, \&
  {Wan}}]{mackey2019b}
{Mackey}, D., {Lewis}, G.~F., {Brewer}, B.~J., {et~al.} 2019{\natexlab{b}},
  \nat, 574, 69, \dodoi{10.1038/s41586-019-1597-1}

\bibitem[{{Marigo} {et~al.}(2013){Marigo}, {Bressan}, {Nanni}, {Girardi}, \&
  {Pumo}}]{PARSEC2013}
{Marigo}, P., {Bressan}, A., {Nanni}, A., {Girardi}, L., \& {Pumo}, M.~L. 2013,
  \mnras, 434, 488, \dodoi{10.1093/mnras/stt1034}

\bibitem[{{Marocco} {et~al.}(2021){Marocco}, {Eisenhardt}, {Fowler},
  {Kirkpatrick}, {Meisner}, {Schlafly}, {Stanford}, {Garcia}, {Caselden},
  {Cushing}, {Cutri}, {Faherty}, {Gelino}, {Gonzalez}, {Jarrett}, {Koontz},
  {Mainzer}, {Marchese}, {Mobasher}, {Schlegel}, {Stern}, {Teplitz}, \&
  {Wright}}]{catwise2021}
{Marocco}, F., {Eisenhardt}, P. R.~M., {Fowler}, J.~W., {et~al.} 2021, \apjs,
  253, 8, \dodoi{10.3847/1538-4365/abd805}

\bibitem[{{Mart{\'\i}nez-Delgado} {et~al.}(2010){Mart{\'\i}nez-Delgado},
  {Gabany}, {Crawford}, {Zibetti}, {Majewski}, {Rix}, {Fliri},
  {Carballo-Bello}, {Bardalez-Gagliuffi}, {Pe{\~n}arrubia}, {Chonis}, {Madore},
  {Trujillo}, {Schirmer}, \& {McDavid}}]{Martinez-Delgado2010}
{Mart{\'\i}nez-Delgado}, D., {Gabany}, R.~J., {Crawford}, K., {et~al.} 2010,
  \aj, 140, 962, \dodoi{10.1088/0004-6256/140/4/962}

\bibitem[{{Massey} {et~al.}(2019){Massey}, {Neugent}, \&
  {Levesque}}]{massey2019}
{Massey}, P., {Neugent}, K.~F., \& {Levesque}, E.~M. 2019, \aj, 157, 227,
  \dodoi{10.3847/1538-3881/ab1aa1}

\bibitem[{{Matheson} {et~al.}(2021){Matheson}, {Stubens}, {Wolf}, {Lee},
  {Narayan}, {Saha}, {Scott}, {Soraisam}, {Bolton}, {Hauger}, {Silva},
  {Kececioglu}, {Scheidegger}, {Snodgrass}, {Aleo}, {Evans-Jacquez}, {Singh},
  {Wang}, {Yang}, \& {Zhao}}]{Matheson2021}
{Matheson}, T., {Stubens}, C., {Wolf}, N., {et~al.} 2021, \aj, 161, 107,
  \dodoi{10.3847/1538-3881/abd703}

\bibitem[{{McConnachie}(2012)}]{McConnachie2012}
{McConnachie}, A.~W. 2012, \aj, 144, 4, \dodoi{10.1088/0004-6256/144/1/4}

\bibitem[{{McConnachie} {et~al.}(2005){McConnachie}, {Irwin}, {Ferguson},
  {Ibata}, {Lewis}, \& {Tanvir}}]{McConnachie2005}
{McConnachie}, A.~W., {Irwin}, M.~J., {Ferguson}, A.~M.~N., {et~al.} 2005,
  \mnras, 356, 979, \dodoi{10.1111/j.1365-2966.2004.08514.x}

\bibitem[{{McConnachie} {et~al.}(2009){McConnachie}, {Irwin}, {Ibata},
  {Dubinski}, {Widrow}, {Martin}, {C{\^o}t{\'e}}, {Dotter}, {Navarro},
  {Ferguson}, {Puzia}, {Lewis}, {Babul}, {Barmby}, {Bienaym{\'e}}, {Chapman},
  {Cockcroft}, {Collins}, {Fardal}, {Harris}, {Huxor}, {Mackey},
  {Pe{\~n}arrubia}, {Rich}, {Richer}, {Siebert}, {Tanvir}, {Valls-Gabaud}, \&
  {Venn}}]{McConnachie2009}
{McConnachie}, A.~W., {Irwin}, M.~J., {Ibata}, R.~A., {et~al.} 2009, \nat, 461,
  66, \dodoi{10.1038/nature08327}

\bibitem[{{McConnachie} {et~al.}(2018){McConnachie}, {Ibata}, {Martin},
  {Ferguson}, {Collins}, {Gwyn}, {Irwin}, {Lewis}, {Mackey}, {Davidge},
  {Arias}, {Conn}, {C{\^o}t{\'e}}, {Crnojevic}, {Huxor}, {Penarrubia},
  {Spengler}, {Tanvir}, {Valls-Gabaud}, {Babul}, {Barmby}, {Bate}, {Bernard},
  {Chapman}, {Dotter}, {Harris}, {McMonigal}, {Navarro}, {Puzia}, {Rich},
  {Thomas}, \& {Widrow}}]{McConnachie2018}
{McConnachie}, A.~W., {Ibata}, R., {Martin}, N., {et~al.} 2018, \apj, 868, 55,
  \dodoi{10.3847/1538-4357/aae8e7}

\bibitem[{{Meisner} {et~al.}(2019){Meisner}, {Lang}, {Schlafly}, \&
  {Schlegel}}]{unWISE5_2019}
{Meisner}, A.~M., {Lang}, D., {Schlafly}, E.~F., \& {Schlegel}, D.~J. 2019,
  \pasp, 131, 124504, \dodoi{10.1088/1538-3873/ab3df4}

\bibitem[{{Meisner} {et~al.}(2017{\natexlab{a}}){Meisner}, {Lang}, \&
  {Schlegel}}]{unWISE3_2017}
{Meisner}, A.~M., {Lang}, D., \& {Schlegel}, D.~J. 2017{\natexlab{a}}, \aj,
  154, 161, \dodoi{10.3847/1538-3881/aa894e}

\bibitem[{{Meisner} {et~al.}(2017{\natexlab{b}}){Meisner}, {Lang}, \&
  {Schlegel}}]{unWISE1_2017}
---. 2017{\natexlab{b}}, \aj, 153, 38, \dodoi{10.3847/1538-3881/153/1/38}

\bibitem[{{Merrett} {et~al.}(2003){Merrett}, {Kuijken}, {Merrifield},
  {Romanowsky}, {Douglas}, {Napolitano}, {Arnaboldi}, {Capaccioli}, {Freeman},
  {Gerhard}, {Evans}, {Wilkinson}, {Halliday}, {Bridges}, \&
  {Carter}}]{Merrett2003}
{Merrett}, H.~R., {Kuijken}, K., {Merrifield}, M.~R., {et~al.} 2003, \mnras,
  346, L62, \dodoi{10.1111/j.1365-2966.2003.07367.x}

\bibitem[{{Merrett} {et~al.}(2006){Merrett}, {Merrifield}, {Douglas},
  {Kuijken}, {Romanowsky}, {Napolitano}, {Arnaboldi}, {Capaccioli}, {Freeman},
  {Gerhard}, {Coccato}, {Carter}, {Evans}, {Wilkinson}, {Halliday}, \&
  {Bridges}}]{Merrett2006}
{Merrett}, H.~R., {Merrifield}, M.~R., {Douglas}, N.~G., {et~al.} 2006, \mnras,
  369, 120, \dodoi{10.1111/j.1365-2966.2006.10268.x}

\bibitem[{{Merrifield} \& {Kuijken}(1998)}]{Merrifield1998}
{Merrifield}, M.~R., \& {Kuijken}, K. 1998, \mnras, 297, 1292,
  \dodoi{10.1046/j.1365-8711.1998.01625.x}

\bibitem[{{Milo{\v{s}}evi{\'c}} {et~al.}(2022){Milo{\v{s}}evi{\'c}},
  {Mi{\'c}i{\'c}}, \& {Lewis}}]{Milosevic2022}
{Milo{\v{s}}evi{\'c}}, S., {Mi{\'c}i{\'c}}, M., \& {Lewis}, G.~F. 2022, \mnras,
  511, 2868, \dodoi{10.1093/mnras/stac249}

\bibitem[{{Mori} \& {Rich}(2008)}]{Mori2008}
{Mori}, M., \& {Rich}, R.~M. 2008, \apjl, 674, L77, \dodoi{10.1086/529140}

\bibitem[{{Myers} {et~al.}(2022){Myers}, {Moustakas}, {Bailey}, {Weaver},
  {Cooper}, {Forero-Romero}, {Abolfathi}, {Alexander}, {Brooks}, {Chaussidon},
  {Chuang}, {Dawson}, {Dey}, {Dey}, {Dhungana}, {Doel}, {Fanning},
  {Gazta{\~n}aga}, {Gontcho}, {Gonzalez-Morales}, {Hahn}, {Herrera-Alcantar},
  {Honscheid}, {Ishak}, {Karim}, {Kirkby}, {Kisner}, {Kremin}, {Lan},
  {Landriau}, {Levi}, {Magneville}, {Martini}, {Meisner}, {Napolitano},
  {Newman}, {Palanque-Delabrouille}, {Percival}, {Poppett}, {Prada},
  {Raichoor}, {Ross}, {Schlafly}, {Schubnell}, {Tan}, {Tarle}, {Wilson},
  {Y{\`e}che}, {Zhou}, {Zhou}, \& {Zou}}]{myers22a}
{Myers}, A.~D., {Moustakas}, J., {Bailey}, S., {et~al.} 2022, arXiv e-prints,
  arXiv:2208.08518.
\newblock \doarXiv{2208.08518}

\bibitem[{{Naidu} {et~al.}(2020){Naidu}, {Conroy}, {Bonaca}, {Johnson}, {Ting},
  {Caldwell}, {Zaritsky}, \& {Cargile}}]{Naidu_etal_2020}
{Naidu}, R.~P., {Conroy}, C., {Bonaca}, A., {et~al.} 2020, \apj, 901, 48,
  \dodoi{10.3847/1538-4357/abaef4}

\bibitem[{{Navarro} {et~al.}(1997){Navarro}, {Frenk}, \&
  {White}}]{NavarroFrenkWhite1997}
{Navarro}, J.~F., {Frenk}, C.~S., \& {White}, S. D.~M. 1997, \apj, 490, 493,
  \dodoi{10.1086/304888}

\bibitem[{{Pillepich} {et~al.}(2018){Pillepich}, {Springel}, {Nelson}, {Genel},
  {Naiman}, {Pakmor}, {Hernquist}, {Torrey}, {Vogelsberger}, {Weinberger}, \&
  {Marinacci}}]{Pillepich2018}
{Pillepich}, A., {Springel}, V., {Nelson}, D., {et~al.} 2018, \mnras, 473,
  4077, \dodoi{10.1093/mnras/stx2656}

\bibitem[{{Pillepich} {et~al.}(2019){Pillepich}, {Nelson}, {Springel},
  {Pakmor}, {Torrey}, {Weinberger}, {Vogelsberger}, {Marinacci}, {Genel}, {van
  der Wel}, \& {Hernquist}}]{Pillepich2019}
{Pillepich}, A., {Nelson}, D., {Springel}, V., {et~al.} 2019, \mnras, 490,
  3196, \dodoi{10.1093/mnras/stz2338}

\bibitem[{{Plummer}(1911)}]{Plummer1911}
{Plummer}, H.~C. 1911, \mnras, 71, 460, \dodoi{10.1093/mnras/71.5.460}

\bibitem[{{Pop} {et~al.}(2018){Pop}, {Pillepich}, {Amorisco}, \&
  {Hernquist}}]{Pop2018}
{Pop}, A.-R., {Pillepich}, A., {Amorisco}, N.~C., \& {Hernquist}, L. 2018,
  \mnras, 480, 1715, \dodoi{10.1093/mnras/sty1932}

\bibitem[{Price-Whelan {et~al.}(2020)Price-Whelan, Sipőcz, Lenz, Greco,
  Starkman, Foreman-Mackey, Lim, Oh, Koposov, \&
  Major}]{adrian_price_whelan_2020_4159870}
Price-Whelan, A., Sipőcz, B., Lenz, D., {et~al.} 2020, adrn/gala: v1.3, v1.3,
  Zenodo, \dodoi{10.5281/zenodo.4159870}

\bibitem[{Price-Whelan(2017)}]{gala}
Price-Whelan, A.~M. 2017, The Journal of Open Source Software, 2,
  \dodoi{10.21105/joss.00388}

\bibitem[{{Robitaille} {et~al.}(2019){Robitaille}, {Beaumont}, {Qian},
  {Borkin}, \& {Goodman}}]{Robitaille2019}
{Robitaille}, T., {Beaumont}, C., {Qian}, P., {Borkin}, M., \& {Goodman}, A.
  2019, {glueviz v0.15.2: multidimensional data exploration}, 0.15.2, Zenodo,
  Zenodo, \dodoi{10.5281/zenodo.3385920}

\bibitem[{{Ruiz-Lara} {et~al.}(2020){Ruiz-Lara}, {Gallart}, {Bernard}, \&
  {Cassisi}}]{Ruiz-Lara2020}
{Ruiz-Lara}, T., {Gallart}, C., {Bernard}, E.~J., \& {Cassisi}, S. 2020, Nature
  Astronomy, 4, 965, \dodoi{10.1038/s41550-020-1097-0}

\bibitem[{{Sadoun} {et~al.}(2014){Sadoun}, {Mohayaee}, \& {Colin}}]{Sadoun2014}
{Sadoun}, R., {Mohayaee}, R., \& {Colin}, J. 2014, \mnras, 442, 160,
  \dodoi{10.1093/mnras/stu850}

\bibitem[{{Sanders} {et~al.}(2012){Sanders}, {Caldwell}, {McDowell}, \&
  {Harding}}]{Sanders2012}
{Sanders}, N.~E., {Caldwell}, N., {McDowell}, J., \& {Harding}, P. 2012, \apj,
  758, 133, \dodoi{10.1088/0004-637X/758/2/133}

\bibitem[{{Sanderson} \& {Helmi}(2013)}]{Sanderson2013}
{Sanderson}, R.~E., \& {Helmi}, A. 2013, \mnras, 435, 378,
  \dodoi{10.1093/mnras/stt1307}

\bibitem[{{Schlafly} {et~al.}(2019){Schlafly}, {Meisner}, \&
  {Green}}]{unWISE5cat_2019}
{Schlafly}, E.~F., {Meisner}, A.~M., \& {Green}, G.~M. 2019, \apjs, 240, 30,
  \dodoi{10.3847/1538-4365/aafbea}

\bibitem[{{Schlafly} {et~al.}(2022)}]{schlafly22a}
{Schlafly}, E.~F., {et~al.} 2022, in prep.

\bibitem[{{Schlaufman} {et~al.}(2012){Schlaufman}, {Rockosi}, {Lee}, {Beers},
  {Allende Prieto}, {Rashkov}, {Madau}, \& {Bizyaev}}]{schlaufman_etal_2012}
{Schlaufman}, K.~C., {Rockosi}, C.~M., {Lee}, Y.~S., {et~al.} 2012, \apj, 749,
  77, \dodoi{10.1088/0004-637X/749/1/77}

\bibitem[{{Smith} {et~al.}(2015){Smith}, {Flynn}, {Candlish}, {Fellhauer}, \&
  {Gibson}}]{smith2015}
{Smith}, R., {Flynn}, C., {Candlish}, G.~N., {Fellhauer}, M., \& {Gibson},
  B.~K. 2015, \mnras, 448, 2934, \dodoi{10.1093/mnras/stv228}

\bibitem[{{Speagle}(2020)}]{speagle2020}
{Speagle}, J.~S. 2020, \mnras, 493, 3132, \dodoi{10.1093/mnras/staa278}

\bibitem[{{Stern} {et~al.}(2012){Stern}, {Assef}, {Benford}, {Blain}, {Cutri},
  {Dey}, {Eisenhardt}, {Griffith}, {Jarrett}, {Lake}, {Masci}, {Petty},
  {Stanford}, {Tsai}, {Wright}, {Yan}, {Harrison}, \&
  {Madsen}}]{SternWISEQSO2012}
{Stern}, D., {Assef}, R.~J., {Benford}, D.~J., {et~al.} 2012, \apj, 753, 30,
  \dodoi{10.1088/0004-637X/753/1/30}

\bibitem[{{Tanaka} {et~al.}(2010){Tanaka}, {Chiba}, {Komiyama}, {Guhathakurta},
  {Kalirai}, \& {Iye}}]{Tanaka2010}
{Tanaka}, M., {Chiba}, M., {Komiyama}, Y., {et~al.} 2010, \apj, 708, 1168,
  \dodoi{10.1088/0004-637X/708/2/1168}

\bibitem[{{Teuben}(1995)}]{Teuben1995}
{Teuben}, P. 1995, in Astronomical Society of the Pacific Conference Series,
  Vol.~77, Astronomical Data Analysis Software and Systems IV, ed. R.~A.
  {Shaw}, H.~E. {Payne}, \& J.~J.~E. {Hayes}, 398

\bibitem[{{van der Marel} {et~al.}(2012){van der Marel}, {Fardal}, {Besla},
  {Beaton}, {Sohn}, {Anderson}, {Brown}, \& {Guhathakurta}}]{vanderMarel2012}
{van der Marel}, R.~P., {Fardal}, M., {Besla}, G., {et~al.} 2012, \apj, 753, 8,
  \dodoi{10.1088/0004-637X/753/1/8}

\bibitem[{{van der Marel} {et~al.}(2019){van der Marel}, {Fardal}, {Sohn},
  {Patel}, {Besla}, {del Pino}, {Sahlmann}, \& {Watkins}}]{vanderMarel2019}
{van der Marel}, R.~P., {Fardal}, M.~A., {Sohn}, S.~T., {et~al.} 2019, \apj,
  872, 24, \dodoi{10.3847/1538-4357/ab001b}

\bibitem[{{Walterbos} \& {Kennicutt}(1987)}]{WalterbosKennicutt1987}
{Walterbos}, R.~A.~M., \& {Kennicutt}, R.~C., J. 1987, \aaps, 69, 311

\bibitem[{{Wang} {et~al.}(2021){Wang}, {Chen}, \& {Ma}}]{Wang2021}
{Wang}, S., {Chen}, B., \& {Ma}, J. 2021, \aap, 645, A115,
  \dodoi{10.1051/0004-6361/202039531}

\bibitem[{{Wenger} {et~al.}(2000){Wenger}, {Ochsenbein}, {Egret}, {Dubois},
  {Bonnarel}, {Borde}, {Genova}, {Jasniewicz}, {Lalo{\"e}}, {Lesteven}, \&
  {Monier}}]{SIMBAD2000}
{Wenger}, M., {Ochsenbein}, F., {Egret}, D., {et~al.} 2000, \aaps, 143, 9,
  \dodoi{10.1051/aas:2000332}

\bibitem[{{Xiang} \& {Rix}(2022)}]{XiangRix2022}
{Xiang}, M., \& {Rix}, H.-W. 2022, \nat, 603, 599,
  \dodoi{10.1038/s41586-022-04496-5}

\bibitem[{{Yuan} {et~al.}(2010){Yuan}, {Liu}, {Huo}, {Zhang}, {Zhao}, {Chen},
  {Bai}, {Zhang}, {Zhang}, {Garc{\'\i}a-Benito}, {Xiang}, {Yan}, {Ren}, {Sun},
  {Zhang}, {Li}, {Lu}, {Wang}, {Ni}, \& {Wang}}]{Yuan2010}
{Yuan}, H.-B., {Liu}, X.-W., {Huo}, Z.-Y., {et~al.} 2010, Research in Astronomy
  and Astrophysics, 10, 599, \dodoi{10.1088/1674-4527/10/7/001}

\end{thebibliography}
